\documentclass[twocolumn,superscriptaddress,amsmath,amssymb,aps,floatfix]{revtex4-1}

\usepackage{graphicx}
\usepackage{slashed}
\usepackage{epstopdf}
\usepackage{bm}
\usepackage{hyperref}
\usepackage{graphicx,color}
\usepackage{multirow}
\usepackage{amsmath,amstext,array}
\usepackage[normalem]{ulem}
\usepackage{url}

\newcommand{\spinas}{SPINAS}
\newcommand{\pdot}[2]{#1 \! \cdot \! #2}

\begin{document}
\title{\spinas: Spinor Amplitude Subroutines for Constructive Diagram Evaluations}
\author{Neil Christensen}
\email{nchris3@ilstu.edu}
\affiliation{Department of Physics, Illinois State University, Normal, IL 61790}

\date{\today}

\begin{abstract}
\spinas{} is a C++ package created for the implementation and numerical computation of phase-space points of constructive amplitudes in particle physics. This package contains a suite of classes and methods for handling particles, propagators, spinor products, and processes. \spinas{} is structured to offer straightforward usability while ensuring maximum efficiency. This is achieved through a design that emphasizes the storage and reuse of intermediate results within amplitude calculations for each phase-space point. We include a user guide describing how to use the components, a complete example of how to use SPINAS for a scattering amplitude, a discussion of the design and implementation useful for those wishing to contribute, and a discussion of our validation of this package, including both a validation of individual components of the package and a comparison of a complete set of Standard Model processes with Feynman diagrams.
\end{abstract}

\maketitle

\setcounter{tocdepth}{2}
\tableofcontents

\section{Introduction}
Feynman diagrams have been indispensable in the calculation of scattering amplitude and their comparison with experiments.  As the need for amplitudes with a greater number of final-state particles increased, and as computational power increased, computer programs increasingly took on the challenge of generating Feynman diagrams and integrating the resulting amplitudes.  An important early milestone in this process was the creation of HELAS: Helicity Amplitude Subroutines for Feynman Diagram Evaluation \cite{Helas}.  This was a set of Fortran routines for each of the vertices and propagating lines in a Feynman diagram and instructions for combining them to form complete Feynman diagrams and amplitudes.  The HELAS routines form the basis of some modern matrix-element generators such as MadGraph \cite{Alwall:2011uj,deAquino:2011ub}.

Among the strengths of Feynman diagrams is that the algorithm for generating them is completely general and their field theory foundation makes the construction of theories that are local, Lorentz invariant and renormalizable straight forward \cite{Weinberg:1995mt}.  On the other hand, they require the addition of unphysical degrees of freedom for massive spin-1 and massless helicity-$\pm1$ bosons, and an accompanying gauge invariance to insure their cancellation.  As a result, each diagram on its own is not typically physically meaningful.  Only gauge invariant sets of diagrams, with significant cancellations between the members of the set, can be considered as physical.  The number of diagrams in a gauge-invariant set increases exponentially with the number of legs.

Over the last several decades, a new form for scattering amplitudes has been discovered that include simpler objects that do not include unphysical degrees of freedom.  In the early years, this was focused on purely massless theories and made use of ``twistors'' \cite{Gastmans:1990xh,Dixon:1996wi}, culminating in results that were radically simpler than their equivalent Feynman-diagram forms \cite{Parke:1986gb}, sometimes with only one simple expression taking the place of thousands or millions of Feynman diagrams.  Moreover, these simple results did not have a gauge parameter and so each part of the expression was physically meaningful and trivially gauge invariant.  Further, an algorithm was found for these twistor amplitudes for any number of particles, so long as the theories were purely massless \cite{Britto:2005fq}.

Twistors are objects that transform under a product of the Lorentz symmetry and the (little-group) helicity symmetry \cite{Wigner:1939cj,Weinberg:1995mt}, and are therefore only appropriate for massless particles.  We call these helicity spinors.  Further progress was made in extending these massless twistors to objects that transform under a product of the Lorentz symmetry and the (little-group) spin symmetry, which we call spin spinors, allowing them to represent massive particles \cite{Arkani-Hamed:2017jhn}.  This laid the foundation for generating scattering amplitudes that bypass fields and Feynman diagrams.  These amplitudes also do not have any unphysical degrees of freedom and do not need or have a gauge parameter and are therefore trivially gauge invariant.  We call these theories ``constructive'' to distinguish them from field-theory Feynman diagrams.

The 3-point vertices in the constructive Standard Model (SM) were found in \cite{Christensen:2018zcq} and some initial amplitudes were validated against Feynman diagrams \cite{Christensen:2019mch}.  However, a challenge was found when considering amplitudes with massive external particles and massless internal helicity-$\pm1$ particles \cite{Christensen:2022nja}, but it was resolved \cite{Lai:2023upa}.  This opened the floodgate for further amplitude calculations in the SM and their validation against Feynman diagrams.

With these advances in constructive amplitudes, it became clear that a general computational package for calculating phase-space points of constructive amplitudes would be beneficial.   It would allow the comparison with Feynman diagrams to be more efficient than comparing them analytically, as well as allowing numerical comparison when analytic comparison became too complicated.  It could also potentially form the foundation for a new generation of matrix-element generators, based on constructive amplitudes in an analogous way to the HELAS package and Feynman diagrams.  To this end, we created this package, \spinas{}, described in this paper.  This C++ package has a collection of classes and methods to support the easy implementation of constructive amplitudes.  In companions to this paper, the 4-point vertices of the constructive SM were determined \cite{Christensen:2024B}, and a comprehensive set of 4-point amplitudes in the constructive SM, validated against Feynman diagrams, was found \cite{Christensen:2024C}.  Indeed, the implementation of a complete set of constructive 4-point amplitudes and their validation against Feynman diagrams formed a critical part of the validation of this package, as we describe in this paper.

In the rest of this paper, we describe the use of this package.  We begin with a ``user guide'' in Sec.~\ref{sec:user-guide}, where we describe each of the components of the \spinas{} package that are directly used when implementing a new constructive amplitude.  In Sec.~\ref{sec:Complete Example}, we give a complete example of using this package to implement the quantum electrodynamics (QED) process $e,\bar{e}\to\mu,\bar{\mu}$.  In Sec.~\ref{sec:design}, we give further detail about the design of the \spinas{} package that would be useful for a person contributing to this code.  In fact, in Sec.~\ref{sec:design:license}, we note that this package is released under the Gnu Public License (GPL) v3, giving users the right to modify and share this package, under the condition ofthe GPLv3.  Our intention is for the future of this package to be determined and driven by the needs of the community.  Finally, in Sec.~\ref{sec:validation}, we describe the validation of this package through a large number of unit tests and a comparison of constructive amplitudes and Feynman diagrams in a comprehensive set of 4-point amplitudes.

\section{\label{sec:user-guide}User Guide}
In this section, we give the details that are most relevant to a typical novice \spinas{} user.  We will consider support, compilation of the \spinas{} source code, compilation of the user's code, data types in \spinas{}, the \spinas{} namespace, the classes, methods and functions typically used, and some tips for troubleshooting the user's code.  For users that want to dive more deeply into the core code, we refer them to Sec.~\ref{sec:design}.  

We have written and validated this package on Linux and our instructions will be for Linux.  We welcome the communities contribution of support for other operating systems.

\subsection{\label{sec:user-guide:support}Support}
Navigating the complexities of writing, compiling, linking, and running source code can be challenging. While the author aims to provide sufficient documentation for \spinas{}, it is not feasible to personally address every support query. To foster a self-sustaining community, users are encouraged to collaboratively use and contribute to several support tools available on our GitHub site \cite{spinas-github}. To streamline support and enhance collaboration, please direct inquiries to these platforms instead of the author's personal email. The author will participate in these community efforts as time permits, but the collective expertise and collaboration of the \spinas{} community are invaluable.

\paragraph{GitHub Wiki:} The \spinas{} Wiki \cite{spinas-wiki} serves as a dynamic platform for the community to create and share supplementary documentation, including FAQs, tips, and other resources that extend beyond this article. Contributions from users of all experience levels are highly encouraged. Those interested in contributing can request editing access to enrich the Wiki with valuable information and insights.

\paragraph{Discussions:} The \spinas{} Discussions page \cite{spinas-discussions} provides an informal space for users to initiate and engage in conversations on a wide range of topics related to \spinas{}. This forum is ideal for asking and answering questions, seeking and offering help, and sharing experiences. We warmly invite members of the community to not only seek assistance here but to also support others, as even insights from newer users can be highly beneficial.

\paragraph{Issue Tracker:} Bugs and feature requests are managed through the \spinas{} Issue Tracker \cite{spinas-issue-tracker}. This tool is not only for reporting issues but also a platform where community members can actively contribute by addressing bugs and enhancing the software. Users at all skill levels, including those new to coding or \spinas{}, are encouraged to participate, starting with simpler tasks and progressively taking on more complex challenges. More details on contributing to the codebase can be found in Sec.~\ref{sec:design}.

Users are advised to utilize the Discussions tool for preliminary discussion of any issues or feature ideas. This preliminary step often helps clarify whether a situation is indeed a bug or if a proposed feature is necessary, potentially resolved through community dialogue.

To increase the likelihood of receiving support and resolving issues, consider the following when reporting a bug:
\begin{itemize}
    \item Isolate the problem to the smallest possible code snippet that reproduces the issue. Large code submissions without specific focus are less likely to be addressed by the community. Responsibility lies on the reporter to provide a concise, reproducible example.
    \item Ensure your bug report is clear and comprehensive, including sufficient code and a detailed explanation to enable easy replication by others. Vague reports with statements like ``It doesn't work" are less likely to receive attention.
\end{itemize}
Adherence to these guidelines greatly enhances your chances of receiving effective support and responses from the community.

\paragraph{AI:} The ability of large language models (LLMs) is astonishing and only improving, especially in the area of coding.  Although they can't yet do everything we would like, nor are they always correct (so-called halucinations are an issue), they can still be extremely helpful.  We encourage users to carefully take advantage of these tools to support them in the use of this code.

\subsection{\label{sec:user-guide:dependencies}Dependencies}
To ensure full functionality and compatibility of this package, it is essential to satisfy the following dependencies. Installation can typically be achieved through system-specific package managers or with assistance from system administrators.

\paragraph{CMake:} For the potential of cross-platform compilation, we utilize CMake (version 3.5 or higher). CMake can be downloaded from the official website \cite{CMake}. Detailed instructions for installation are available there.

\paragraph{C++ Compiler:} As the package is developed in C++, an appropriate C++ compiler supporting the ISO/IEC 14882:2011 standard (commonly known as C++11) or later is required. Our testing exclusively employed the GNU Compiler Collection \cite{GNU-C}.  However, other compilers  should also be compatible.

\paragraph{Boost Unit Test Framework:} For module testing, we employ the Boost Unit Test framework (version 1.71 or higher) \cite{Boost}. This framework was selected for its extensive features and compatibility with various C++ environments, contributing significantly to the robust testing of our package.

\subsection{\label{sec:user-guide:download}Package Download}

The \spinas{} package is readily available for download at \cite{spinas-website} as a \url{.tar.gz} file.  For those with experience in software development or version control systems, the package's source code is also accessible directly through our GitHub repository \cite{spinas-github}. We actively encourage and welcome contributions from the community, as detailed in Sec.~\ref{sec:design}.

Upon downloading the \spinas{} package, users should move the downloaded file to their preferred working directory. This directory will serve as the primary location for storing and working with the \spinas{} code.

In order to unpack the package, the user should open a terminal and navigate to the directory containing the file and execute \texttt{tar -xzvf spinas-vX.X.X.tar.gz}, where \texttt{X.X.X} represents the version number. This command will extract the package contents into the current directory.

\subsection{\label{sec:user-guide:compiling SPINAS}\spinas{} Compilation}
Upon successfully unpacking the \spinas{} package, the user can initiate the setup by first changing into the \spinas{} root directory.   You can verify you're in the root directory of \spinas{} by listing the contents with \url{ls} and ensuring the presence of files like \url{CMakeLists.txt}, \url{LICENSE}, and \url{README}, along with directories such as \url{include}, \url{SM}, \url{source}, \url{tests}, and \url{user-dir}.

In the root directory of \spinas{}, the user should create a separate \url{build} directory using:
\begin{verbatim}
    mkdir build
    cd build
\end{verbatim}
where the second step changes into the \url{build} directory. This directory will host all the compilation files, keeping the source code intact in the root directory.

The build process is initiated with the command:
\begin{verbatim}
    cmake ..
\end{verbatim}
This step configures the build environment within the \url{build} directory, leaving the root directory unchanged. You may clear the \url{build} directory at any time to reset the compilation process without affecting the source code.

Next, we compile the package using:
\begin{verbatim}
    make [-jN]
\end{verbatim}
The optional \url{-jN} flag enables parallel compilation using N CPUs, enhancing its speed (\url{N} should be replaced with the number of CPUs to use). Absence of this flag defaults to single-CPU compilation. This process compiles the \spinas{} package but, again, does not alter any files in the root directory.

After the compilation is complete and successful, the user should run the Boost Unit Tests by executing:
\begin{verbatim}
    ./test_spinas
\end{verbatim}
This step runs extensive tests on each package component. Occasional errors may arise due to numerical precision limits.  However, if they do not persist upon retesting, the compilation should normally be considered successful.

If the Unit Tests are successful, the user should run the further SM process tests with the command:
\begin{verbatim}
    ./test_SM
\end{verbatim}
This will run a comprehensive set of $2\to2$ processes in the SM, comparing their results with those from Feynman diagrams. Under normal circumstances, all these tests should pass.

With these steps, the \spinas{} package is fully compiled and operational. It is advisable to refrain from modifying files in both the \url{build} and root directories to ensure stability and integrity of the package.

\subsection{\label{sec:user-guide:compilation}Compiling User Code with \spinas{}:}
Before the user writes code for their process, they should create a new directory to contain their source code and binaries outside the \spinas{} root directory, in order to keep the source code clean.  We will call this directory \url{user-dir} and assume that the user has used \url{cd} to get into \url{user-dir}, so that all terminal instructions in this subsection are assumed to be from inside this directory.

In order to write source code for a new process, the user should open a text editor and create a new file in the \url{user-dir}.  We will call this file \url{user-file.cpp}.  Basic no-frills text editors are \url{emacs} or \url{vi} on Linux.  Once \url{user-file.cpp} is open in the text editor, the user can create their code.  A complete example of creating this file as well as compiling and running it can be found in Sec.~\ref{sec:Complete Example}.  Here, we will assume the file is created and saved in the \url{user-dir}.  

Our next step is to compile the new code and can be done with a command such as:
\begin{verbatim}
    g++ -o user-bin user-file.cpp 
        -I/path/to/spinas/include 
        -L/path/to/spinas/build -lspinas 
        -std=c++11 -DWITH_LONG_DOUBLE -O3
\end{verbatim}
all on one line.  \url{g++} refers to the C++ compiler, and it should be the same compiler that was used to compile the \spinas{} package.  The file \url{user-bin} is the name of the binary and can be chosen by the user.  \url{user-file.cpp} is the source file to compile.  \url{/path/to/spinas} should be replaced by the path to the \spinas{} root directory.  \url{-I/path/to/spinas/include} and \url{-L/path/to/spinas/build} give the paths to the \spinas{} include files and the \spinas{} library, respectively.  \url{-lspinas} tells the compiler to link against the \spinas{} library.  \url{-std=c++11} tells the compiler to use the C++11 standard, and is required to make your program consistent with the \spinas{} package.  A later standard can also be used.  The flag \url{-DWITH\_LONG\_DOUBLE} is required to ensure that the precision of the user binary is the same as the \spinas{} package.  The optional flag \url{-O3} tells the compiler to optimize the binary, making it more efficient.  If errors are encountered in the compilation, the user is encouraged to begin with the first error and work their way down.  Sometimes clearing up earlier bugs resolves later ones. 

Once the user's file is compiled, it can be run with a command such as
\begin{verbatim}
    ./user-bin
\end{verbatim}
on Linux.

\subsection{\label{sec:user-guide:data-types}Data Types in \spinas}
Precision consistency is crucial in numerical calculations, particularly in physics simulations. To achieve this in the \spinas{} package, we have introduced two specific data types. The first is \url{ldouble} for real values, and the second is \url{cdouble} for complex values. We strongly encourage users to consistently use these types in their code. This practice not only maintains high precision but also prevents type errors that can arise from mismatched data types.

\paragraph{ldouble:} The \url{ldouble} type is designed for variables holding real numbers, typically floating-point numbers. By default, \url{ldouble} is defined as \url{long} \url{double}, which typically occupies 16 bits on most modern computer architectures. This provides a higher precision compared to standard \url{double}. Here are a couple of usage examples:
\begin{verbatim}
    ldouble MW = 80.385;  
    ldouble three = 3;    
\end{verbatim}
In these examples, \url{MW} is assigned the mass of the W boson in GeV, and \url{three} is a floating-point representation of the integer 3.  They can now be used in amplitude expressions, while maintaining a consistently high level of precision.

\paragraph{cdouble:} For handling complex numbers, the \url{cdouble} type is required. It stores a pair of \url{ldouble} values representing the real and imaginary parts of a complex number. This allows for precise and accurate handling of complex algebra. Here are a few examples:
\begin{verbatim}
    cdouble amplitude = cdouble(0,0);  
    cdouble two = cdouble(2,0);        
    cdouble i = cdouble(0,1);          
\end{verbatim}
In these examples, \url{amplitude} initializes a complex number to zero, \url{two} represents the integer 2, and \url{i} represents the imaginary unit. These complex variables can also be used in amplitude calculations and, as before, the precision is kept consistent and high.

It's also important to note that amplitude calculations in \spinas{} can involve both \url{ldouble} and \url{cdouble} types. Their interactions, such as sums and products, are consistently handled to ensure precision. For example:
\begin{verbatim}
    amplitude = three/(two*MW)*... + ...
\end{verbatim}
This expression combines both types and is perfectly valid in \spinas{}.

However, mixing raw integers or explicit floating-point values directly in expressions can lead to precision issues. Therefore, it's advisable to store such values in variables of type \url{ldouble} or \url{cdouble} before using them in calculations. For instance, the expression \url{3/(2.0*MW)} might be less precise or even cause compilation errors compared to \url{three/(two*MW)}. This approach not only facilitates successful compilation but also ensures the precision and consistency of the calculations.

Finally, as outlined in the previous subsection, users must include the \url{-DWITH_LONG_DOUBLE} flag in their compilation commands. This flag is a default in \spinas{} compilation and ensures that user code aligns with the precision standards set by the package.

\subsection{\label{sec:user-guide:namespace}The \spinas{} Namespace}

To maintain a clear organizational structure and avoid potential naming conflicts with core C++ code or other packages, the \spinas{} package encapsulates its components within the \url{spinas} namespace. This namespace includes all essential classes, functions, and data types that form the backbone of \spinas{}'s functionalities. Utilizing a dedicated namespace is a standard practice in software development to segregate package-specific elements from the global scope.

Each class and function described in this documentation, pertaining to the \spinas{} package, resides within the \url{spinas} namespace. Consequently, their usage in user code necessitates a prefix of \url{spinas::} to denote their affiliation with the \spinas{} package. For instance:
\begin{verbatim}
    #include "spinas.h"
    ...
    spinas::particle p1;
    ...
    p1.set_mass(MW);
\end{verbatim}
In this example, after including the main header file of \spinas{}, a particle object \url{p1} is declared using the \url{particle} class from the \spinas{} package. By prefixing \url{spinas::} to \url{particle}, we explicitly indicate that the class is a member of the \spinas{} namespace, distinguishing it from any similarly named classes in other libraries or the standard C++ library.

It is important to note that this namespacing convention applies only when referencing class or function names from the \spinas{} package. Once an object, such as \url{p1} in the example, is instantiated, it can be used directly without the namespace prefix, as demonstrated in the object's method call \url{p1.set_mass(MW)}. This practice ensures clarity in code while leveraging the organizational benefits of a dedicated namespace.

\subsection{\label{sec:user-guide:particle}Class: particle}
In \spinas{}, the \url{particle} class plays a crucial role in representing external particles involved in physical processes. This includes both incoming and outgoing particles. Each \url{particle} object encapsulates the particle's properties such as its momentum and its representation in spinor form, crucial for calculating spinor products. It is important to create a distinct \url{particle} object for each external particle in a process, even if they represent the same Standard Model (SM) particle. For instance, if multiple electrons are part of the process, each should have its own \url{particle} object.

The constructor of the \url{particle} class requires a single parameter of type \url{ldouble} which represents the mass of the particle. Consider the following code snippet:
\begin{verbatim}
  spinas::particle p1(me);  
  spinas::particle p2(me);  
  spinas::particle p3(mm);  
  spinas::particle p4(mm);  
\end{verbatim}
Here, \url{p1} and \url{p2} are initialized as electrons with mass \url{me}, while \url{p3} and \url{p4} are muons with mass \url{mm}.

To modify the mass of a particle after its instantiation, the \url{set_mass} method is provided. This method also takes a single \url{ldouble} argument, the new mass value. For example:
\begin{verbatim}
    p1.set_mass(me_new);  
    p2.set_mass(me_new);  
\end{verbatim}
Here, \url{me_new} is the updated mass value for the electrons.

Setting the momentum of the particle is essential for computing the amplitude at a specific phase-space point. This is achieved using the \url{set_momentum} method. It requires a reference to a 4-dimensional array of type \url{ldouble}, representing the momentum components. For instance:
\begin{verbatim}
  ldouble mom1[4] = {std::sqrt(me*me+2.0*2.0), 
                     0.0, 0.0, 2.0};
  p1.set_momentum(mom1);  
\end{verbatim}
In this example, \url{mom1} is defined as a 4-dimensional array and initialized with appropriate values, including energy computed using the square root function from the standard C++ library.

Additionally, the \url{particle} class offers the \url{dot} method for calculating the inner product of the momenta of two particles. This method accepts another \url{particle} object as its argument. Therefore, the inner product \( p_1 \cdot p_2 \) can be calculated as follows:
\begin{verbatim}
    p1.dot(p2);  
\end{verbatim}

An essential aspect of utilizing the \url{particle} class is the setting of particle momentum for each phase-space point. This class is designed to automatically handle the computation of associated spinors and the 2x2 complex momentum matrices, which are integral in spinor product calculations. As such, the correct and timely use of the \url{set_momentum} method becomes imperative. It is crucial to invoke this method for every particle at each phase-space point \textit{before} initiating any related calculations.  Failure to update the particle's momentum for a new phase-space point prior to calculations could lead to incorrect results, as the computations would be based on outdated data.

\subsection{\label{sec:user-guide:propagator}Class: propagator}
The \url{propagator} class in \spinas{} is designed to represent particles on internal lines in constructive diagrams. This class is primarily responsible for providing the propagator denominator. A key feature of the \url{propagator} class is that it stores the mass and width of the particle it represents, but it does not internally store the momentum.  Therefore, one of the practical aspects of using the \url{propagator} class is that a single object can be reused for the same particle appearing on multiple internal lines, provided the particle's mass and width remain constant. However, on the other hand, distinct \url{propagator} objects must be created for internal particles with different masses and widths.

The constructor of the \url{propagator} class accepts two parameters: the mass and the width of the particle. For example:
\begin{verbatim}
  spinas::propagator propZ(MZ,WZ);
\end{verbatim}

Should there be a need to modify the mass or width of a propagator, the \url{propagator} class offers the \url{set_mass} and \url{set_width} methods. These methods allow the user to update the mass and width of the particle, respectively:
\begin{verbatim}
  propZ.set_mass(MZ);
  propZ.set_width(WZ);
\end{verbatim}

The core functionality of the \url{propagator} class is encapsulated in the \url{denominator} method. This method calculates the propagator denominator and requires a single argument: a reference to a 4-dimensional array of type \url{ldouble}, representing the momentum of the propagator. Since the momentum of an internal particle is typically a combination of external momenta, it must be computed before invoking \url{denominator}. For instance, if we wanted the propagator denominator
\begin{equation}
    \left[\left(p_1+p_2\right)^2-M_Z^2-i M_Z \Gamma_Z\right],
\end{equation}
we could do:
\begin{verbatim}    
  ldouble propP[4];
  for(int j=0;j<4;j++)
    propP[j] = mom1[j]+mom2[j];
  cdouble pDenS = propZ.denominator(propP);
\end{verbatim}
In this example, \url{propP} is initialized as a 4-dimensional array to store the propagator momentum. It is calculated by combining the relevant components of external momenta. Finally, the \url{denominator} method computes the propagator denominator, returning a complex number of type \url{cdouble}, suitable for subsequent amplitude calculations.

An additional utility of the \url{propagator} class is its application in calculating Mandelstam variables. These variables, such as the $s$-parameter, can be conveniently computed using a \url{propagator} object. To do this, create a \url{propagator} instance with zero mass and width, and then utilize its denominator. For example:
\begin{verbatim}
    spinas:propagator prop0(0,0);
    ...
    cdouble M12 = prop0.denominator(propP);
\end{verbatim}
In this snippet, \url{prop0} is a \url{propagator} object with no mass or width. The \url{...} indicates the process of calculating \url{propP}, as demonstrated in the previous example. The result, \url{M12}, represents the \(s\)-parameter, calculated using the denominator of \url{prop0}.

The choice of naming this parameter \url{M12} is intentional to avoid confusion with the naming conventions typically used for spinor products. For example, a name like \url{s12} would closely resemble \url{s12s}, a common notation for the spinor product $\lbrack\mathbf{12}\rbrack$. While these names can be chosen at the user's discretion, we recommend this naming scheme for its clarity and to prevent potential ambiguities in complex calculations.

\subsection{\label{sec:user-guide:sproduct}Class: sproduct}
In constructive field theories, spinor products are essential for calculating amplitudes. To facilitate this, the \url{sproduct} class in \spinas{} is designed to store references to particles involved in a spinor product and to compute these products efficiently. This class manages the Lorentz indices, ensuring correct matching. Once a \url{sproduct} object is constructed, the user needs to update it only when the phase-space point changes, and then use its methods to derive the spinor product for the desired spin configuration.

Constructing an object of the \url{sproduct} class involves passing multiple arguments. The first argument specifies the type of the left spinor, either an angle spinor (\url{ANGLE}) or a square spinor (\url{SQUARE}). The second argument is a reference to the particle corresponding to this left spinor. Subsequent arguments can include up to six intermediate momenta, represented by references to their respective particles, sandwiched between the left and right spinors. The final argument is a reference to the particle associated with the right spinor. For example, to define spinor products such as $\langle\mathbf{13}\rangle$, $\lbrack\mathbf{2}4\rbrack$, $\langle4\mathbf{2}\rangle$, $\langle\mathbf{1}\lvert p_4\rvert\mathbf{2}\rbrack$, and $\lbrack4\lvert p_1p_2\rvert4\rbrack$, the declarations in the header file might look like:
\begin{verbatim}
  spinas::sproduct a13a, s24s, a42a, 
                   a142s, s4124s;
\end{verbatim}
These objects can then be instantiated in the process constructor as follows:
\begin{verbatim}
a13a=spinas::sproduct(ANGLE,&p1,&p3);
s24s=spinas::sproduct(SQUARE,&p2,&p4);
s42s=spinas::sproduct(ANGLE,&p4,&p2);
a142s=spinas::sproduct(ANGLE,&p1,&p4,&p2);
s4124s=spinas::sproduct(SQUARE,&p4,&p1,&p2,&p4);
\end{verbatim}
In this context, specifying whether the spinors are helicity or spin spinors is unnecessary, as this information is deduced from the masses of the particles, which are inherent to the particle objects. Additionally, the type of the right-end spinor (angle or square) is inferred based on the left spinor type and the number of intermediate momenta. It is also important to note that by default, the spinors are assumed to have upper spin indices if they are massive. This assumption aligns with the conventional treatment of uncontracted spin indices in amplitude formulas. The method to specify lower indices for spin contractions will be covered later in this subsection.

Moreover, the \url{sproduct} objects incorporate references to the \url{particle} objects, which inherently include their momenta. Consequently, it is unnecessary to explicitly supply phase-space points to the \url{sproduct} instances; this information is implicitly derived from the associated particle objects. However, it is important to note that the \url{sproduct} class retains the calculations performed for previous phase-space points, to enhance efficiency of the amplitude calculation. Therefore, in order to update these calculations for new phase-space configurations, the \url{update} method must be used. This update should occur \textit{after} setting the new momenta for the particles, but \textit{before} calculating the amplitude expression. For instance, when a new phase-space point is selected, the user should first update the momenta of all particles and then refresh all the spinor products as follows:
\begin{verbatim}
  p1.set_momentum(mom1);
  p2.set_momentum(mom2);
  p3.set_momentum(mom3);
  p4.set_momentum(mom4);
  a13a.update();
  s24s.update();
  a42a.update();
  a142s.update();
  s4124s.update();
\end{verbatim}

For the practical application of updated spinor products in amplitude calculations, the \url{sproduct} class provides the \url{v} method, which stands for ``value". This naming choice is intended to keep amplitude expressions compact, thereby enhancing their readability and simplifying debugging of the amplitude expression. The \url{v} method returns a complex number of type \url{cdouble} and is designed to be directly integrated into amplitude expressions. Furthermore, this method supports four different spin combinations, adapting to the nature of the spinors involved. If both spinors represent massless particles and are thus helicity spinors without spin indices, the \url{v} method requires no arguments. For example, in a scenario where particle 4 is massless, the spinor product $\lbrack4\lvert p_1p_2\rvert4\rbrack$ would be integrated into an amplitude expression as follows:
\begin{verbatim}
  amp =  ...*s4124s.v()*...
\end{verbatim}

When dealing with spinor products involving both massive and massless spinors, the \url{sproduct} class adapts to represent their distinct spin properties. In such cases, where one spinor is massive (with a spin index) and the other is massless (without a spin index), the spinor product has only one spin index. Consequently, the \url{v} method in this scenario takes a single integer argument. This argument represents twice the spin of the massive particle's spinor and accepts only two values: $+1$ for spin $+\frac{1}{2}$ and $-1$ for spin $-\frac{1}{2}$. No other values are permissible. It's important to note that particles with higher spin require symmetric combinations of spinors, which will be discussed further in Sec.~\ref{sec:user-guide:process}. The user does not need to specify which spinor the spin index refers to; this is automatically determined based on the particles' properties, with the spin index pertaining to the massive particle's spinor. For instance, if we aim to compute the spin-up component for particle 2 in the spinor products $\lbrack\mathbf{2}4\rbrack$ and $\langle4\mathbf{2}\rangle$, we might define \url{ds2=1} (double the spin of particle 2) and use it as follows:
\begin{verbatim}
  amp = ...*s24s.v(ds2)*... 
        + ...*a42a.v(ds2)*...
\end{verbatim}
Here, because particle $4$ is massless, the \url{sproduct} object interprets the spin argument appropriately for each term, recognizing it as pertaining to the massive spinor in the product (the left spinor in the first term and the rightspinor in the second term).

In cases where both spinors are massive and thus each has a spin index, the \url{v} method requires two arguments. Each argument corresponds to twice the spin of each respective spinor. For example, let's say we want to evaluate spinor products with a specific spin combination defined by \url{ds1=1}, \url{ds2=-1}, and \url{ds3=-1}. The spinor products $\langle\mathbf{13}\rangle$ and $\langle\mathbf{1}\lvert p_4\rvert\mathbf{2}\rbrack$ in an amplitude expression would be represented as:
\begin{verbatim}
  amp = ...*a13a.v(ds1,ds3)*...
        + ...*a142s.v(ds1,ds2)*...
\end{verbatim}
In this context, the user must input the spin arguments in the order corresponding to the spinors in the product. The design of the \url{sproduct} class and the convention for naming spinor-product variables and spin indices are intended to assist in maintaining this order, thus ensuring clarity and accuracy in the representation of spinor products in amplitude calculations.

In most situations, the constructors and methods previously described for the \url{sproduct} class will suffice. Typically, spin contractions should simplified during the preliminary stages of amplitude calculation, meaning that by the time one reaches the phase-space point calculations, no spin contractions should remain. However, there is an exception to this rule, particularly relevant in scenarios involving production processes with spin-correlated decays. In such cases, it becomes necessary to contract the spins of the final states in the production process with the spins of the initial states in the decays. This requirement implies that one spin index should be upper, as has been the default assumption so far, and the other should be lower. To accommodate this, the \url{sproduct} class includes constructors that allow specifying the position of the spin index.

It is essential to note that when the position of a spin index is not explicitly stated, it is conventionally assumed to be upper. Therefore, the constructor examples provided earlier, which do not specify the spin index position, are all assumed to have upper indices. However, if a scenario requires the left spinor to have a lower index, the argument \url{LOWER} is added immediately after the particle reference for the left spinor. For instance:
\begin{verbatim}
a13a=spinas::sproduct(ANGLE,&p1,LOWER,&p3);
s24s=spinas::sproduct(SQUARE,&p2,&p4);
a142s=spinas::sproduct(ANGLE,&p1,LOWER,&p4,&p2);
\end{verbatim}
In these examples, the left spinor is set to have a lower index in the first and third cases (\url{a13a} and \url{a142s}), while it remains upper in the second case (\url{s24s}). For all three examples, the right spinor is assumed to be upper since its position was not specified.

When the intention is to designate the right spinor as lower, the \url{LOWER} keyword should be placed after the reference to the right spinor. For example:
\begin{verbatim}
a13a=spinas::sproduct(ANGLE,&p1,&p3);
s42s=spinas::sproduct(ANGLE,&p4,&p2,LOWER);
a142s=spinas::sproduct(ANGLE,&p1,&p4,&p2,LOWER);
\end{verbatim}
In these examples, the right spin index is set to lower for the second and third constructs (\url{s42s} and \url{a142s}), while it remains upper for the first one (\url{a13a}). The left spinor maintains an upper index in all three cases.

Additionally, it is possible to configure both the left and right spinors as lower. This is achieved by adding the \url{LOWER} argument after both particle references. For instance:
\begin{verbatim}
a13a=spinas::sproduct(ANGLE,&p1,LOWER,
                      &p3,LOWER);
a142s=spinas::sproduct(ANGLE,&p1,LOWER,&p4,
                      &p2,LOWER);
\end{verbatim}
In this configuration, both sides of the spinor products are set to have lower indices.

It is important to note that the specification of the spin index (upper or lower) is exclusively determined in the constructor of the \url{sproduct} class. All other methods, including \url{update()} and \url{v(...)}, function identically regardless of this specification. The \url{update()} method does not require any arguments, while the \url{v(...)} method accepts between zero and two arguments, corresponding to twice the spin of the spinors. The distinction between upper and lower indices is established solely by the constructor, streamlining the process and maintaining consistency across different methods of the class.

\subsection{\label{sec:user-guide:process}Class: process}
The \url{process} class is a fundamental component of \spinas{}, designed to underpin process calculations. Its structure allows for expansion with additional methods in future updates. The primary function of this class is to act as a base class, enabling users to define their specific process classes by inheriting from \url{process}. For instance, in simulating the process \( e,\bar{e} \rightarrow \mu,\bar{\mu} \), a class named \url{eemm} could be declared to inherit from \url{process}:
\begin{verbatim}
  class eemm : public spinas::process {
    ...
  };
\end{verbatim}
Inheriting from the \url{process} class endows \url{eemm} with a pre-defined set of methods, augmenting the user’s custom functionalities.

The methods in \url{process} currently fall into two categories. The first includes methods for handling spin combinations for particles of spin 1, such as the $W$ and $Z$ bosons. The second category consists of methods for validating squared amplitude calculations against established Feynman-diagram results, specifically in $2\to2$ processes. These methods are crucial for ensuring the accuracy and reliability of user-generated calculations.

\subsubsection{\label{sec:user-guide:process:higher spins}Higher-Spin Spinors}
In amplitude calculations involving spin-1 massive bosons, such as the \( W \) and \( Z \) bosons, each boson is represented by a pair of spin-\(\frac{1}{2}\) spinors. While manually incorporating the necessary loops and normalization factors is feasible for a single, or even two, spin-1 bosons, this task becomes  more challenging as their number increases. To aid in this process, \spinas{} provides several functions designed to streamline these calculations.

Understanding these functions requires revisiting how we represent the spin states of spin-1 particles through the combination of two spin-$\frac{1}{2}$ spinors. The three spin states are constructed as follows:  
\begin{align}
    \lvert1,+1\rangle &= \lvert\frac{1}{2},+\frac{1}{2}\rangle\lvert\frac{1}{2},+\frac{1}{2}\rangle 
    \label{eq:user-guide:1,1}\\
    \lvert1,0\rangle &= \frac{1}{\sqrt{2}}\left(
    \lvert\frac{1}{2},+\frac{1}{2}\rangle\lvert\frac{1}{2},-\frac{1}{2}\rangle
    + \lvert\frac{1}{2},-\frac{1}{2}\rangle\lvert\frac{1}{2},+\frac{1}{2}\rangle
    \right) 
    \label{eq:user-guide:1,0}\\
    \lvert1,-1\rangle &= \lvert\frac{1}{2},-\frac{1}{2}\rangle\lvert\frac{1}{2},-\frac{1}{2}\rangle
    \label{eq:user-guide:1,-1}
\end{align}
Depending on the spin state, different approaches are required. For instance, a loop over combinations is needed for the \(0\) spin state [as in Eq.~(\ref{eq:user-guide:1,0})], while it is not required for the \(+1\) and \(-1\) states [as in Eqs.~(\ref{eq:user-guide:1,1}) and (\ref{eq:user-guide:1,-1})]. Normalization factors, such as \(1/\sqrt{2}\) in Eq.~(\ref{eq:user-guide:1,0}), might also be necessary. Additionally, integers representing the double spin indices for each spinor are required to obtain symmetric combinations as in Eq.~(\ref{eq:user-guide:1,0}).

To facilitate amplitude calculations with spin-1 massive bosons, \spinas{} provides several methods. The first method, \url{get_num_spin_loops}, determines the number of iterations required in order to include all the spinor combinations. The number of arguments is equal to the number of spin-1 particles in the process, with each argument being twice the spin component of the spin-1 particle (values of $-2, 0$, or $+2$). For instance, if particle 3 is the sole massive spin-1 boson in the process, the number of required spin combinations can be obtained as follows:
\begin{verbatim}
  int nCombinations = get_num_spin_loops(ds3);
\end{verbatim}
Here, \url{ds3} represents double the spin component of particle 3 ($-2,0$, or $+2$). For a process involving two massive spin-1 bosons, such as particles 2 and 4, the method is called with both their spin components:
\begin{verbatim}
  int nCombinations = 
         get_num_spin_loops(ds2,ds4);
\end{verbatim}
The return value of this method varies depending on the spin components. For example, it returns $1$ for \url{ds2=+2} and \url{ds4=-2}, $2$ for \url{ds2=0} and \url{ds4=-2}, and $4$ for \url{ds2=0} and \url{ds4=0}. As another example, if particles 1,3 and 4 were massive spin-1 bosons, 
\begin{verbatim}
  int nCombinations = 
         get_num_spin_loops(ds1,ds3,ds4);
\end{verbatim}
and so on.  This method supports up to six massive spin-1 bosons. For processes involving more than six bosons, a method is provided taking an array of spins and a length, as in:
\begin{verbatim} 
  int dsList[5] = {ds1,ds3,ds4,ds6,ds8};
  int dsLength = 5;
  int nCombinations = 
         get_num_spin_loops(dsList, dsLength);
\end{verbatim}
In this example, particles 1, 3, 4, 6 and 8 are spin-1 bosons, with double spins \url{ds1, ds3, ds4, ds6} and \url{ds8}.

The second method, \url{get_spin_normalization}, calculates the normalization factor for the spinor combination. It takes the same arguments as \url{get_num_spin_loops} and returns a value of type \url{ldouble}. For a process with two massive spin-1 bosons, the normalization factor is determined as follows:
\begin{verbatim}
  ldouble normFactor=
         get_spin_normalization(ds2,ds4);
\end{verbatim}
Using the same examples as before, this function would return $1$ for \url{ds2=+2} and \url{ds4=-2}, $1/\sqrt{2}$ for \url{ds2=0} and \url{ds4=-2}, and $1/2$ for \url{ds2=0} and \url{ds4=0}.  Once again, this method works for up to six spin-1 bosons.  If more than six external particles are spin 1, a general form is supplied taking the same form as for \url{get_num_spin_loops}.  For example:

\begin{verbatim}
  ldouble normFactor=
     get_spin_normalization(dsList, dsLength);
\end{verbatim}

In scenarios involving multiple iterations through spin loops, with massive spin-1 bosons, it is essential to determine the specific spins for each spinor. Consider a situation where particles 1, 3, and 4 are massive spin-1 bosons. We would require pairs of integer variables to represent the double spins of the spinors for each boson, as shown below:
\begin{verbatim}
  int ds1a, ds1b, ds3a, ds3b, ds4a, ds4b;
\end{verbatim}
These variables, which can take values of either $-1$ or $+1$, will be assigned based on the current iteration of the spin loop. For instance, the assignment process for three such spin-1 particles can be implemented as follows:
\begin{verbatim}
  for(int i=0;i<nCombinations;i++){
    get_spinor_spins(ds1,ds1a,ds1b, 
             ds3,ds3a,ds3b, ds4,ds4a,ds4b, i);
      ...
  }
\end{verbatim}
In this method, the arguments include the double spin component (values of $-2, 0$, or $+2$) for each spin-1 boson and the corresponding pairs of double spins for the spinors. The iteration variable, \url{i} in this case, is also passed. This method assigns appropriate values to the double spins of the spinors (\url{ds1a}, \url{ds1b}, \url{ds3a}, \url{ds3b}, \url{ds4a}, and \url{ds4b}) based on the spin states of the bosons and the current iteration, without modifying the double spins of the bosons (\url{ds1}, \url{ds3}, \url{ds4}) or the iteration variable.

With these methods, all possible spin combinations are considered, and normalization factors are correctly applied. For example, the amplitude calculation within the loop might include expressions like:
\begin{verbatim}
  amp += normFactor*s13s.v(ds1a,ds3a)*
      a34a.v(ds3b,ds4a)*s124a.v(ds1b,ds4b)*...
\end{verbatim}
Here, \url{amp} is initialized to zero before the loop and accumulates the amplitude contribution from each iteration. A detailed example involving one massive spin-1 boson can be found in Sec.~\ref{sec:Complete Example:Spin-1}, with additional examples for varying numbers of spin-1 bosons available in the \url{SM} directory, as discussed in Sec.~\ref{sec:user-guide:SM}.

Once again, this method is defined for up to six spin-1 bosons. If a greater number of spin-1 bosons are present, the general method should be used.  For example:
\begin{verbatim}
  for(int i=0;i<nCombinations;i++){
    get_spinor_spins(dsList, dsaList, dsbList, 
             dsLength, i);
      ...
  }    
\end{verbatim}
Normally, we anticipate that these general methods for more than six spin-1 bosons would not be necessary for hand-written code.  On the other hand, computer-generated code would likely use them.

The methodology outlined above is tailored specifically for massive spin-1 bosons and is not required for massless helicity-$\pm1$ bosons, such as photons or gluons. These particles lack a spin index, as each helicity state is directly represented by a product of two helicity spinors. Consequently, the spinor products for these particles do not possess a spin index. Detailed examples that showcase the implementation of amplitudes involving external photons and gluons are available in the \url{SM} directory of the \spinas{} package, with further discussion in Sec.~\ref{sec:user-guide:SM}.

It's important to note that, although the \spinas{} package can be used for higher-spin particles, the current methods in the \url{process} class do not extend to massive particles with spins higher than 1.  For these particles, the user will currently have to determine the number of loops, normalization factor and double-spin indices on their own.  Expansion of these capabilities to encompass higher-spin particles is an area open for future development by the scientific community. Researchers interested in contributing to this aspect can find foundational information and guidance in Sec.~\ref{sec:design:process} and the source code for this class, which lays out the groundwork necessary for such an extension.

\subsubsection{\label{sec:user-guide:process:testing}Testing $2\to2$ Processes}
The \spinas{} package supports processes of any multiplicity.  However, this initial version is focused on successfully achieving and testing $2\to2$ processes against Feynman rules.  For this reason, we currently have a set of functions that support the comparison of constructive amplitudes against Feynman rules for $2\to2$ processes.  We intend to add support for testing of higher-multiplicity amplitudes in future versions, including a more general testing framework, and we invite collaboration with the community as we work towards this goal.  We are also open to the community adding other useful testing methods.

In the current version, we have four methods, namely: \url{test_2to2_amp2}, \url{test_2to2_amp2_rotations}, \url{test_2to2_amp2_boosts}, and \url{test_2to2_amp2_boosts_and_rotations}.  All four functions perform a series of tests of the squared amplitude and return the number of tests where the discrepancy is greater than the allowed threshold.  The tests are considered successful if $0$ is returned.  If any comparisons with Feynman diagrams are greater than the threshold, these functions also print a message with details about the comparison.  The arguments for all four methods are exactly the same.  We will first compare the methods and when to use them.  Afterwards, we will describe the arguments of these methods.

The fundamental testing method is \url{test_2to2_amp2}, which evaluates the squared amplitude in the center of momentum (CM) frame across 20 evenly distributed phase-space points. Specifically, it considers scenarios where particle 1 travels along the \(z\)-direction, testing polar-angle points \(\cos\theta_3 \in \{-0.95, -0.85, ..., 0.85, 0.95\}\) with a fixed azimuthal angle \(\phi_3 = 0\). Further details on the method's arguments will be discussed later. Generally, this method suffices for assessing squared amplitudes, summed over spins. It has been observed that when this method yields successful results, the subsequent methods typically also succeed, and vice versa. However, there is an exception to this, which we will address later in this subsection. Users are advised to first achieve agreement with this fundamental test before proceeding to the additional methods.

Lorentz invariance is a fundamental property of the squared amplitude and, therefore, it must remain constant under both rotations and boosts, after summing over spins. To rigorously test this invariance in $2 \to 2$ processes, \spinas{} includes three additional methods beyond the base \url{test_2to2_amp2}.

The method \url{test_2to2_amp2_rotations} extends the base test by applying random rotations. Starting from the same equally spaced phase-space points, it generates a random spatial rotation for each point and applies it to the momenta of all four particles ($p_1, p_2, p_3$, and $p_4$). The squared amplitude is then recalculated and compared against Feynman diagram results. This comparison is performed for 10 random rotations at each initial polar angle.

For testing boosts, the method \url{test_2to2_amp2_boosts} is employed. Similar to the rotation test, it begins with the phase-space points from the base test but introduces random Lorentz boosts to the CM-frame momenta. The method tests the invariance of the squared amplitude under these boosts, again comparing the results with those from Feynman diagrams for 10 random boosts per polar angle. Before moving on, it is crucial to note that in certain high-energy processes with significant diagram cancellations at high energies, such as $W,\bar{W} \to W,\bar{W}$, numerical precision limits can cause slight differences in this and the following test, despite analytical verifications of the cancellations \cite{Christensen:2024B}.  This issue should be considered when evaluating the test results that include random boosts.

The final method, \url{test_2to2_amp2_boosts_and_rotations}, combines both rotations and boosts. For each polar angle, it iterates through 10 sets of random rotations and boosts applied to the CM-frame momenta. The resulting squared amplitudes are then compared to the expected results from Feynman diagrams, testing the comprehensive Lorentz invariance of the amplitude.

It is important to emphasize that for these tests to yield meaningful results, the particle spins must be summed over in the squared amplitude calculations. Also, these tests serve dual purposes: they validate the user's implementation of amplitudes using the \spinas{} package and simultaneously verify the package's underlying algorithms. The success of these tests underpins our confidence in the correctness and reliability of \spinas{}, as further elaborated in Sec.~\ref{sec:validation:2to2}.

The testing methods in \spinas{} for \(2 \to 2\) processes, as previously mentioned, share the same set of arguments. The first argument is a lambda function, which references the user's squared amplitude function. This approach allows flexibility in testing different forms of the squared amplitude. For instance, in processes involving massless helicity-\(\pm1\) particles like photons or gluons, users might want to test the squared amplitude both after and before summing over the particle's helicity. The lambda function enables specifying the desired form of the squared amplitude for each test.

Consider the process $ e,\gamma^{\pm} \to e,h $ as an example. The squared amplitude's Lorentz invariance holds irrespective of whether the photon's helicity is summed over. Consequently, it is beneficial to test both scenarios – with and without helicity summation. This can be implemented as follows:
\begin{verbatim}
  test_2to2_amp2(
    [&](){ return eAehAmp.amp2(); }, ...);
  test_2to2_amp2(
    [&](){ return eAehAmp.amp2_Aplus(); }, ...);
\end{verbatim}
In this case, the user's process class (e.g., \url{eAehAmp}) should have methods like \url{amp2} and \url{amp2_Aplus}, with the former summing over helicities and the latter focusing on a specific helicity.  

The subsequent arguments for these testing methods include the masses of the particles, the spatial momentum of the incoming particle in the CM frame, and the Feynman-diagram data for comparison. For example:
\begin{verbatim}
  test_2to2_amp2(
    ...,  me,0,me,mh,pspatial,dataFD);
  test_2to2_amp2(
    ...,  me,0,me,mh,pspatial,dataFD_Aplus);
\end{verbatim}
The momenta for these tests are determined by the methods and given by:
\begin{align}
    p_1^\mu &= \left(E_1,0,0,p_{in}\right)
    \label{eq:user-guide:p1 from pin}\\
    p_2^\mu &= \left(E_2,0,0,-p_{in}\right)
    \label{eq:user-guide:p2 from pin}\\
    p_3^\mu &= \left(E_3,p_{out}\sin(\theta),0,p_{out}\cos(\theta)\right)
    \label{eq:user-guide:p3 from pin}\\
    p_4^\mu &=
    \left(E_4,-p_{out}\sin(\theta),0,-p_{out}\cos(\theta)\right)
    \label{eq:user-guide:p4 from pin}
\end{align}
The energy \(E_i\) is determined as $E_i = \sqrt{m_i^2 + \lvert\vec{p}_i\rvert^2}$. The incoming momentum is determined by the argument of \url{test_2to2_amp2} that we called \url{pspatial} in the example above.  That is, $p_{in}$ is equal to \url{pspatial}. The outgoing momentum \(p_{out}\) is calculated using the formula:
\begin{equation}
    p_{out} = \frac{1}{2\sqrt{s}}\sqrt{[s-(m_3+m_4)^2][s-(m_3-m_4)^2]},
    \label{eq:user-guide:pout from pin}
\end{equation}
where $s = (p_1 + p_2)^2$. Each phase-space point is defined by $p_{in}$ and the polar angle, with twenty polar angles chosen as $\cos(\theta) \in \{-0.95, -0.85, ..., 0.85, 0.95\}$. 

The Feynman-rule data arrays \url{dataFD} and \url{dataFD_Aplus} are 20-dimensional arrays of type \url{ldouble}, containing values corresponding to these angles.  For this example, they look like:
\begin{verbatim}
  ldouble dataFD[20] = {2.022591814703232E-15,
         ...,1.486097950409267E-11};
\end{verbatim}
and should contain high-precision, in order for the tests to be successful.  
For this particlar process, it turns out that \url{dataFD_Aplus} has the same values as \url{dataFD} because the squared amplitude is the same for both helicities and is averaged over in the case where we sum over the helicities.  In other cases, this may not be the case.  For further examples, see the process files in the \url{SM} directory.  

It is important to note that, in order to achieve agreement between the squared amplitudes from Feynman diagrams and constructive calculations, the width must treated exactly the same.  This could be non-trivial since the width is treated in a program-specific way.  For example, in CalcHEP \cite{CalcHEP}, by default, the width is turned off in t- and u-channel diagrams and only kept near the resonance in s-channel diagrams.  This is to reduce violations of gauge invariance.  For our $2\to2$ process tests, we have found it best to set the widths to zero for all diagrams and to avoid the propagators going on shell when choosing phase-space points.  

It is also important to ensure the amplitude calculations have other factors treated the same.  For example, if testing the squared amplitude, whether the spins and colors are summed/averaged must be treated the same as also the symmetry factor.   Furthermore, exactly the same coupling constants must be used.  Any running of the couplings must be taken into account.

\subsection{\label{sec:user-guide:SM}SM Directory}
The \url{SM} directory within \spinas{} provides the implementation of a complete set of $2 \to 2$ processes in the Standard Model (SM). This directory serves as a valuable resource for users encountering difficulties in implementing their processes, offering a range of similar examples for reference. Additionally, these examples can be utilized as templates for developing processes that extend beyond the Standard Model (BSM). In such cases, users are advised to create copies of these files in a separate directory for modification, to maintain the integrity of the original examples.

Many of the 4-point amplitudes are related to each other by crossing symmetry and are grouped into crossing-symmetry groups of amplitudes.  From each of these groups, we have included at least two processes related by crossing symmetry.  A list of all these processes can be found in Table~\ref{tab:user-guide:SM Processes}.  
\begin{table*}
  \begin{center}
    \renewcommand{\arraystretch}{1.25}
    \setlength{\tabcolsep}{10pt}
    \begin{tabular}{|lll|}
       \hline \multicolumn{3}{|l|}{$q,\bar{q}\to q,\bar{q}$ \ Neutral } \\\hline
       $u,\bar{u}\to c,\bar{c}$ & $u,c\to u,c$ &\\
       $u,\bar{u}\to u,\bar{u}$ & $u,u\to u,u$ &\\
       $u,\bar{u}\to s,\bar{s}$ & $u,s\to u,s$ &\\
       $u,\bar{u}\to d,\bar{d}$ & $u,d\to u,d$ &\\
       $d,\bar{d}\to s,\bar{s}$ & $d,s\to d,s$ &\\
       $d,\bar{d}\to d,\bar{d}$ & $d,d\to d,d$ &\\
       \hline \multicolumn{3}{|l|}{$q,\bar{q}\to l,\bar{l}$ \ Neutral }\\\hline
       $u,\bar{u}\to e,\bar{e}$ & $u,e\to u,e$ &\\
       $u,\bar{u}\to \nu_e,\bar{\nu}_e$ & $u, \nu_\mu\to \nu_\mu,u$ &\\
       $d,\bar{d}\to e,\bar{e}$ & $d,e\to e,d$ &\\
       $d,\bar{d}\to \nu_e,\bar{\nu}_e$ & $d,\nu_\mu\to \nu_\mu,d$ &\\
       \hline \multicolumn{3}{|l|}{$l,\bar{l}\to l,\bar{l}$ \ Neutral }\\\hline
       $e,\bar{e}\to\mu,\bar{\mu}$ & $e,\mu\to e,\mu$ &\\
       $e,\bar{e}\to e,\bar{e}$ & $e,e\to e,e$ &\\
       $e,\bar{e}\to \nu_\mu,\bar{\nu}_\mu$ & $e,\nu_\mu\to e,\nu_\mu$ &\\
       $e,\bar{e}\to \nu_e,\bar{\nu}_e$ & $e,\nu_e\to\nu_e,e$ &\\
       $\nu_e,\bar{\nu}_e\to\nu_\mu,\bar{\nu}_\mu$ & $\nu_e,\nu_\mu\to\nu_e,\nu_\mu$ &\\
       $\nu_e,\bar{\nu}_e\to\nu_e,\bar{\nu}_e$ & $\nu_e,\nu_e\to\nu_e,\nu_e$ &\\
       \hline \multicolumn{3}{|l|}{$f,\bar{f}\to f,\bar{f}$ \ Charged }\\\hline
       $u,\bar{d}\to t,\bar{b}$ & $u,b\to d,t$ &\\
       $u,\bar{d}\to\nu_\tau,\bar{\tau}$ & $u,\tau\to\nu_\tau,d$ &\\
       $\mu,\bar{e}\to\bar{\nu}_e,\nu_\mu$ & $\mu,\nu_e\to e,\nu_\mu$ &\\
       \hline \multicolumn{3}{|l|}{$l,\bar{l}\to b,\bar{b}$ \ Neutral }\\\hline
       $e,\bar{e}\to h,h$ & $e,h\to e,h$ &\\
       $e,\bar{e}\to\gamma,h$ & $e,\gamma\to e,h$ &\\
       $e,\bar{e}\to Z,h$ & $e,Z\to e,h$ &\\
       $e,\bar{e}\to\gamma,\gamma$ & $e,\gamma\to\gamma,e$ & $\gamma,\gamma\to e,\bar{e}$ \\
       $\gamma,Z\to\bar{e},e$ & $\gamma,e\to Z,e$ &\\
       $e,\bar{e}\to Z,Z$ & $e,Z\to Z,e$ &\\
       $e,\bar{e}\to W,\bar{W}$ & $e,W\to W,e$ &\\
       $\nu_e,\bar{\nu}_e\to Z,h$ & $\nu_e,Z\to\nu_e,h$ &\\
       $\nu_e,\bar{\nu}_e\to Z,Z$ & $\nu_e,Z\to Z,\nu_e$ &\\
       $\nu_e,\bar{\nu}_e\to W,\bar{W}$ & $\nu_e,W\to W,\nu_e$ &\\
       \hline \multicolumn{3}{|l|}{$q,\bar{q}\to b,\bar{b}$ \ Neutral without $g$}\\\hline
       $u,\bar{u}\to h,h$ & $u,h\to u,h$ &\\
       $u,\bar{u}\to\gamma,h$ & $u,\gamma\to u,h$  &\\
       $u,\bar{u}\to Z,h$ & $u,Z\to u,h$ &\\
       $u,\bar{u}\to\gamma,\gamma$ & $\gamma,\gamma\to u,\bar{u}$ & $u,\gamma\to\gamma,u$ \\
       $\gamma,Z\to u,\bar{u}$ & $\gamma,u\to Z,u$ & \\
       \hline
    \end{tabular}
    \begin{tabular}{|lll|}
       \hline \multicolumn{3}{|l|}{$q,\bar{q}\to b,\bar{b}$ \ Neutral without $g$ Continued }\\\hline
       $u,\bar{u}\to Z,Z$ & $u,Z\to Z,u$ & \\
       $u,\bar{u}\to W,\bar{W}$ & $u,W\to W,u$ & \\
       $d,\bar{d}\to h,h$ & $d,h\to d,h$ & \\
       $d,\bar{d}\to\gamma,h$ & $d,\gamma\to d,h$ & \\
       $d,\bar{d}\to Z,h$ & $d,Z\to d,h$ & \\
       $d,\bar{d}\to\gamma,\gamma$ & $\gamma,\gamma\to d,\bar{d}$ & $d,\gamma\to \gamma,d$ \\
       $\gamma,Z \to d,\bar{d}$ & $\gamma,d\to Z,d$ & \\
       $d,\bar{d}\to Z,Z$ & $d,Z\to Z,d$ & \\
       $\bar{d},d\to W,\bar{W}$ & $W,d\to W,d$ & \\
       \hline \multicolumn{3}{|l|}{$q,\bar{q}\to b,\bar{b}$ \ Neutral with $g$}\\\hline
       $u,g\to u,h$ & $h,g\to u,\bar{u}$ & \\
       $g,Z\to u,\bar{u}$ & $g,u\to Z,u$ & \\
       $g,\gamma\to u,\bar{u}$ & $g,u\to \gamma,u$ & \\
       $g,g\to u,\bar{u}$ & $g,u\to g,u$ & \\
       $d,g\to d,h$ & $h,g\to d,\bar{d}$ & \\
       $g,Z\to d,\bar{d}$ & $g,d\to Z,d$ & \\
       $g,\gamma\to d,\bar{d}$ & $d,\gamma\to g,d$ & \\
       $g,g\to d,\bar{d}$ & $\bar{d},g\to g,\bar{d}$ & \\
       \hline \multicolumn{3}{|l|}{$f,\bar{f}\to b,\bar{b}$ \ Charged} \\\hline
       $\bar{e},\nu_e\to W,h$ & $h,\nu_e\to W,e$ & \\
       $\bar{e},\nu_e\to \gamma,W$ & $\gamma,\nu_e\to W,e$ & \\
       $\bar{e},\nu_e\to Z,W$ & $Z,\nu_e\to W,e$ & \\
       $u,\bar{d}\to W,h$ & $u,h\to W,d$ & \\
       $\bar{u},d\to \gamma,\bar{W}$ & $\gamma,W\to u,\bar{d}$ & \\
       $d,W\to g,u$ & $g,W\to u,\bar{d}$ & \\
       $\bar{u},d\to Z,\bar{W}$ & $W,d\to Z,u$ & \\
       \hline \multicolumn{3}{|l|}{$b,\bar{b}\to b,\bar{b}$}  \\\hline
       $h,h\to h,h$ & & \\
       $h,h\to Z,Z$ & $h,Z\to Z,h$ & \\
       $h,h\to W,\bar{W}$ & $h,W\to W,h$ & \\
       $\gamma,h\to W,\bar{W}$ & $\gamma,W\to W,h$ & \\
       $Z,h\to W,\bar{W}$ & $Z,W\to W,h$ & \\
       $\gamma,\gamma\to W,\bar{W}$ & $\gamma,W\to\gamma,W$ & \\
       $\gamma,Z\to W,\bar{W}$ & $\gamma,W\to Z,W$ & \\
       $Z,Z\to Z,Z$ & & \\
       $Z,Z\to W,\bar{W}$ & $Z,W\to Z,W$ & \\
       $W,W\to W,W$ & $W,\bar{W}\to W,\bar{W}$ & \\
       $g,g\to g,g$ & & \\
       &&\\
       \hline
    \end{tabular}
  \end{center}
  \caption{This is a list of all the SM processes implemented and found in the \url{SM} directory.  Each line contains the implemented processes related by crossing symmetry.}
  \label{tab:user-guide:SM Processes}
\end{table*}
For each crossing-symmetry group, we have included notes in one of the process files, describing how its amplitude is obtained from the other in the group.  We have also included notes in many of these processes describing how the $2\to2$ process with two outgoing particles is obtained from the all-ingoing process amplitude.  These methods can also be found in \cite{Christensen:2024B,Christensen:2024C}.

Our processes are grouped in the table with all four-fermion processes, followed by two-fermion and then all-boson processes.  The fermions in our processes only include separate generations when doing so includes distinct diagrams not present with another generation.  Thus, we have $e,\bar{e}\to e,\bar{e}$ in addition to $e,\bar{e}\to \mu,\bar{\mu}$ because the first has diagrams that are not present in the second.  However, we did not include $e,\bar{e}\to \tau,\bar{\tau}$ because it was exactly the same as the process $e,\bar{e}\to \mu,\bar{\mu}$, except for a change to a mass.  Any $2\to2$ process in the SM, no matter what generation or crossing form, should be obtainable from these with a simple set of changes: a rearrangement of particles, a switch between ingoing and outgoing momenta, and a change in masses and possibly the addition of CKM elements if desired.  This should serve as a solid foundation for users creating their own processes.

This set of processes is also a foundational part of our validation, as described in Sec.~\ref{sec:validation:2to2}.

\section{\label{sec:Complete Example}Complete Example of \spinas{} Use}

In this section, we give a complete, detailed walk through of implementing the Quantum Electrodynamics (QED) process $e,\bar{e}\to\mu,\bar{\mu}$, focusing on the photon contribution for simplicity.  At the end, we will give some details about implementing an amplitude with massive spin-1 particles.  This is designed to be a practical guide for users crafting their own process files.  Tackling more complex process amplitudes can pose significant challenges, particularly for newcomers, with the accurate alignment of relative signs and factors being a common stumbling block. To assist users in this endeavor, we have compiled a comprehensive set of Standard Model (SM) processes and put them in the SM directory. These can serve as valuable references for further exploration. By understanding the concepts presented in this and the preceding section, and after successfully compiling and running the examples, users will be well-equipped to comprehend the code used in other SM processes. Our aim is to lay a robust foundation for the effective utilization of this package.

\subsection{Setting Up the Files and Directories}
Before beginning, the user should create a directory for their code.  In this example, we will call this directory \url{user-dir/} and we will assume that all files, compilations and running of the binary are done within this directory.  We recommend against adding the user's file directly in one of the \spinas{} directories to prevent potential conflicts and ensure easier updates and maintenance of the \spinas{} code base.  Inside this directory, the user would usually also create a subdirectory for their header file.  We will call this directory \url{user-dir/include/}.  For this example, we will create two files.  The first is the header file, which we will call \url{eemm.h}, and we will put this in the \url{user-dir/include/} directory.  The second is the source code file for this process.  We will call this \url{eemm.cpp} and put it in the \url{user-dir/} directory.  

\subsection{The Header File: eemm.h}
The header file is where we declare the classes, their variables and methods and any other user-defined functions.  We will call our class \url{eemm} and it can be declared with:
\begin{verbatim}
class eemm : public spinas::process {
    //class variables
    //class methods    
};
\end{verbatim}
\texttt{: public spinas::process} declares it to be a subclass of \url{spinas::process}, giving it some useful built-in methods, such as normalization routines when including massive spin-1 bosons in the incoming and/or outgoing states (see Sec.~\ref{sec:user-guide:process} for further details), and some built-in tests to compare your squared amplitude with the results from Feynman diagrams when the process is $2\to2$.  We will include an example of using the built-in tests for this process later in this section.  We note the use of \url{spinas::} to indicate that \url{process} is inside the \url{spinas} namespace.  We leave it to expert users to understand how and when to modify this.  For this example, we will always include \url{spinas::} when referring to \spinas{} classes and methods. 

The variables for this class should include all the coupling constants, masses, particle variables, propagators, sproducts and any other variables the user needs to implement their process.  For example, for this process, we include:
\begin{verbatim}
private:
    ldouble e;//Electric Charge
    ldouble me,mm;//Mass of e and mu
    spinas::particle p1,p2,p3,p4;
    spinas::propagator prop;
    cdouble pDenS;
    spinas::sproduct a13a, s13s, a14a, s14s, 
             a23a, s23s, a24a, s24s;
\end{verbatim}
We recommend to declare all class variables to be private and emphasize the importance of encapsulation in object-oriented programming, which enhances the security and robustness of the code by preventing external access to internal states.
We declare the following set of variables for this class.  \url{e} is the electric charge of the positron, \url{me} is the mass of the electron and \url{mm} is the mass of the muon.  Note that we always use the built-in type \url{ldouble} when declaring real variables to avoid type issues in the amplitude expressions.  Next, we declare a variable for each of the external particles.  We prefer to call these particles \texttt{p1, p2, p3} and \texttt{p4}, where the \url{p} stands for particle and the integer represents which particle they are in the process.  These are declared to be of type \url{particle}, the built-in class for particles in \spinas, which contains all the methods required for calculating spinors and momenta for the particles.  

Our next line contains \url{prop} declared as type \url{propagator}.  In this case, we only have one propagator, but in other processes, we need to declare separate objects of this type for each internal particle.  Note that if the same particle is present on multiple internal lines, only one propagator object needs to be declared for each particle, but can be reused for each internal line containing that particle.  Although not required, we often find it convenient to declare a variable of type \url{cdouble} to store the complex value of the propagator for a particular phase-space point.  If the same particle is present in multiple different internal lines, unique variables of this type will be necessary, one for each line.  Just as we always use \url{ldouble} for real variables, we always use \url{cdouble} for complex variables to ensure no type issues when writing the amplitude expressions.  

Finally, we declare all the spinor products we will need for this process in the line \texttt{spinas::sproduct a13a, s13s, ... s24s;}.  \texttt{sproduct} is the class that contains the particles in a product of spinors and momenta.  We need a separate object of this type for each spinor product appearing in our amplitude.  Although the naming can be anything the user likes, we find it convenient to add an \texttt{a} at the beginning and/or end of the object's name if the spinors at those ends are angle spinors and an \texttt{s} if they are square spinors.  The integers represent the external particles for the spinors on the left and right end and, if there are any momenta in between the spinors, the integers in the middle would represent those momenta.  For example, for $\langle\mathbf{13}\rangle$, we create the object \texttt{a13a} in this example.  If we required the spinor product $\langle\mathbf{1}\lvert p_4\rvert\mathbf{3}\rbrack$, we would use the object name \texttt{a143s}, for convenience and clarity of notation.  

We next move on to declare the methods, beginning with the constructor for this process.  The arguments of this constructor should include the values of coupling constants and masses and anything else about this process that does not depend on the phase-space point.  In this example, it is just the electric charge and the masses.  We can do this with a declaration such as
\begin{verbatim}
public:
    eemm(const ldouble& echarge, 
            const ldouble& masse, 
            const ldouble& massmu);
\end{verbatim}
Notice that, in the first line, we declare these methods to be public.  This allows them to be used outside the object.  
Adding \texttt{\&} to the end of \texttt{ldouble} is not required but allows the value to be passed more efficiently to the method.  Using the keyword \texttt{const} is also not required, but should be used along with pass by reference to remove the possibility that this class change the values of the variables being passed.  We will use this method of passing arguments throughout this example.  

We sometimes want to calculate the same amplitude for different mass values.  In order to do this, we can create the method
\begin{verbatim}
    void set_masses(const ldouble& masse, 
                    const ldouble& massmu);
\end{verbatim}
In order to calculate different phase-space points, we must have a method that updates the momenta of all the particles, propagators and spinor products.  Therefore, we must declare a method along the lines of
\begin{verbatim}
    void set_momenta(const ldouble mom1[4], 
                     const ldouble mom2[4], 
                     const ldouble mom3[4], 
                     const ldouble mom4[4]);
\end{verbatim}
In this example, the momenta of the external particles (the phase-space point) are passed as separate 4-dimensional arrays of type \texttt{ldouble}.  The name of this method and the way the phase-space point is passed in is up to the user.  However, if the user would like to use the built-in tests for $2\to2$ processes, this is the declaration that is expected.  Moreover, the user is free to overload this method or define different methods of passing the phase-space point, as long as they are consistent and as long as they faithfully update all the particles, propagators and spinor products appropriately.   This method, whatever its declaration, must be called every time the phase-space point is changed.  

Typically, the user would declare a method to calculate and return the amplitude for a particular spin combination, for example
\begin{verbatim}
    cdouble amp(const int& ds1, const int& ds2, 
                const int& ds3, const int& ds4);
\end{verbatim}
where \texttt{ds1, ds2, ds3} and \texttt{ds4} are double the spins of particles $1,2,3$ and $4$.  The spins are doubled so that integers can be used, removing issues with the way fractions are stored as real variables.  The name of this method is free for the user to choose but the return value should be of type \texttt{cdouble}, since the amplitude is complex.

In addition to this, we often want the squared amplitude, so can declare
\begin{verbatim}
    ldouble amp2();
\end{verbatim}
This method does not need arguments because it will iterate through the spins of the squared amplitude and return the sum.  Since squaring removes the complex nature, this method returns type \texttt{ldouble}.  As before, this naming scheme is convenient, but not required.  

This is all that is required in the header, but the user can supplement this with class methods that are useful for their particular calculation.  

Finally, rigorously testing your process is critical. We suggest creating a function outside the class declaration. This external test simulates how an external function would interact with your class, ensuring that your implementation behaves as expected when integrated into larger applications.
So, after all the class methods are declared, and the closing curly brace of the class is closed, the user can declare a function such as
\begin{verbatim}
class eemm : public spinas::process {
    //class variables
    //class methods    
};
int test_eemm();
\end{verbatim}
The user can name this test function as they like.  It will create objects of class \texttt{eemm} and run its methods.  In particular, it is important to test the amplitude against known values.  Both the arguments and the return value are up to the user.  We will give an example using this declaration below.

\subsection{The Source File: eemm.cpp}
Our next task is to create the source file, where we define all the constructors, methods and functions.  The first few lines of the source file are the include statements:
\begin{verbatim}
#include <iostream>
#include "spinas.h"
#include "include/eemm.h"
\end{verbatim}
We have included \texttt{iostream} so that our test function, \texttt{test\_eemm}, can output messages to the user.  \texttt{spinas.h} is the header file that includes all the \spinas{} declarations and \texttt{eemm.h} is the header file that we created in the previous subsection.  Any other headers that the user requires can be included here.

Our next block of code defines the constructor for this class and looks like
\begin{verbatim}
eemm::eemm(const ldouble& echarge, 
           const ldouble& masse, 
           const ldouble& massmu){
  e=echarge;
  me=masse;
  mm=massmu;
  prop=spinas::propagator(0,0);
  p1=spinas::particle(me);
  p2=spinas::particle(me);
  p3=spinas::particle(mm);
  p4=spinas::particle(mm);
  a13a=spinas::sproduct(ANGLE,&p1,&p3);
  s13s=spinas::sproduct(SQUARE,&p1,&p3);
  a14a=spinas::sproduct(ANGLE,&p1,&p4);
  s14s=spinas::sproduct(SQUARE,&p1,&p4);
  a23a=spinas::sproduct(ANGLE,&p2,&p3);
  s23s=spinas::sproduct(SQUARE,&p2,&p3);
  a24a=spinas::sproduct(ANGLE,&p2,&p4);
  s24s=spinas::sproduct(SQUARE,&p2,&p4);
}
\end{verbatim}
After the initial line, which we recognize from the declaration in the header, we set the values of the class variables \texttt{e}, \texttt{me} and \texttt{mm} to the values given by the user when initializing this class.  The class object \texttt{prop} is set equal to an object of type \texttt{spinas::propagator} with the arguments \texttt{0,0} since the internal photon has no mass and no width.  The four particles in this class are set equal to objects of type \texttt{spinas::particle}, each with its mass.  The spinor products \texttt{a13a} through \texttt{s24s} are set equal to objects of type \texttt{spinas::sproduct}.  The first argument of the \texttt{sproduct} constructor is whether the spinor on the left end is of type \texttt{ANGLE} or \texttt{SQUARE}.  We do not specify the type of the spinor on the right end because it is inferred from the left end and the number of particles in the spinor product.  In this case, since all the spinor products have an even number of particles, the spinor at the right end is of the same type.  Note that we do not need to specify whether the spinors are massless helicity spinors or massive spin spinors since this is inferred from the properties of the particle.  The user is free to perform other calculations here as necessary.  

Our next block of code is a method to change the masses of the particles and looks like
\begin{verbatim}
void eemm::set_masses(const ldouble& masse, 
                      const ldouble& massmu){
  me=masse;
  mm=massmu;
  p1.set_mass(me);
  p2.set_mass(me);
  p3.set_mass(mm);
  p4.set_mass(mm);
}
\end{verbatim} 
Once again, we need to include \texttt{eemm::} before the name to specify the class that this method belongs to.  The first couple of lines set the masses to the arguments of the function.  This is followed calling the \texttt{set\_mass} method of each of the particles.  If the internal particle were massive and its mass was being changed, we would also have to set the mass of the propagator.  Since in this case, the internal particle is a photon, that is unnecessary here.  Note that we do not need to do anything to the spinor products here because their mass is automatically inferred from the properties of these particles.  Once again, this can be adjusted according to the needs of the user as long as the particles and the propagators are updated appropriately.

The next block of code updates the momenta for the process.
\begin{verbatim}
void eemm::set_momenta(const ldouble mom1[4], 
                       const ldouble mom2[4], 
                       const ldouble mom3[4], 
                       const ldouble mom4[4]){
  //Particles
  p1.set_momentum(mom1);
  p2.set_momentum(mom2);
  p3.set_momentum(mom3);
  p4.set_momentum(mom4);
  a13a.update();
  s13s.update();
  a14a.update();
  s14s.update();
  a23a.update();
  s23s.update();
  a24a.update();
  s24s.update();
  //Propagator Momentum
  ldouble propP[4];
  for(int j=0;j<4;j++)
    propP[j] = mom1[j]+mom2[j];
  pDenS = prop.denominator(propP);
}
\end{verbatim}
We first see that the \texttt{set\_momentum} method must be called for each particle.  Following this, all the spinor products must be reset with the new momenta.  To do this, we call the \texttt{update} method for each spinor product.  This is important because the \texttt{sproduct} objects remember their values for different spins in order to speed up the calculation.  Calling the \texttt{update} method resets all the values so that they must be recalculated for the new momenta.  Following this, we update the propagator denominator.  To do this, we calculate the momentum of the internal line.  In this case, because it is an S-channel diagram, we find the sum of the momenta for particles $1$ and $2$.  We then set \texttt{pDenS} equal to \texttt{prop.denominator(propP)}.  This uses the propagator object that we declared in the header \texttt{prop} and initialized in the constructor above with no mass and no width.  We use its \texttt{denominator} method to calculate the propagator denominator with these values.  This method must be called every time the phase-space point is changed.  The user is free to change this method's name and the arguments and method that the phase-space point is communicated to it.  However, if the user would like to use the built-in testing functions for $2\to2$ processes, this is the form that is expected.  

Our next method calculates the amplitude and returns it.
\begin{verbatim}
cdouble eemm::amp(const int& ds1, 
            const int& ds2, const int& ds3, 
            const int& ds4){
  //-2e^2(<13>[24]+<14>[23]+[13]<24>+[14]<23>)
  constexpr ldouble two = 2.0;
  return -two*e*e*(
            a13a.v(ds1,ds3)*s24s.v(ds2,ds4) 
            + a14a.v(ds1,ds4)*s23s.v(ds2,ds3) 
            + s13s.v(ds1,ds3)*a24a.v(ds2,ds4) 
            + s14s.v(ds1,ds4)*a23a.v(ds2,ds3)
            )/pDenS;
}
\end{verbatim}
In the first line, we have a comment with the amplitude in human-readable form.  This is, of course, not required, but is convenient for checking the expression against the code.  The next line initializes a constant of type \texttt{ldouble}.  This is to avoid errors associated with algebra that includes different types and is recommended.  After the electric charge squared is included, we have the sum of spinor products.  Because all the spinors are massive and have two spins, we see that the \texttt{v} method for each \texttt{sproduct} is called with two arguments.  These values are always twice the spin of the spin spinor, namely either \texttt{+1} or \texttt{-1}.  These are the only values allowed.  (For a massive particle of spin $1$, there will always be two spinors in each term for that particle and their spins will add to give the spin of the higher-spin particle.  We will see an example of this in Sec.~\ref{sec:Complete Example:Spin-1}.)  The method is called \texttt{v} for ``value" and was purposely made as short as possible to enhance the readability of these expressions.  We can see that twice the spins of particles, \texttt{ds1} through \texttt{ds4}, appears once in each term and in the appropriate position in the appropriate spinor.  This must be done carefully in order to achieve correct results.  Further, although the overall sign does not matter, the relative signs between terms and any relative factors certainly do matter and must be accounted for correctly.  Once we are done with the numerator, we divide by the propagator denominator.  Before moving on, we note again that the \texttt{set\_momenta} method must be called before this \texttt{amp} method is called any time the phase-space point changes.  

We note that the \texttt{sproduct} class is designed to reduce the number of times a spinor product is calculated, thereby increasing the efficiency of the calculation.  In particular, if the same \texttt{sproduct} is called again with the same spins, and the momenta have not been changed, it will simply return the previously calculated value.  It will not calculate it again.  For example, in the first term, we have \texttt{a13a.v(ds1,ds3)*s24s.v(ds2,ds4)}.  Suppose, after calling \texttt{set\_momenta} with a new phase-space point, we first call \texttt{amp} with the values $-1, -1, -1, -1$ for the double spins \texttt{ds1, ds2, ds3} and \texttt{ds4}.  On this pass, both \texttt{a13a.v(ds1,ds3)} and \texttt{s24s.v(ds2,ds4)} will be calculated and stored before returning their values.  However, suppose that later the values $-1, -1, -1, +1$ are called (without calling \texttt{set\_momenta} again).  In this case, \url{a13a.v(ds1,ds3)} will simply return the previously calculated value since it has seen these double spins before.  \url{s24s.v(ds2,ds4)}, on the other hand, will calculate the spinor product fresh since it hasn't seen these double spin values yet.  Each \texttt{sproduct} object will store the value for all four spin combinations (for two massive spinors) and reuse them until its \texttt{update} method is called (in the \texttt{set\_momenta} method). 

Although some users may only need the amplitude, very often, we would like to square the amplitude and sum (average) it over the spins.  For this, we create
\begin{verbatim}
ldouble eemm::amp2(){
  constexpr ldouble four = 4.0;
  ldouble amp2 = 0;
  cdouble M;
  
  //Sum over spins
  for(int ds1=-1;ds1<=1;ds1+=2)
    for(int ds2=-1;ds2<=1;ds2+=2)
      for(int ds3=-1;ds3<=1;ds3+=2)
	        for(int ds4=-1;ds4<=1;ds4+=2){
	           M = amp(ds1,ds2,ds3,ds4);
	           amp2 += std::pow(std::abs(M),2);
	    }
  //Average over initial spins 1/2*1/2 = 1/4
  return amp2/four;
}
\end{verbatim}
We begin by creating a constant variable \texttt{four}, a variable \texttt{amp2} of type \texttt{ldouble} to store the accumulated squared amplitude and a variable \texttt{M} of type \texttt{cdouble} to store the amplitude for a particular spin combination.  We next set up loops for each double spin which, in this case, switches between \texttt{-1} and \texttt{+1}.  For each double-spin combination, the \texttt{amp} method that we just defined is called and stored in \texttt{M}.  In the following line, its absolute value is taken and squared and then added to \texttt{amp2}.  Once the loop is complete, we return it divided by $4$ since we are averaging over the initial spins.  As usual, the name and other properties can be chosen by the user, but if they want to use the built-in testing features, this is the form expected.  In the case that different diagrams have different QCD color combinations, they will have to be calculated separately, squared separately and added with their appropriate color factors.  Interested users should consult one of the QCD examples in the \texttt{SM} directory.  

This is all that is required to calculate both the amplitude and the squared amplitude.  However, it is essential that the code be tested.  For this, we add one more function that runs over multiple phase-space points and compares the result with known values.  For example, for this process we could create
\begin{verbatim}
int test_eemm(){
  int n=0;
  std::cout<<"\t* e , E  -> m , M  (QED)\n";
  {
    int i=0;
    ldouble e=0.31333, me=0.0005, mmu=0.105;
    eemm eemmAmp = eemm(e,me,mmu);
    ldouble pspatial=250;
    ldouble data[20] = {...};
    i += eemmAmp.test_2to2_amp2(
        [&](){return eemmAmp.amp2(); }
        , me,me,mmu,mmu,pspatial,data);
    ...
    n+=i;
  }
  
  return n;
}
\end{verbatim}
We first note that this time we did not use \url{eemm::} at the beginning of this function name.  That is because this is not a method of the \url{eemm} class.  It is a function outside of this class, intended to call this class in the same way as an external function in a larger calculation.  In the next line, we create a variable of type \url{int} to store the number of times our tests fail.  This is not necessary, but we find it helpful.  We next print a message to standard out, telling the user that we are testing the process \url{e , E -> m , M (QED)}.  This is not required and it can be modified as convenient.  This line is the reason we included the \url{iostream} header at the beginning of this file.  After creating a new block of code with the curly braces \texttt{\{} and \texttt{\}} for convenience, we create a new integer called \url{i}
 to keep track of the number of failed tests for this mass combination.  We then create variables for the charge and masses and use them to initialize a new object of type \url{eemm} that we call \url{eemmAmp}.  We note that we cannot use the same name for the object as for the class.  We also note that these variables \url{e, me} and \url{mmu} are not the same as the variables inside our object \url{eemmAmp}, so changing these variables will not automatically change the charge and masses inside \url{eemmAmp}.  For that, we would need to call \url{eemmAmp.set_masses(me,mmu);} with the new values.  We next define the variable \url{pspatial} to store the spatial part of the incoming momentum, where we are assuming we are in the center of mass frame.  Following this we create an array of size $20$ and type \url{ldouble} to store the expected values for the squared amplitude at the angles given by $\cos(\theta)=0.95, 0.85, \cdots -0.85, -0.95$.  There should be 20 values, equally spaced, altogether if using this built-in test.  The test is run on the next line with the built-in method \url{test\_2to2\_amp2}.  The first argument of this method is the function for the amplitude squared.  We chose to use a lambda expression to allow different squared amplitude methods to be called.  An example of when we find this useful is when we have photons or gluons in the initial states and want to test the squared amplitude with specific helicity combinations as well as the fully summed squared amplitude.  For examples, see the photon and gluon processes in the \url{SM} directory.  Finally, we include the masses of each of the four particles, the spatial part of the incoming momentum and the already-known data for comparison.  On the next line we have \url{...} representing other tests.  We emphasize that the user does not have to use the built-in tests and can write any test code here that validates their amplitude code.

Once the source file for the process is complete, we need a main function to create a binary to run the tests.  We could put it in the same file, but typically this process will be just one of multiple for a given calculation.  So, we will create a file called \url{test.cpp} in the \url{user-dir/} and add the following lines to it:
\begin{verbatim}				    
#include "spinas.h"
#include "include/eemm.h"

int main(){ 
  test_eemm();
}
\end{verbatim}
After including the header for \spinas{} and for this process, we create a main function.  In our case, the only purpose of the main function is to run our tests, but the user could add other code here as desired.

\subsection{Compiling and Running}
Once the files are created, we compile the code with a command (on Linux) such as
\begin{verbatim}
    g++ -o test test.cpp eemm.cpp -I../include 
        -L../build -lspinas -std=c++11 
        -DWITH_LONG_DOUBLE -O3
\end{verbatim}
all on one line.
An explanation of the parts of this command can be found in Sec.~\ref{sec:user-guide:compilation}.  We are assuming the \url{user-dir/} is inside the main \spinas{} directory and that this command is being run from inside \url{user-dir/}.  If the user's directory is located somewhere else, \url{-I../include} and \url{-L../build} will have to be modified to represent the paths to the \spinas{} include and build directories.

Once the user's code is compiled, it can be run with a command (on Linux)
\begin{verbatim}
    ./test
\end{verbatim}
If the user created the files with exactly the same contents as described here, the output will be \texttt{* e , E  -> m , M  (QED)}.  
If there were any discrepancies, the output should have a list of the differences between the user's code and the Feynman-diagram results.  If the user has success here, we recommend they modify the amplitude in some way and see what happens with this test to get a better feel for what they would see if the amplitude expression were not correct.  For example, try switching a relative sign between terms, or try changing the overall factor.  Of course, the user can add further tests and more output as they choose.  This just gives a starting point.

\subsection{\label{sec:Complete Example:Spin-1}Some Notes on massive Spin-1 amplitudes}
Each process will come with its own challenges and we have included a complete set of processes from the SM in the \url{SM} directory.  Before leaving this section, however, we would like to outline the unique code when dealing with spin-1 particles.  Our spinors are all spin-$\frac{1}{2}$ and only have two possible values, one for spin $+\frac{1}{2}$ and one for spin $-\frac{1}{2}$.  Therefore, the spin-1 object is obtained by a symmetric combination of these spin-$\frac{1}{2}$ objects, in the usual product-representation way.  See Sec.~\ref{sec:user-guide:process:higher spins} for further details.  In this subsection, we will demonstrate using the functions involved with these spins in a couple of processes. 

Our first examle involves the process $A^{+},Z\to e,\bar{e}$.  The amplitude for this process is given by
\begin{align}
    \mathcal{M} &= pre\frac{\lbrack1\lvert p_3p_4\rvert1\rbrack (g_{Re} \lbrack24\rbrack \langle23\rangle + g_{Le} \langle24\rangle \lbrack23\rbrack}{(t-m_e^2) (u-m_e^2)}\nonumber\\
	&+ pre\lbrack12\rbrack \left( \frac{g_{Re} \lbrack14\rbrack \langle23\rangle}{(u-m_e^2)} - \frac{g_{Le} \lbrack13\rbrack \langle24\rangle}{(t-m_e^2)} \right),
\end{align}
where $pre$ is the constant prefactor.
We see that, in this example, the square bracket $\lvert1\rbrack$ appears twice in every term.  This is the photon and is a massless helicity spinor and already gives helicity $+1$ for the photon.  On the other hand, we see that every term has two massive spin spinors for the $Z$ boson, namely a $\lvert\mathbf{2}\rangle$ and a $\lvert\mathbf{2}\rbrack$.  These massive spin spinors each have two spin choices.  If the spin of the $Z$ boson is $\pm1$, then there is only one combination.  However, if the spin ofthe $Z$ boson is $0$, then there are two combinations that must be added, and this must be followed by a normalization factor.  

To make this more straight forward, we have created the functions \url{get_num_spin_loops}, \url{get_spin_normalization} and \url{get_spinor_spins}.  Here is an example for this process.
\begin{verbatim}
cdouble AZee::amp(const int& ds1, 
        const int& ds2, const int& ds3, 
        const int& ds4){
  cdouble amplitude(0,0);
  int ds2a, ds2b;

  int nCombs=get_num_spin_loops(ds2);
  ldouble normFactor=get_spin_normalization(ds2);
    
  for(int i=0;i<nCombs;i++){
    get_spinor_spins(ds2,ds2a,ds2b, i);
      
    if(ds1>0){	
	      amplitude += normFactor*pre*s1341s.v()*
        (gR*s24s.v(ds2a,ds4)*a23a.v(ds2b,ds3) 
        +gL*s23s.v(ds2a,ds3)*a24a.v(ds2b,ds4))
        /pDenT/pDenU;
	      ...
    }
    else if(ds1<0){
      ...
    }      
  }
   
  return amplitude;
}
\end{verbatim}
After creating a complex variable, called \url{amplitude} to store the amplitude, we introduce two new integers, \url{ds2a} and \url{ds2b}.  These store double the spin of each massive spin spinor that make up the spin-1 $Z$ boson.  Each of them will take either the value $+1$ or $-1$.  We next create an integer called \url{nCombs}, which stores the number of different spin combinations that must be added to achieve the double spin given by \url{ds2}, which is double the Z-boson spin.  It's argument is the double spin of the Z boson.  For example, if the double spin \url{ds2} is either $+2$ or $-2$, then, \url{nCombs} will be equal to 1.  If, on the other hand, the double spin \url{ds2} is $0$, then \url{nCombs} will be 2.  On the next line, we create a variable of type \url{ldouble} which contains the normalization for the spin.  It's argument is, once again, the double spin of the Z boson.  It gives $1$ when \url{ds2} is $\pm2$ and $1/\sqrt{2}$ when \url{ds2} is $0$.  

At this point, we are ready to iterate through the different spin combinations, so we create a \url{for} loop that adds the contribution from each spin combination.  Before we actually calculate the contribution to the amplitude for this spin combination, we must determine what the double spins \url{ds2a} and \url{ds2b} are.  For this, we have the line \url{get_spinor_spins}, which takes as arguments, the double spin of the $Z$ boson \url{ds2}, the spinor double spins \url{ds2a} and \url{ds2b} and the current iteration \url{i}.  The values of \url{ds2a} and \url{ds2b} are updated according to the value of \url{ds2} and \url{i}.  With this, we are ready to calculate this contribution to the amplitude.  After entering the first loop if the photon is helicity $+1$, we add the expression to \url{amplitude}.  We first multiply the entire term by \url{normFactor} described above, the prefactor \url{pre} (containing the coupling constants and anything that could be factored out) and the sproduct \url{s1341s.v()}, which represents the value of $\lbrack1\lvert p_3p_4\rvert1\rbrack$.  Since the photon is massless, there are no spin choices and, therefore, no arguments to this value.  Next, in parentheses, we add two pieces for the left- and right-chiral pieces of the amplitude contribution.  For the right chirality, we have \url{s24s.v(ds2a,ds4)*a23a.v(ds2b,ds3)}.  We see that the first of these spinor products uses the argument \url{sp2a} and the second uses the argument \url{sp2b}.  The order does not matter since it is symmetrized, as long as every term has each of them once.  The \url{...} represent other terms that are not shown, but can be seen in the file \url{AZee.cpp} in the \url{SM} directory.  Finally, at the end, we return the amplitude.

Using these functions may seem overkill since the previous example was so simple.  However, these functions work on amplitudes with greater number of external massive spin-1 particles, where it becomes increasingly more complicated.  Let us give one more example.  Consider the process $A^+,Z\to W,\bar{W}$.  This process contains three massive spin-1 bosons and each spin combination for the $Z, W$ and $\bar{W}$ will have a different number of sub-spins, different normalization and different combination of sub-spins for each iteration.  This can become quite complicated.  For this reason, we have created our functions to work for any number of spin-1 particles, with the caveat that if the process contains more than six massive spin spinors, the user should use the more general form of the functions described in Sec.~\ref{sec:user-guide:process:higher spins}.  Here is the outline of the code that can be seen in \url{SM/AZWW.cpp},
\begin{verbatim}
    
cdouble AWZW::amp(const int& ds1, 
        const int& ds2, const int& ds3, 
        const int& ds4)
  {
    cdouble amplitude(0,0);
    int ds3a, ds3b, ds4a, ds4b, ds2a, ds2b;

    int nCombs=get_num_spin_loops(ds3,ds4,ds2);
    
    ldouble normFactor=
        get_spin_normalization(ds3,ds4,ds2);
    
    for(int i=0;i<nCombs;i++){
      get_spinor_spins(ds3,ds3a,ds3b, 
        ds4,ds4a,ds4b, ds2,ds2a,ds2b, i);

      if(ds1>0){
      amplitude += normFactor*pre*(
        + CW*CW*s13s.v(ds3a)*s13s.v(ds3b)
          * a24a.v(ds2a,ds4a)*a24a.v(ds2b,ds4b)
        + ...
        )/pDenU/pDenS;	
    }
    else if(ds1<0){
	      ...
    }    
  }    
  return amplitude;
}
\end{verbatim} 
We focus here on the parts that contain the spins of the massive spin-1 particles.  Shortly after opening the function, we declare six new integers \url{ds3a}, \url{ds3b}, \url{ds4a}, \url{ds4b}, \url{ds2a} and \url{ds2b}, two for each particle.  After this, we set the number of combinations \url{nCombs} by use of \url{get_num_spin_loops}, however, this time it uses all three double spins to determine the number of loops that are appropriate.  The normalization factor 
\url{normFactor} is obtained from \url{get_spin_normalization}, again with all three double spins as arguments.  For both of these functions, the order of the double spins does not matter.  Once we begin iterating through the spin combinations, we set the sub-spins by use of the function \url{get_spinor_spins}.  This time, it contains the spins and sub-spins of each particle in turn.  The order of the particles in this function does not matter, but for each particle, the double spin of the particle must be first followed by its sub-spins.  Finally, when we add the contribution to the amplitude, each sub-spin is present once per term.  The order doesn't matter here.

\subsection{\label{sec:Complete Example:SM}Complete Set of SM Processes}
While this section presents one complete example of a process implementation, the \spinas{} codebase offers a comprehensive collection of $2\to2$ SM processes. These can be found in the \url{SM} directory. A full listing of these processes is provided in Sec.~\ref{sec:user-guide:SM}, serving as a practical resource for users. By examining a process similar to their research interest, users can gain valuable insights into applying \spinas{} in their own work.

\section{\label{sec:design}Design and Implementation}
This section delves into the design and implementation aspects of the \spinas{} package, primarily aimed at individuals interested in contributing to its development. While the typical user may not require this level of detail, these insights are valuable for those looking to understand the inner workings of \spinas{} and contribute effectively. Contributions from the community are highly encouraged and are integral to the evolution of this software.

\subsection{\label{sec:design:license}License}
\spinas{} is distributed under the GNU General Public License (GPL) Version 3, aligning with our commitment to openness and collaborative development within the scientific community. This license ensures that all enhancements and modifications to the software remain freely accessible. It grants users the freedom to run, study, share, and modify \spinas{}. The full GPL V3 license text is available in the software repository and detailed at \cite{GPLv3}, providing comprehensive information about the associated rights and responsibilities.

\subsection{\label{sec:design:github}\spinas{} Repository and Contributing}
The \spinas{} project is hosted on GitHub, offering an accessible platform for downloading the software, contributing to its development, and engaging with the user community. The repository, found at:
\begin{verbatim}
  https://github.com/neilthec/spinas
\end{verbatim}
which serves as a hub for the latest stable release, issue tracking, feature requests, and viewing the development history.

We actively encourage community contributions to \spinas{}, encompassing various forms of participation:
\begin{itemize}
    \item \textbf{Community Support:} Contributing through discussions, answering queries, and sharing experiences with \spinas{}.
    \item \textbf{Documentation:} Enhancing and expanding documentation to facilitate user understanding and engagement.
    \item \textbf{Issue Reporting:} Utilizing GitHub’s issue tracker for bug reports and feature suggestions, aiding in continuous improvement.
    \item \textbf{Code Contributions:} Developing new features, optimizing code, and resolving bugs.
\end{itemize}

\spinas{}'s future development relies on community involvement, with a vision for collaborative growth that extends beyond our team to include valuable user contributions. This community-driven model is crucial for maintaining \spinas{} as a state-of-the-art tool in particle physics research.

\subsection{\label{sec:design:Compilation}Build System}
The \spinas{} package currently employs CMake as its build system, selected for its wide acceptance, cross-platform support, and adaptability in handling complex build scenarios. CMake streamlines the compilation process, ensuring a user-friendly and flexible platform suitable for various computational requirements. While CMake meets the current needs effectively, we remain open to the possibility of adopting a different build system in the future. Any such transition would be guided by the goal of further enhancing the user experience, particularly in simplifying and optimizing the build process across different platforms.

A central aspect of the \spinas{} compilation strategy is the strategic use of compiler flags. These flags play a crucial role in tailoring the software to specific needs. In the current iteration of \spinas{}, compiler flags are primarily utilized to determine the precision of calculations, as described in Sec.~\ref{sec:design:types}.

\subsection{\label{sec:design:directories}Directories}

The \spinas{} package is organized into several directories, each serving a distinct purpose in the framework’s structure and functionality. This organization facilitates ease of navigation and maintenance of the codebase.

\begin{itemize}
    \item \textbf{Source Directory:} Located in \url{source}, this directory contains all the source files of \spinas{}. These files comprise the core functionalities and algorithms that drive the package.
    \item \textbf{Include Directory:} The \url{include} directory is dedicated to all the core header files of \spinas{}. 
    \item \textbf{Tests Directory:} All unit tests for \spinas{} components, leveraging the Boost testing framework, are housed in the \url{tests} directory. 
    \item \textbf{SM Processes Directory:} The \url{SM} directory contains all the $2 \to 2$ processes within the Standard Model (SM). As we expand to include higher-multiplicity SM processes, we anticipate subdividing this directory into more specialized categories. 
    \item \textbf{User Directory:} The \url{user-dir} directory serves as a template for users looking to develop their own code using \spinas{}. It includes the complete example, as detailed in Sec.~\ref{sec:Complete Example}, providing a practical guide and starting point for user-implemented projects.
\end{itemize}

This directory structure is designed not only for the current state of \spinas{} but also with an eye towards its future development.

\subsection{\label{sec:design:spinas.h}The \spinas{} Header File}
In order that the user does not need to specify each individual header file, we have created the \url{spinas.h} header file which includes all the others.

\subsection{\label{sec:design:types}Data Types}
To facilitate consistent precision management across the package and in user-implemented code, we have introduced a \url{types.h} header file. This file defines the central floating-precision and complex types, controlled by a compiler flag, which currently sets the precision to either \url{long} \url{double} or \url{double}. Within this file, types such as \url{ldouble} and \url{cdouble} are defined to reflect the desired precision. By employing these types throughout the \spinas{} codebase and encouraging users to do the same, precision modifications can be implemented uniformly, simply by adjusting the relevant compiler flag.

In an early, non-public version of \spinas{}, support for arbitrary precision was explored. Developers interested in this feature or looking to contribute towards its integration into future versions are encouraged to contact the author. Such collaboration aligns with our ongoing commitment to enhancing \spinas{}'s capabilities in response to evolving user needs and scientific advancements.

\subsection{\label{sec:design:cvector and cmatrix}Complex Vectors and Complex Matrices}
In the realm of constructive theory, 2-dimensional complex vectors and 2x2 complex matrices are indispensable components. To this end, the \spinas{} package incorporates these elements through the implementation of two classes: \url{cvector} and \url{cmatrix}. It's important to note that these classes primarily serve as internal tools within \spinas{}, facilitating higher-level operations rather than being intended for direct use by end-users. Their design is streamlined, encompassing only those properties and methods that are essential for the specific demands of constructive calculations.

In addition to directly specifying the components of these objects, \url{cmatrix} can also be constructed with a momentum and a choice of whether the Lorentz indices are upper or lower.  Additionally, these classes have some standard methods, such as \url{get_conjugate} for \url{cvector} and \url{get_det} for \url{cmatrix} to obtain the determinant, corresponding with $p^2$ for the momentum. 

A key feature of these classes is that they overload standard algebraic operators such as \url{+}, \url{-}, \url{*}, and \url{/}. This overloading simplifies the implementation of mathematical operations, making the code more intuitive and aligned with conventional mathematical expressions. Such design choices in \url{cvector} and \url{cmatrix} contribute to the efficiency and readability of computations within the \spinas{} package.

\subsection{\label{sec:design:particle}Particles}
A central role in \spinas{} is played by the \url{particle} class, which implements many properties of the particles related to the spinor algebra and is one of the classes directly used by the end user.  

The construction of a \url{particle} object requires the particle's mass as the sole argument. Users have the flexibility to modify the mass post-construction using \url{set_mass} and can retrieve the current mass value with \url{get_mass}. Setting the particle's momentum is accomplished via \url{set_momentum} (details in Sec.~\ref{sec:user-guide:particle}), and the momentum can be accessed using \url{get_momentum}. It's important to note that the momentum should be set after specifying the mass, as the \url{set_momentum} method updates numerous other internal variables. This sequence aligns with the typical order followed in amplitude calculations for phase-space points. Additionally, the \url{dot} method provides the inner product of the momenta of this particle and another.  The other methods, described below, are primarily intended for internal use.

One of the main design principles of \spinas{} was to maximally store calculations so that if they were needed multiple times within an amplitude calculation, they would not be recalulated, but rather simply returned.  Consequently, the \url{particle} class maintains private variables that store a particle's properties and representations. These include the momentum's magnitude and its polar and azimuthal angles, recalculated whenever the momentum is altered. The other particle representations are only calculated when they are first needed.  This means that aspects of the particle that are not needed in a calculation are never calculated.  Each property is calculated when it is first needed and then stored for later reuse.  They are not reset and recalculated until after the momentum is changed.

The other private variables include the 2x2 complex matrix of upper and lower Lorentz indices in \url{cmatrix} objects and are obtained with the methods \url{umat} and \url{lmat}, respectively.  The spinor forms of the particles are stored in 2-dimensional complex vectors of type \url{cvector}.  If the particle is massless, the left- and right-angle helicity spinors are obtained with the methods \url{langle} and \url{rangle}, respectively, while the left- and right-square helicity spinors are obtained with the methods \url{lsquare} and \url{rsquare}, all with no arguments.  If the particle is massive and the spinor has an upper spin index, which is the default, the left- and right-angle spinors can be obtained with the methods \url{langle} and \url{rangle}, respectively, while the left- and right-square helicity spinors can be obtained with the methods \url{lsquare} and \url{rsquare}, this time with one argument, which is double the desired spin (either $\pm1$).  Finally, if the particle is massive and the spinor has a lower spin index, the same methods can be called, this time with two arguments.  The first argument is double the spin (either $\pm1$) and the second argument is \url{LOWER}.  Strictly speaking, the case with upper indices could be obtained in the same way, where the second argument is \url{UPPER}.  The values \url{LOWER} and \url{UPPER} are set to boolean values in the header files in order to simplify and reduce mistake in implementation.

\subsection{\label{sec:design:sproduct}Spinor Products}
At the heart of amplitude calculations in \spinas{} lies the \url{sproduct} class, which handles the product of two spinors with an intervening set of momenta. The constructor of this class, detailed in Sec.~\ref{sec:user-guide:sproduct}, allows users to specify the nature of the left spinor as either an angle spinor (\url{ANGLE}) or a square spinor (\url{SQUARE}), with the right spinor determined by the class. Additionally, users can determine the position of the spin indices as either upper (\url{UPPER}) or lower (\url{LOWER}), as elaborated in the preceding subsection.  As discussed in the previous subsection, we have created these compile-time constants in order to simplify the use of these classes.  

A key design philosophy of \spinas{} is the efficient reuse of calculations, a principle that is extensively applied in the \url{sproduct} class. This class is constructed with references to the participating particles, leveraging the fact that spinors and momentum matrices are properties of these particles. By extracting spinor and momentum matrix information directly from the \url{particle} class, \url{sproduct} avoids redundant calculations. For instance, if the same particle spinor (e.g., $\langle\mathbf{1}\rvert$) is used in multiple spinor products such as $\langle\mathbf{1}\lvert p_2\rvert\mathbf{3}\rbrack$ and $\langle\mathbf{14}\rangle$, it is computed only once and then reused in both instances. This approach extends to momentum matrices as well, ensuring that any matrix, such as that for $p_2$, is computed just once irrespective of its frequency of use in different spinor products. As a result, the more complex the amplitude, the greater the efficiency gains from this implementation.

Additionally, \url{sproduct} stores the computed value of each spinor product and reuses these values in subsequent calculations. For example, if $\langle\mathbf{12}\rangle$ occurs multiple times within the same or different diagrams, it is calculated only once, with the result being reused in each future use. This efficiency underscores the importance of the \url{update} method. When a new phase-space point is set, invoking \url{update} resets the \url{sproduct} object, clearing its memory of previous calculations and ensuring accuracy in the new context.

\subsection{\label{sec:design:propagator}Propagators}

The propagator class is designed to calculate the propagator denominator.  It stores the mass and width of a particle but it does not store the momentum.  Our reason for this was so that the same propagator object could be used for multiple lines, containing the same particle.  This is partly because the momenta passing through the propagators are combinations of the external momenta and need to be calculated for each phase-space point.  In order to simplify the user's code, we encourage the user to create a variable of \url{cdouble} type for each propagator denominator and set its value after updating the phase-space point.  

We acknowledge that this design choice is subject to the evolving needs of the \spinas{} user community. Therefore, we are open to feedback and suggestions on the future development of this class, aiming to align it closely with user requirements and advancements in the field.

\subsection{\label{sec:design:process}Processes}
The \url{process} class was designed to store other methods useful to general processes.  It currently contains methods for testing $2\to2$ processes as well as methods to determine spin properties for massive spin-1 particles.  In the future, support for testing higher-multiplicity amplitudes and for higher-spin particles is likely to be added.  Furthermore, the community may decide to add other properties and methods.

\subsection{\label{sec:design:utilities}Other Functions}
Within the utilities file of the \spinas{} package, we have incorporated several auxiliary functions that extend beyond the scope of specific classes. These functions were created to facilitate validation and testing, but are available for other uses.

One such function is \url{rotate_momentum}, which performs the rotation of a 4-dimensional momentum. It requires three arguments: the 4-dimensional momentum array to be rotated, a 3-dimensional normalized array defining the rotation axis, and the rotation angle. While we currently do not offer a function for rotation based on Euler angles, we encourage the community to contribute such a feature if deemed beneficial.

Another important function is \url{boost_momentum}, which applies a Lorentz boost to a 4-momentum. This function takes two arguments: the 4-dimensional momentum array undergoing the boost and a 3-dimensional array representing the boost's velocity. Similar to rotation, we have not implemented a function for boosting based on rapidity, but we welcome community contributions in this area.

Additionally, we have included functions designed to randomly generate 4-momenta. The first, \url{choose_random_momentum}, randomly generates a 4-momentum for a particle with mass, ensuring that the energy component satisfies the condition $E>\lvert\vec{p}\rvert$. This function also randomly determines the mass of the particle. The second function, \url{choose_random_massless_momentum}, generates a massless 4-momentum. Both functions share the same set of arguments: a 4-dimensional array to store the generated 4-momentum and the minimum and maximum values for the momentum vector components.

\section{\label{sec:validation}Testing and Validation}
Testing and validation are crucial components in the development of any software, particularly for a complex system like \spinas{} that deals with particle physics simulations. In SPINAS, two primary testing methodologies are employed: Boost Unit Tests and a comprehensive set of Standard Model 2→2 process tests. These tests not only ensure the correctness of the implementation but also validate the results against established benchmarks.  If a bug is discovered, we will add a test for the bug before fixing it.

\subsection{\label{sec:validation:boost}Boost Unit Tests}
Boost Unit Tests provide a robust framework for performing unit testing in C++ applications. In \spinas{}, these tests are used to validate the functionality of individual components and modules.  In this subsection, we will describe the tests we have performed for the components of the \spinas{} package.  For many of these tests, we have repeated the test one hundred times, each time with new randomly generated points.  When generating random values, we generally generate them between $-50$ and $50$ if either sign is allowed or between $0$ and $50$ if non-negative.  Occasionally, a random point is chosen that pushes the precision too far and the test reports test failures.  However, this is usually a case of loss of precision in the tail of randomly generated tests.  Further evidence of this is the testing of the SM processes described in Sec.~\ref{sec:validation:2to2}.  If this occurs, we encourage the user to rerun the Boost tests.  If the error does not occur again and the SM tests pass, the issue is likely not a sign of a problem with the package, but rather a sign of the limitations of the precision.  

\subsubsection{\label{sec:validation:boost:cvector}cvector}
We have tested complex vectors in the \url{cvector} class in the \url{tests/cvector.cpp} file.  We test the following things:
\begin{itemize}
    \item the constructor;
    \item the conjugation method; and
    \item multiplication of two \url{cvector} objects as well as a \url{cvector} object and a \url{cmatrix} object in either order.
\end{itemize}

\subsubsection{\label{sec:validation:boost:cmatrix}cmatrix}
We have tested complex matrices in the \url{cmatrix} class in the \url{tests/cmatrix.cpp} file.  We test:
\begin{itemize}
    \item the constructor;
    \item the determinant;
    \item addition of two complex matrices;
    \item subtraction of two complex matrices;
    \item multiplication of two complex matrices (matrix multiplication).
\end{itemize}

\subsubsection{\label{sec:validation:boost:propagator}propagator}
We have tested the \url{propagator} class in \url{tests/propagator.cpp} file.  We test:
\begin{itemize}
    \item the constructor, and
    \item the denominator.
\end{itemize}

\subsubsection{\label{sec:validation:boost:particle}particle}
We have tested the \url{particle} class in \url{tests/particle.cpp} file.  We test:
\begin{itemize}
    \item the constructor;
    \item the \url{set_momentum} method;
    \item momentum matrices with upper and lower Lorentz indices;
\end{itemize}

Furthermore, we have checked various identities for the spinors.  We describe these identities in App.~\ref{app:validation identities}.  We begin with massless helicity spinors.  For these, we test:
    \begin{itemize}
        \item $\lvert i\rangle=\lbrack i\rvert^*$,
        \item $\langle i\rvert=\lvert i\rbrack^*$,
        \item $p_i\lvert i\rangle=0$,  
        \item $\lbrack i\rvert p_i=0$,
        \item $\langle ii\rangle=0$,
        \item $\lbrack ii\rbrack=0$,
        \item $\lvert i\rangle\lbrack i\rvert=p_i$,
        \item $\lvert i\rbrack\langle i\rvert=p_i$,
        \item $p_i\lvert i\rbrack=0$,         \item $\langle i\rvert p_i =0$,
    \end{itemize}
and
    \begin{itemize}
        \item $\mathcal{J}^{(3)}\lvert i\rangle = -\frac{1}{2}\lvert i\rangle$,
        \item $\mathcal{J}^{(\pm)}\lvert i\rangle = 0$,
        \item $\mathcal{J}^{(3)}\langle i\rvert = -\frac{1}{2}\langle i\rvert$,
        \item $\mathcal{J}^{(\pm)}\langle i\rvert = 0$,
        \item $\mathcal{J}^{(3)}\lbrack i\rvert = +\frac{1}{2}\lbrack i\rvert$,
        \item $\mathcal{J}^{(\pm)}\lbrack i\rvert = 0$,
        \item $\mathcal{J}^{(3)}\lvert i\rbrack = +\frac{1}{2}\lvert i\rbrack$, and
        \item $\mathcal{J}^{(\pm)}\lvert i\rbrack = 0$,
    \end{itemize}
where the $\mathcal{J}^{(\pm)}$ identities include two identities that were each tested.

For massive spinors, we check the related identities, taking into account the spin indices:
    \begin{itemize}
        \item $\lvert\mathbf{i}\rangle^{\mathrm{I}}=\left(\lbrack\mathbf{i}\rvert_{\mathrm{I}}\right)^*$,
        \item $\langle\mathbf{i}\rvert^{\mathrm{I}}=\left(\lvert\mathbf{i}\rbrack_{\mathrm{I}}\right)^*$,
        \item $\lvert\mathbf{i}\rangle_{\mathrm{I}}=-\left(\lbrack\mathbf{i}\rvert^{\mathrm{I}}\right)^*$,
        \item $\langle\mathbf{i}\rvert_{\mathrm{I}}=-\left(\lvert\mathbf{i}\rbrack^{\mathrm{I}}\right)^*$,
        \item $\langle\mathbf{ii}\rangle^{\mathrm{I J}}=-m_i\epsilon^{\mathrm{IJ}}$,
        \item $\langle\mathbf{ii}\rangle^{\mathrm{I}}_{\mathrm{J}}=-m_i\delta^{\mathrm{I}}_{\mathrm{J}}$,
        \item $\langle\mathbf{ii}\rangle_{\mathrm{I}}^{\mathrm{J}}=m_i\delta_{\mathrm{I}}^{\mathrm{J}}$,
        \item $\langle\mathbf{ii}\rangle_{\mathrm{I J}}=m_i\epsilon_{\mathrm{I J}}$,
        \item $\lbrack\mathbf{ii}\rbrack_{\mathrm{I J}}=-m_i\epsilon_{\mathrm{I J}}$,
        \item $\lbrack\mathbf{ii}\rbrack_{\mathrm{I}}^{\mathrm{J}}=-m_i\delta_{\mathrm{I}}^{\mathrm{J}}$,
        \item $\lbrack\mathbf{ii}\rbrack^{\mathrm{I}}_{\mathrm{J}}=m_i\delta^{\mathrm{I}}_{\mathrm{J}}$,
        \item $\lbrack\mathbf{ii}\rbrack^{\mathrm{I J}}=m_i\epsilon^{\mathrm{I J}}$,
        \item $\lvert\mathbf{i}\rangle^{\mathrm{I}}\lbrack\mathbf{i}\rvert_{\mathrm{I}} = p_i$,
        \item $\lvert\mathbf{i}\rangle_{\mathrm{I}}\lbrack\mathbf{i}\rvert^{\mathrm{I}}=-p_i$,
        \item $\lvert\mathbf{i}\rbrack_{\mathrm{I}}\langle\mathbf{i}\rvert^{\mathrm{I}}=p_i$,
        \item $\lvert\mathbf{i}\rbrack^{\mathrm{I}}\langle\mathbf{i}\rvert_{\mathrm{I}}=-p_i$,
        \item $p_i\lvert\mathbf{i}\rangle^{\mathrm{I}}=-m_i\lvert\mathbf{i}\rbrack^{\mathrm{I}}$,
        \item $\left(\langle\mathbf{i}\rvert^{\mathrm{I}}\right) p_i=m_i\lbrack\mathbf{i}\rvert^{\mathrm{I}}$,
        \item $p_i\lvert\mathbf{i}\rangle_{\mathrm{I}}=-m_i\lvert\mathbf{i}\rbrack_{\mathrm{I}}$,
        \item $\left(\langle\mathbf{i}_{\mathrm{I}}\right)p_i=m_i\lbrack\mathbf{i}\rvert_{\mathrm{I}}$,
        \item $p_i\lvert\mathbf{i}\rbrack^{\mathrm{I}}=-m_i\lvert\mathbf{i}\rangle^{\mathrm{I}}$,
        \item $\left(\lbrack\mathbf{i}\rvert^{\mathrm{I}}\right)p_i=m_i\langle\mathbf{i}\rvert^{\mathrm{I}}$,
        \item $p_i\lvert\mathbf{i}\rbrack_{\mathrm{I}}=-m_i\lvert\mathbf{i}\rangle_{\mathrm{I}}$,
        \item $\left(\lbrack\mathbf{i}\rbrack_{\mathrm{I}}\right)p_i=m_i\langle\mathbf{i}\rvert_{\mathrm{I}}$,
    \end{itemize}
and
    \begin{itemize}
        \item $\mathcal{J}\lvert \mathbf{i}\rangle^{\mathrm{I}} = \lvert \mathbf{i}\rangle^{\mathrm{J}} J_{\mathrm{J}}^{\ \mathrm{I}}$,
        \item $\mathcal{J}\langle \mathbf{i}\rvert^{\mathrm{I}} = \langle \mathbf{i}\rvert^{\mathrm{J}} J_{\mathrm{J}}^{\ \mathrm{I}}$,
        \item $\mathcal{J}\lvert \mathbf{i}\rangle_{\mathrm{I}} = \lvert \mathbf{i}\rangle_{\mathrm{J}} J_{\ \mathrm{I}}^{\mathrm{J}}$,
        \item $\mathcal{J}\langle \mathbf{i}\rvert_{\mathrm{I}} = \langle \mathbf{i}\rvert_{\mathrm{J}} J_{\ \mathrm{I}}^{\mathrm{J}}$,
        \item $\mathcal{J}\lvert \mathbf{i}\rbrack^{\mathrm{I}} = \lvert \mathbf{i}\rbrack^{\mathrm{J}} J_{\mathrm{J}}^{\ \mathrm{I}}$,
        \item $\mathcal{J}\lbrack \mathbf{i}\rvert^{\mathrm{I}} = \lbrack \mathbf{i}\rvert^{\mathrm{J}} J_{\mathrm{J}}^{\ \mathrm{I}}$,
        \item $\mathcal{J}\lvert \mathbf{i}\rbrack_{\mathrm{I}} = \lvert \mathbf{i}\rbrack_{\mathrm{J}} J_{\ \mathrm{I}}^{\mathrm{J}}$, and
        \item $\mathcal{J}\lbrack \mathbf{i}\rvert_{\mathrm{I}} = \lbrack \mathbf{i}\rvert_{\mathrm{J}} J_{\ \mathrm{I}}^{\mathrm{J}}$,
    \end{itemize}
where each of these rows corresponds with three identities which were each tested.  They were with $\mathcal{J}^{(3)}$ and $j^{(3)}$ or $\mathcal{J}^{(+)}$ and $j^{(+)}$ or $\mathcal{J}^{(-)}$ and $j^{(-)}$.

\subsubsection{\label{sec:validation:boost:sproduct}sproduct}
We have tested a significant number of spinor product identities in the \url{sproduct} class in \url{tests/sproduct.cpp} file.  We have tested these identities for a variety of randomly generated masses and momenta, including both massive cases and massless cases (when appropriate).  

In our first series of tests, we have considered spinor-product identities where all the spin indices on the left are contracted so that the right side of the equality does not have any spinor indices.  For these, not only do we randomly choose the momenta and masses for each of the tests, but we also test after random rotations and boosts.

For our first batch of tests, we have taken the product of two spinor products with no intermediate momenta.  We have tested:
\begin{itemize}
    \item $\lbrack \mathrm{i j}\rbrack\langle \mathrm{j i}\rangle=2\pdot{p_i}{p_j}$ in the following ways:
    \begin{itemize}
        \item $\lbrack \mathrm{i j}\rbrack\langle \mathrm{j i}\rangle=2\pdot{p_i}{p_j}$ with $m_i=0$ and $m_j=0$;
        \item $\lbrack \mathrm{i} \mathbf{j}^{\mathrm{J}}\rbrack\langle \mathbf{j}_{\mathrm{J}} \mathrm{i}\rangle=2\pdot{p_i}{p_j}$ with $m_i=0$ and $m_j\neq0$;
        \item $\lbrack \mathbf{i}^{\mathrm{I}} \mathrm{j}\rbrack\langle \mathrm{j} \mathbf{i}_{\mathrm{I}}\rangle=2\pdot{p_i}{p_j}$ with $m_i\neq0$ and $m_j=0$; and
        \item $\lbrack \mathbf{i}^{\mathrm{I}} \mathbf{j}^{\mathrm{J}}\rbrack\langle \mathbf{j}_{\mathrm{J}} \mathbf{i}_{\mathrm{I}}\rangle=2\pdot{p_i}{p_j}$ with $m_i\neq0$ and $m_j\neq0$;
    \end{itemize}
    \item $\lbrack\mathrm{i j}\rbrack\lbrack\mathrm{j i}\rbrack=-2m_im_j$ in the following ways:
    \begin{itemize}
        \item $\lbrack\mathrm{i j}\rbrack\lbrack\mathrm{j i}\rbrack=-2m_im_j$ with $m_i=0$ and $m_j=0$;
        \item $\lbrack \mathrm{i} \mathbf{j}^{\mathrm{J}}\rbrack\lbrack \mathbf{j}_{\mathrm{J}} \mathrm{i}\rbrack=-2m_i m_j$ with $m_i=0$ and $m_j\neq0$;
        \item $\lbrack \mathbf{i}^{\mathrm{I}} \mathrm{j}\rbrack\lbrack \mathrm{j} \mathbf{i}_{\mathrm{I}}\rbrack=-2m_im_j$ with $m_i\neq0$ and $m_j=0$; and
        \item $\lbrack \mathbf{i}^{\mathrm{I}} \mathbf{j}^{\mathrm{J}}\rbrack\lbrack \mathbf{j}_{\mathrm{J}} \mathbf{i}_{\mathrm{I}}\rbrack=-2m_i m_j$ with $m_i\neq0$ and $m_j\neq0$;
    \end{itemize}
    \item $\langle \mathrm{i j}\rangle\langle\mathrm{j i}\rangle=-2m_im_j$ in the following ways:
    \begin{itemize}
        \item $\langle \mathrm{i j}\rangle\langle\mathrm{j i}\rangle=-2m_im_j$ with $m_i=0$ and $m_j=0$;
        \item $\langle \mathrm{i} \mathbf{j}^{\mathrm{J}}\rangle\langle \mathbf{j}_{\mathrm{J}} \mathrm{i}\rangle=-2m_im_j$ with $m_i=0$ and $m_j\neq0$;
        \item $\langle \mathbf{i}^{\mathrm{I}} \mathrm{j}\rangle\langle \mathrm{j} \mathbf{i}_{\mathrm{I}}\rangle=-2m_im_j$ with $m_i\neq0$ and $m_j=0$; and
        \item $\langle \mathbf{i}^{\mathrm{I}} \mathbf{j}^{\mathrm{J}}\rangle\langle \mathbf{j}_{\mathrm{J}} \mathbf{i}_{\mathrm{I}}\rangle=-2m_im_j$ with $m_i\neq0$ and $m_j\neq0$;
    \end{itemize}
\end{itemize}

We also tested products of two spinor products with one intermediate momentum:
\begin{itemize}
    \item $\lbrack\mathbf{i}^{\mathrm{I}} \mathrm{j}\rbrack\langle\mathrm{j}\lvert p_k\rvert\mathbf{i}_{\mathrm{I}}\rbrack = -2m_i\pdot{p_j}{p_k}$ with the properties:
    \begin{itemize}
        \item $m_i\neq0$ since this identity does not apply otherwise;
        \item $m_j=0$; and 
        \item $m_k=0$ and $m_k\neq0$;
    \end{itemize}
    \item $\lbrack\mathbf{i}^{\mathrm{I}} \mathbf{j}^{\mathrm{J}}\rbrack\langle\mathbf{j}_{\mathrm{J}}\lvert p_k\rvert\mathbf{i}_{\mathrm{I}}\rbrack = -2m_i\pdot{p_j}{p_k}$ with the properties:
    \begin{itemize}
        \item $m_i\neq0$ since this identity does not apply otherwise;
        \item $m_j\neq0$; and
        \item $m_k=0$ and $m_k\neq0$;
    \end{itemize}
\end{itemize}

Our next batch of tests has two momenta both in the same spinor product:
\begin{itemize}
    \item $\lbrack\mathbf{i}^{\mathrm{I}}\mathbf{j}^{\mathrm{J}}\rbrack\lbrack\mathbf{j}_{\mathrm{J}}\lvert p_k p_l\rvert\mathbf{i}_{\mathrm{I}}\rbrack=-2m_im_j \pdot{p_k}{p_l}$ with the properties:
    \begin{itemize}
        \item $m_i\neq0$ and $m_j\neq0$ since this identity does not apply otherwise;
        \item $m_k=0$ and $m_k\neq0$; and 
        \item $m_l=0$ and $m_l\neq0$;
    \end{itemize}
    \item $\mbox{Re}\left(\lbrack\mathrm{i j}\rbrack\langle\mathrm{j}\lvert p_k p_l\rvert\mathrm{i}\rangle\right) = 2\pdot{p_i}{p_j} \pdot{p_k}{p_l} - 2\pdot{p_i}{p_k} \pdot{p_j}{p_l} + 2\pdot{p_i}{p_l} \pdot{p_j}{p_k}$ with the properties:
    \begin{itemize}
        \item $m_i=0$ and $m_j=0$;
        \item $m_k=0$ and $m_k\neq0$; and
        \item $m_l=0$ and $m_l\neq0$;
    \end{itemize}
    \item $\mbox{Re}\left(\lbrack\mathbf{i}^{\mathrm{I}} \mathrm{j}\rbrack\langle\mathrm{j}\lvert p_k p_l\rvert\mathbf{i}_{\mathrm{I}}\rangle\right) = 2\pdot{p_i}{p_j} \pdot{p_k}{p_l} - 2\pdot{p_i}{p_k} \pdot{p_j}{p_l} + 2\pdot{p_i}{p_l} \pdot{p_j}{p_k}$ with the properties:
    \begin{itemize}
        \item $m_i\neq0$ and $m_j=0$;
        \item $m_k=0$ and $m_k\neq0$; and
        \item $m_l=0$ and $m_l\neq0$;
    \end{itemize}
    \item $\mbox{Re}\left(\lbrack\mathrm{i} \mathbf{j}^{\mathrm{J}}\rbrack\langle\mathbf{j}_{\mathrm{J}}\lvert p_k p_l\rvert\mathrm{i}\rangle\right) = 2\pdot{p_i}{p_j} \pdot{p_k}{p_l} - 2\pdot{p_i}{p_k} \pdot{p_j}{p_l} + 2\pdot{p_i}{p_l} \pdot{p_j}{p_k}$ with the properties:
    \begin{itemize}
        \item $m_i=0$ and $m_j\neq0$;
        \item $m_k=0$ and $m_k\neq0$; and
        \item $m_l=0$ and $m_l\neq0$;
    \end{itemize}
    \item $\mbox{Re}\left(\lbrack\mathbf{i}^{\mathrm{I}} \mathbf{j}^{\mathrm{J}}\rbrack\langle\mathbf{j}_{\mathrm{J}}\lvert p_k p_l\rvert\mathbf{i}_{\mathrm{I}}\rangle\right) = 2\pdot{p_i}{p_j} \pdot{p_k}{p_l} - 2\pdot{p_i}{p_k} \pdot{p_j}{p_l} + 2\pdot{p_i}{p_l} \pdot{p_j}{p_k}$ with the properties:
    \begin{itemize}
        \item $m_i\neq0$ and $m_j\neq0$;
        \item $m_k=0$ and $m_k\neq0$; and
        \item $m_l=0$ and $m_l\neq0$;
    \end{itemize}
\end{itemize}

We follow this with a series of tests where the two momenta are in separate spinor products:
\begin{itemize}
    \item $\lbrack\mathbf{i}^{\mathrm{I}}\lvert p_l\rvert\mathbf{j}^{\mathrm{J}}\rangle\langle\mathbf{2}_{\mathrm{J}}\lvert p_k\rvert\mathbf{i}_{\mathrm{I}}\rbrack=2m_im_j\pdot{p_k}{p_l}$ with the properties:
    \begin{itemize}
        \item $m_i\neq0$ and $m_j\neq0$ since this identity does not otherwise apply;
        \item $m_k=0$ and $m_k\neq0$; and
        \item $m_l=0$ and $m_l\neq0$;
    \end{itemize}
    \item $\mathrm{Re}\left(\lbrack\mathrm{i}\lvert p_l\rvert\mathrm{j}\rangle\lbrack\mathrm{j}\lvert p_k\rvert\mathrm{i}\rangle\right) = -2\pdot{p_i}{p_l} \pdot{p_j}{p_k} + 2\pdot{p_i}{p_j} \pdot{p_k}{p_l} - 2\pdot{p_i}{p_k} \pdot{p_j}{p_l}$ with the properties that:
    \begin{itemize}
        \item $m_i=0$ and $m_j=0$;
        \item $m_k=0$ and $m_k\neq0$; and
        \item $m_l=0$ and $m_l\neq0$;
    \end{itemize}
    \item $\mathrm{Re}\left(\lbrack\mathbf{i}^{\mathrm{I}}\lvert p_l\rvert\mathrm{j}\rangle\lbrack\mathrm{j}\lvert p_k\rvert\mathbf{i}_{\mathrm{I}}\rangle\right) = -2\pdot{p_i}{p_l} \pdot{p_j}{p_k} + 2\pdot{p_i}{p_j} \pdot{p_k}{p_l} - 2\pdot{p_i}{p_k} \pdot{p_j}{p_l}$ with the properties that:
    \begin{itemize}
        \item $m_i\neq0$ and $m_j=0$;
        \item $m_k=0$ and $m_k\neq0$; and
        \item $m_l=0$ and $m_l\neq0$;
    \end{itemize}
    \item $\mathrm{Re}\left(\lbrack\mathrm{i}\lvert p_l\rvert\mathbf{j}^{\mathrm{J}}\rangle\lbrack\mathbf{j}_{\mathrm{J}}\lvert p_k\rvert\mathrm{i}\rangle\right) = -2\pdot{p_i}{p_l} \pdot{p_j}{p_k} + 2\pdot{p_i}{p_j} \pdot{p_k}{p_l} - 2\pdot{p_i}{p_k} \pdot{p_j}{p_l}$ with the properties that:
    \begin{itemize}
        \item $m_i=0$ and $m_j\neq0$;
        \item $m_k=0$ and $m_k\neq0$; and
        \item $m_l=0$ and $m_l\neq0$;
    \end{itemize}
    \item $\mathrm{Re}\left(\lbrack\mathbf{i}^{\mathrm{I}}\lvert p_l\rvert\mathbf{j}^{\mathrm{J}}\rangle\lbrack\mathbf{j}_{\mathrm{J}}\lvert p_k\rvert\mathbf{i}_{\mathrm{I}}\rangle\right) = -2\pdot{p_i}{p_l} \pdot{p_j}{p_k} + 2\pdot{p_i}{p_j} \pdot{p_k}{p_l} - 2\pdot{p_i}{p_k} \pdot{p_j}{p_l}$ with the properties that:
    \begin{itemize}
        \item $m_i=0$ and $m_j\neq0$;
        \item $m_k=0$ and $m_k\neq0$; and
        \item $m_l=0$ and $m_l\neq0$;
    \end{itemize}
\end{itemize}

We also test with two momenta in one spinor product and one momentum in the other spinor product:
\begin{itemize}
    \item $\mathrm{Re}\left(\lbrack\mathrm{i}\lvert p_l\rvert\mathbf{j}^{\mathrm{J}}\rangle\langle\mathbf{j}_{\mathrm{J}}\lvert p_mp_k\rvert\mathrm{i}\rangle\right) = -2m_j(\pdot{p_i}{p_l} \pdot{p_k}{p_m}-\pdot{p_i}{p_m} \pdot{p_k}{p_l}+\pdot{p_i}{p_k} \pdot{p_l}{p_m})$ with the properties that:
    \begin{itemize}
        \item $m_j\neq0$ since the identity does not apply otherwise;
        \item $m_i=0$;
        \item $m_k=0$ and $m_k\neq0$;
        \item $m_l=0$ and $m_l\neq0$; and
        \item $m_m=0$ and $m_m\neq0$;
    \end{itemize}
    \item $\mathrm{Re}\left(\lbrack\mathbf{i}^{\mathrm{I}}\lvert p_l\rvert\mathbf{j}^{\mathrm{J}}\rangle\langle\mathbf{j}_{\mathrm{J}}\lvert p_mp_k\rvert\mathbf{i}_{\mathrm{I}}\rangle\right) = -2m_j(\pdot{p_i}{p_l} \pdot{p_k}{p_m}-\pdot{p_i}{p_m} \pdot{p_k}{p_l}+\pdot{p_i}{p_k} \pdot{p_l}{p_m})$ with the properties that:
    \begin{itemize}
        \item $m_j\neq0$ since the identity does not apply otherwise;
        \item $m_i\neq0$;
        \item $m_k=0$ and $m_k\neq0$;
        \item $m_l=0$ and $m_l\neq0$; and
        \item $m_m=0$ and $m_m\neq0$;
    \end{itemize}
  \end{itemize}

Next, we have tested spinor products where not all the spin indices are summed over, so that there are spinor products on both sides of the equation.  We begin with tests that contract an angle and a square spin spinor to obtain a momentum.  We have tested:
  \begin{itemize}
    \item $\lbrack\mathrm{i}\mathbf{j}_{\mathrm{J}}\rbrack\langle\mathbf{j}^{\mathrm{J}}\mathrm{k}\rangle=\lbrack\mathrm{i}\lvert p_j\rvert\mathrm{k}\rangle$ with:
    \begin{itemize}
        \item $m_j\neq0$, $m_i=0$ and $m_k=0$;
    \end{itemize}
    \item $\lbrack\mathbf{i}^{\mathrm{I}}\mathbf{j}_{\mathrm{J}}\rbrack\langle\mathbf{j}^{\mathrm{J}}\mathrm{k}\rangle=\lbrack\mathbf{i}^{\mathrm{I}}\lvert p_j\rvert\mathrm{k}\rangle$ with:
    \begin{itemize}
        \item $m_j\neq0$, $m_i\neq0$ and $m_k=0$;
    \end{itemize}
    \item $\lbrack\mathrm{i}\mathbf{j}_{\mathrm{J}}\rbrack\langle\mathbf{j}^{\mathrm{J}}\mathbf{k}^{\mathrm{K}}\rangle=\lbrack\mathrm{i}\lvert p_j\rvert\mathbf{k}^{\mathrm{K}}\rangle$ with:
    \begin{itemize}
        \item $m_j\neq0$, $m_i=0$ and $m_k\neq0$;
    \end{itemize}
    \item $\lbrack\mathbf{i}^{\mathrm{I}}\mathbf{j}_{\mathrm{J}}\rbrack\langle\mathbf{j}^{\mathrm{J}}\mathbf{k}^{\mathrm{K}}\rangle=\lbrack\mathbf{i}^{\mathrm{I}}\lvert p_j\rvert\mathbf{k}^{\mathrm{K}}\rangle$ with:
    \begin{itemize}
        \item $m_j\neq0$, $m_i\neq0$ and $m_k\neq0$;
    \end{itemize}
    \item $\lbrack\mathrm{i}\mathbf{j}^{\mathrm{J}}\rbrack\langle\mathbf{j}_{\mathrm{J}}\mathrm{k}\rangle=-\lbrack\mathrm{i}\lvert p_j\rvert\mathrm{k}\rangle$ with:
    \begin{itemize}
        \item $m_j\neq0$, $m_i=0$ and $m_k=0$;
    \end{itemize}
    \item $\lbrack\mathbf{i}^{\mathrm{I}}\mathbf{j}^{\mathrm{J}}\rbrack\langle\mathbf{j}_{\mathrm{J}}\mathrm{k}\rangle=-\lbrack\mathbf{i}^{\mathrm{I}}\lvert p_j\rvert\mathrm{k}\rangle$ with:
    \begin{itemize}
        \item $m_j\neq0$, $m_i\neq0$ and $m_k=0$;
    \end{itemize}
    \item $\lbrack\mathrm{i}\mathbf{j}^{\mathrm{J}}\rbrack\langle\mathbf{j}_{\mathrm{J}}\mathbf{k}^{\mathrm{K}}\rangle=-\lbrack\mathrm{i}\lvert p_j\rvert\mathbf{k}^{\mathrm{K}}\rangle$ with:
    \begin{itemize}
        \item $m_j\neq0$, $m_i=0$ and $m_k\neq0$;
    \end{itemize}
    \item $\lbrack\mathbf{i}^{\mathrm{I}}\mathbf{j}^{\mathrm{J}}\rbrack\langle\mathbf{j}_{\mathrm{J}}\mathbf{k}^{\mathrm{K}}\rangle=-\lbrack\mathbf{i}^{\mathrm{I}}\lvert p_j\rvert\mathbf{k}^{\mathrm{K}}\rangle$ with:
    \begin{itemize}
        \item $m_j\neq0$, $m_i\neq0$ and $m_k\neq0$;
    \end{itemize}
    \item $\langle\mathrm{i}\mathbf{j}^{\mathrm{J}}\rangle\lbrack\mathbf{j}_{\mathrm{J}}\mathrm{k}\rbrack=\langle\mathrm{i}\lvert p_j\rvert\mathrm{k}\rbrack$ with:
    \begin{itemize}
        \item $m_j\neq0$, $m_i=0$ and $m_k=0$;
    \end{itemize}
    \item $\langle\mathbf{i}^{\mathrm{I}}\mathbf{j}^{\mathrm{J}}\rangle\lbrack\mathbf{j}_{\mathrm{J}}\mathrm{k}\rbrack=\langle\mathbf{i}^{\mathrm{I}}\lvert p_j\rvert\mathrm{k}\rbrack$ with:
    \begin{itemize}
        \item $m_j\neq0$, $m_i\neq0$ and $m_k=0$;
    \end{itemize}
    \item $\langle\mathrm{i}\mathbf{j}^{\mathrm{J}}\rangle\lbrack\mathbf{j}_{\mathrm{J}}\mathbf{k}^{\mathrm{K}}\rbrack=\langle\mathrm{i}\lvert p_j\rvert\mathbf{k}^{\mathrm{K}}\rbrack$ with:
    \begin{itemize}
        \item $m_j\neq0$, $m_i=0$ and $m_k\neq0$;
    \end{itemize}
    \item $\langle\mathbf{i}^{\mathrm{I}}\mathbf{j}^{\mathrm{J}}\rangle\lbrack\mathbf{j}_{\mathrm{J}}\mathbf{k}^{\mathrm{K}}\rbrack=\langle\mathbf{i}^{\mathrm{I}}\lvert p_j\rvert\mathbf{k}^{\mathrm{K}}\rbrack$ with:
    \begin{itemize}
        \item $m_j\neq0$, $m_i\neq0$ and $m_k\neq0$;
    \end{itemize}
    \item $\langle\mathrm{i}\mathbf{j}_{\mathrm{J}}\rangle\lbrack\mathbf{j}^{\mathrm{J}}\mathrm{k}\rbrack=-\langle\mathrm{i}\lvert p_j\rvert\mathrm{k}\rbrack$ with:
    \begin{itemize}
        \item $m_j\neq0$, $m_i=0$ and $m_k=0$;
    \end{itemize}
    \item $\langle\mathbf{i}^{\mathrm{I}}\mathbf{j}_{\mathrm{J}}\rangle\lbrack\mathbf{j}^{\mathrm{J}}\mathrm{k}\rbrack=-\langle\mathbf{i}^{\mathrm{I}}\lvert p_j\rvert\mathrm{k}\rbrack$ with:
    \begin{itemize}
        \item $m_j\neq0$, $m_i\neq0$ and $m_k=0$;
    \end{itemize}
    \item $\langle\mathrm{i}\mathbf{j}_{\mathrm{J}}\rangle\lbrack\mathbf{j}^{\mathrm{J}}\mathbf{k}^{\mathrm{K}}\rbrack=-\langle\mathrm{i}\lvert p_j\rvert\mathbf{k}^{\mathrm{K}}\rbrack$ with:
    \begin{itemize}
        \item $m_j\neq0$, $m_i=0$ and $m_k\neq0$;
    \end{itemize}
    \item $\langle\mathbf{i}^{\mathrm{I}}\mathbf{j}_{\mathrm{J}}\rangle\lbrack\mathbf{j}^{\mathrm{J}}\mathbf{k}^{\mathrm{K}}\rbrack=-\langle\mathbf{i}^{\mathrm{I}}\lvert p_j\rvert\mathbf{k}^{\mathrm{K}}\rbrack$ with:
    \begin{itemize}
        \item $m_j\neq0$, $m_i\neq0$ and $m_k\neq0$;
    \end{itemize}
  \end{itemize}

  We have also tested contractions of two square brackets and two angle brackets, which give a mass.  We have done the following tests:
  \begin{itemize}
    \item $\lbrack\mathrm{i}\mathbf{j}_{\mathrm{J}}\rbrack\lbrack\mathbf{j}^{\mathrm{J}}\mathrm{k}\rbrack=m_j\lbrack\mathrm{i k}\rbrack$ with:
    \begin{itemize}
        \item $m_j\neq0$, $m_i=0$ and $m_k=0$;
    \end{itemize}
    \item $\lbrack\mathbf{i}^{\mathrm{I}}\mathbf{j}_{\mathrm{J}}\rbrack\lbrack\mathbf{j}^{\mathrm{J}}\mathrm{k}\rbrack=m_j\lbrack\mathbf{i}^{\mathrm{I}} \mathrm{k}\rbrack$ with:
    \begin{itemize}
        \item $m_j\neq0$, $m_i\neq0$ and $m_k=0$;
    \end{itemize}
    \item $\lbrack\mathrm{i}\mathbf{j}_{\mathrm{J}}\rbrack\lbrack\mathbf{j}^{\mathrm{J}}\mathbf{k}^{\mathrm{K}}\rbrack=m_j\lbrack\mathrm{i} \mathbf{k}^{\mathrm{K}}\rbrack$ with:
    \begin{itemize}
        \item $m_j\neq0$, $m_i=0$ and $m_k\neq0$;
    \end{itemize}
    \item $\lbrack\mathbf{i}^{\mathrm{I}}\mathbf{j}_{\mathrm{J}}\rbrack\lbrack\mathbf{j}^{\mathrm{J}}\mathbf{k}^{\mathrm{K}}\rbrack=m_j\lbrack\mathbf{i}^{\mathrm{I}} \mathbf{k}^{\mathrm{K}}\rbrack$ with:
    \begin{itemize}
        \item $m_j\neq0$, $m_i\neq0$ and $m_k\neq0$;
    \end{itemize}
    \item $\lbrack\mathrm{i}\mathbf{j}^{\mathrm{J}}\rbrack\lbrack\mathbf{j}_{\mathrm{J}}\mathrm{k}\rbrack=-m_j\lbrack\mathrm{i k}\rbrack$ with:
    \begin{itemize}
        \item $m_j\neq0$, $m_i=0$ and $m_k=0$;
    \end{itemize}
    \item $\lbrack\mathbf{i}^{\mathrm{I}}\mathbf{j}^{\mathrm{J}}\rbrack\lbrack\mathbf{j}_{\mathrm{J}}\mathrm{k}\rbrack=-m_j\lbrack\mathbf{i}^{\mathrm{I}} \mathrm{k}\rbrack$ with:
    \begin{itemize}
        \item $m_j\neq0$, $m_i\neq0$ and $m_k=0$;
    \end{itemize}
    \item $\lbrack\mathrm{i}\mathbf{j}^{\mathrm{J}}\rbrack\lbrack\mathbf{j}_{\mathrm{J}}\mathbf{k}^{\mathrm{K}}\rbrack=-m_j\lbrack\mathrm{i} \mathbf{k}^{\mathrm{K}}\rbrack$ with:
    \begin{itemize}
        \item $m_j\neq0$, $m_i=0$ and $m_k\neq0$;
    \end{itemize}
    \item $\lbrack\mathbf{i}^{\mathrm{I}}\mathbf{j}^{\mathrm{J}}\rbrack\lbrack\mathbf{j}_{\mathrm{J}}\mathbf{k}^{\mathrm{K}}\rbrack=-m_j\lbrack\mathbf{i}^{\mathrm{I}} \mathbf{k}^{\mathrm{K}}\rbrack$ with:
    \begin{itemize}
        \item $m_j\neq0$, $m_i\neq0$ and $m_k\neq0$;
    \end{itemize}
    \item $\langle\mathrm{i}\mathbf{j}^{\mathrm{J}}\rangle\langle\mathbf{j}_{\mathrm{J}}\mathrm{k}\rangle=m_j\lbrack\mathrm{i k}\rbrack$ with:
    \begin{itemize}
        \item $m_j\neq0$, $m_i=0$ and $m_k=0$;
    \end{itemize}
    \item $\langle\mathbf{i}^{\mathrm{I}}\mathbf{j}^{\mathrm{J}}\rangle\langle\mathbf{j}_{\mathrm{J}}\mathrm{k}\rangle=m_j\lbrack\mathbf{i}^{\mathrm{I}} \mathrm{k}\rbrack$ with:
    \begin{itemize}
        \item $m_j\neq0$, $m_i\neq0$ and $m_k=0$;
    \end{itemize}
    \item $\langle\mathrm{i}\mathbf{j}^{\mathrm{J}}\rangle\langle\mathbf{j}_{\mathrm{J}}\mathbf{k}^{\mathrm{K}}\rangle=m_j\lbrack\mathrm{i} \mathbf{k}^{\mathrm{K}}\rbrack$ with:
    \begin{itemize}
        \item $m_j\neq0$, $m_i=0$ and $m_k\neq0$;
    \end{itemize}
    \item $\langle\mathbf{i}^{\mathrm{I}}\mathbf{j}^{\mathrm{J}}\rangle\langle\mathbf{j}_{\mathrm{J}}\mathbf{k}^{\mathrm{K}}\rangle=m_j\lbrack\mathbf{i}^{\mathrm{I}} \mathbf{k}^{\mathrm{K}}\rbrack$ with:
    \begin{itemize}
        \item $m_j\neq0$, $m_i\neq0$ and $m_k\neq0$;
    \end{itemize}
    \item $\lbrack\mathrm{i}\mathbf{j}^{\mathrm{J}}\rbrack\lbrack\mathbf{j}_{\mathrm{J}}\mathrm{k}\rbrack=-m_j\lbrack\mathrm{i k}\rbrack$ with:
    \begin{itemize}
        \item $m_j\neq0$, $m_i=0$ and $m_k=0$;
    \end{itemize}
    \item $\lbrack\mathbf{i}^{\mathrm{I}}\mathbf{j}^{\mathrm{J}}\rbrack\lbrack\mathbf{j}_{\mathrm{J}}\mathrm{k}\rbrack=-m_j\lbrack\mathbf{i}^{\mathrm{I}} \mathrm{k}\rbrack$ with:
    \begin{itemize}
        \item $m_j\neq0$, $m_i\neq0$ and $m_k=0$;
    \end{itemize}
    \item $\lbrack\mathrm{i}\mathbf{j}^{\mathrm{J}}\rbrack\lbrack\mathbf{j}_{\mathrm{J}}\mathbf{k}^{\mathrm{K}}\rbrack=-m_j\lbrack\mathrm{i} \mathbf{k}^{\mathrm{K}}\rbrack$ with:
    \begin{itemize}
        \item $m_j\neq0$, $m_i=0$ and $m_k\neq0$;
    \end{itemize}
    \item $\lbrack\mathbf{i}^{\mathrm{I}}\mathbf{j}^{\mathrm{J}}\rbrack\lbrack\mathbf{j}_{\mathrm{J}}\mathbf{k}^{\mathrm{K}}\rbrack=-m_j\lbrack\mathbf{i}^{\mathrm{I}} \mathbf{k}^{\mathrm{K}}\rbrack$ with:
    \begin{itemize}
        \item $m_j\neq0$, $m_i\neq0$ and $m_k\neq0$;
    \end{itemize}
\end{itemize}

In the following, we have tested an identity under momentum conservation.  That is, we took $p_l=-p_i-p_j-p_k$, numerically.  We then plugged it in to $\lbrack\mathrm{i}\lvert p_l\rvert\mathrm{j}\rangle$ and compared it with $m_j\lbrack\mathrm{i}\mathrm{j}\rbrack-m_i\langle\mathrm{i}\mathrm{j}\rangle-\lbrack\mathrm{i}\lvert p_k\rvert\mathrm{j}\rangle$. We did this for a variety of masses.  In each case, $m_l$ was determined by momentum conservation.  We tested:
\begin{itemize}
    \item $\lbrack\mathrm{i}\lvert p_l\rvert\mathrm{j}\rangle=-\lbrack\mathrm{i}\lvert p_k\rvert\mathrm{j}\rangle$ with:
    \begin{itemize}
        \item $m_i=0$ and $m_j=0$; and
        \item $m_k=0$ and $m_k\neq0$;
    \end{itemize}
    \item $\lbrack\mathbf{i}^{\mathrm{I}}\lvert p_l\rvert\mathrm{j}\rangle=-m_i\langle\mathbf{i}^{\mathrm{I}}\mathrm{j}\rangle-\lbrack\mathbf{i}^{\mathrm{I}}\lvert p_k\rvert\mathrm{j}\rangle$ with:
    \begin{itemize}
        \item $m_i\neq0$ and $m_j=0$; and
        \item $m_k=0$ and $m_k\neq0$;
    \end{itemize}
    \item $\lbrack\mathrm{i}\lvert p_l\rvert\mathbf{j}^{\mathrm{J}}\rangle=m_j\lbrack\mathrm{i}\mathbf{j}^{\mathrm{J}}\rbrack-\lbrack\mathrm{i}\lvert p_k\rvert\mathbf{j}^{\mathrm{J}}\rangle$ with:
    \begin{itemize}
        \item $m_i=0$ and $m_j\neq0$; and
        \item $m_k=0$ and $m_k\neq0$;
    \end{itemize}
    \item $\lbrack\mathbf{i}^{\mathrm{I}}\lvert p_l\rvert\mathbf{j}^{\mathrm{J}}\rangle=m_j\lbrack\mathbf{i}^{\mathrm{I}}\mathbf{j}^{\mathrm{J}}\rbrack-m_i\langle\mathbf{i}^{\mathrm{I}}\mathbf{j}^{\mathrm{J}}\rangle-\lbrack\mathbf{i}^{\mathrm{I}}\lvert p_k\rvert\mathbf{j}^{\mathrm{J}}\rangle$ with:
    \begin{itemize}
        \item $m_i\neq0$ and $m_j\neq0$; and
        \item $m_k=0$ and $m_k\neq0$;
    \end{itemize}
\end{itemize}

Finally, another test of momentum conservation was performed where $p_l=-p_i-p_j-p_k$ was calculated and used in $\lbrack\mathrm{i}\lvert p_l\rvert\mathrm{j}\rangle+\lbrack\mathrm{j}\lvert p_l\rvert\mathrm{i}\rangle=-\lbrack\mathrm{i}\lvert p_k\rvert\mathrm{j}\rangle-\lbrack\mathrm{j}\lvert p_k\rvert\mathrm{i}\rangle$ where $m_i=m_j$.  The mass $m_l$ was determined from momentum conservation.  We tested:
\begin{itemize}
    \item $\lbrack\mathrm{i}\lvert p_l\rvert\mathrm{j}\rangle+\lbrack\mathrm{j}\lvert p_l\rvert\mathrm{i}\rangle=-\lbrack\mathrm{i}\lvert p_k\rvert\mathrm{j}\rangle-\lbrack\mathrm{j}\lvert p_k\rvert\mathrm{i}\rangle$ with:
    \begin{itemize}
        \item $m_i=0$ and $m_j=0$; and
        \item $m_k=0$ and $m_k\neq0$;
    \end{itemize}
    \item $\lbrack\mathbf{i}^{\mathrm{I}}\lvert p_l\rvert\mathbf{j}^{\mathrm{J}}\rangle+\lbrack\mathbf{j}^{\mathrm{J}}\lvert p_l\rvert\mathbf{i}^{\mathrm{I}}\rangle=-\lbrack\mathbf{i}^{\mathrm{I}}\lvert p_k\rvert\mathbf{j}^{\mathrm{J}}\rangle-\lbrack\mathbf{j}^{\mathrm{J}}\lvert p_k\rvert\mathbf{i}^{\mathrm{I}}\rangle$ with:
    \begin{itemize}
        \item $m_i\neq0$ and $m_j\neq0$; and
        \item $m_k=0$ and $m_k\neq0$;
    \end{itemize}
\end{itemize}

\subsection{\label{sec:validation:2to2}SM $\mathbf{2\to2}$ Process Tests}
\spinas{} complements its unit tests with a robust suite of tests for \(2 \to 2\) processes in the Standard Model (SM), further validating the package's implementation of spinors, our understanding of the relationship of processes related by crossing symmetry and its simulation capabilities in particle physics. 

For testing, we utilize methods  \url{test\_2to2\_amp2}, \url{test\_2to2\_amp2\_rotations}, \url{test\_2to2\_amp2\_boosts}, and \url{test\_2to2\_amp2\_boosts\_and\_rotations}, detailed in Sec.~\ref{sec:user-guide:process:testing}. The squared amplitudes used for comparison are generated using CalcHEP \cite{CalcHEP}.

As mentioned at the end of Sec.~\ref{sec:user-guide:process:testing}, we have set the widths to zero for all our tests.  Furthermore, since CalcHEP gives the squared amplitude after summing (averaging) the spins and colors in the final (initial) states, we have done the same.  We have also included symmetry factors to match the same done in CalcHEP.  Obviously, the couplings must also be the same.  Typically, that means turning off any running of the coupling during the comparison to ensure identical values.

The full list of \(2 \to 2\) SM processes tested can be found in Sec.~\ref{sec:user-guide:SM}. We present the masses and momenta used in each test, as well as the file name for the process in App.~\ref{app:testing:SM}.  The momentum given is the magnitude of the vector momentum of the incoming particle in the CM frame, denoted as \(p_{\text{in}}\). The relations between each momentum component and \(p_{\text{in}}\) are given in Eqs.~(\ref{eq:user-guide:p1 from pin}) through (\ref{eq:user-guide:pout from pin}), with more details on the testing methodology provided in Sec.~\ref{sec:user-guide:process:testing}.  

We have used the same electric coupling constant, strong coupling constant, and Weinberg angle for each process.  All dimensionful variables are given in GeV.  We chose a variety of masses and momenta, including cases where the momentum was near threshold and far above threshold.  We also considered mass hierarchies that were SM like, as well as when the masses were similar and when they were inverted.  The exact mass and momentum values are less important than covering a range of combinations that change the importance of different diagrams within the process.  In order to keep this list reasonable, we will only show the values from the first process in each set.  The values for the second and/or third processes of the set were mostly identical, with small changes when the momentum is near threshold.  The specifics for each process follow the order in Table~\ref{tab:user-guide:SM Processes}.  To keep the main part of this document clean, we have included these values in App.~\ref{app:testing:SM}.

\section{Conclusion and Future Work}
This paper has presented \spinas{}, a versatile C++ package designed for efficient calculation of scattering amplitudes at specific phase-space points. Utilizing spin and helicity spinors, \spinas{} offers a complementary approach to traditional Feynman diagram calculations, setting the stage for comparative efficiency analysis between these two methods in future research endeavors.

Section~\ref{sec:user-guide} detailed each high-level component of \spinas{} and how to use it in an amplitude calculation.  It culminates with a comprehensive list of implemented Standard Model (SM) processes. These examples, cover at least two members of every crossing-symmetry classes of $2\to2$ processes in the SM, up to changes of mass for different fermion generations and serve as practical templates for users to base their custom implementations.

We followed this in Sec.~\ref{sec:Complete Example} with a complete example of using this package to calculate the $e,\bar{e}\to\mu,\bar{\mu}$ amplitude in QED.  In both of these sections, we have focused on a novice user, attempting to give all the required details.  

Section~\ref{sec:design} shifted focus to the design principles behind \spinas{}, targeting potential contributors. We discussed the package's release under the GNU Public License v3 and encouraged active community participation in its development. We envision \spinas{} as a collaborative, community-driven project and eagerly anticipate its future evolution as guided by user contributions and feedback.

In Section~\ref{sec:validation}, we outlined the rigorous validation processes applied to \spinas{}. This included comprehensive testing of each class and method, as well as validations of every \(2 \to 2\) process against established Feynman diagram results, across a range of masses and momenta. These tests not only affirm the robustness of \spinas{} but also validate the implemented amplitude calculations.
 
Looking ahead, a key objective is to develop an algorithm for the automatic generation of constructive amplitudes across any process and multiplicity within any constructive model. This ambitious goal marks our future direction, promising to significantly advance the field of amplitude calculations.

%\section*{Acknowledgments}
% Acknowledgments section.

\appendix
\section{\label{app:validation identities}Spinor Algebra for Boost Validation}
In this appendix, we review some of the spinor algebra that was used to implement the spinors and to develop the tests described in Secs.~\ref{sec:validation:boost:particle} and \ref{sec:validation:boost:sproduct}.

\subsection{Definitions}

The epsilon tensors that raises and lowers Lorentz indices (and spin indices where the Lorentz indices are replaced with spin indices) are
\begin{equation}
\epsilon_{\alpha\beta}=-\epsilon^{\alpha\beta}=\epsilon_{\dot{\alpha}\dot{\beta}}=-\epsilon^{\dot{\alpha}\dot{\beta}}=\left(\begin{array}{cc}0&-1\\1&0\end{array}\right).
\end{equation}
The momenta are written with Lorentz SL(2,$\mathcal{C}$) indices as,
\begin{equation}
    p_{\alpha\dot{\beta}}=%p_\mu\sigma^\mu_{\alpha\dot{\beta}}=
    \left(\begin{array}{cc}p^0+p^3&p^1-i p^2\\p^1+i p^2&p^0-p^3\end{array}\right),
    \label{eq:p_alphabetadot}
\end{equation}
\begin{equation}
    p^{\dot{\alpha}\beta}=%p_\mu\bar{\sigma}^{\mu\dot{\alpha}\beta}=
    \left(\begin{array}{cc}p^0-p^3&-p^1+i p^2\\-p^1-i p^2&p^0+p^3\end{array}\right),
    \label{eq:p^alphadotbeta}
\end{equation}
where the determinant of both of these satisfies,
\begin{equation}
    \det(p_{\alpha\dot{\beta}})=\det(p^{\dot{\alpha}\beta}) = p^2=m^2.
\end{equation}

The anticommutator of two momenta gives
\begin{eqnarray}
    p_{k}^{\dot{\alpha}\beta}p_{l\beta\dot{\omega}}+p_{l}^{\dot{\alpha}\beta}p_{k\beta\dot{\omega}} &=& 2p_k\cdot p_l \delta^{\dot{\alpha}}_{\dot{\omega}} \, ,\\
    p_{k\alpha\dot{\beta}}p_l^{\dot{\beta}\omega}+p_{l\alpha\dot{\beta}}p_k^{\dot{\beta}\omega} &=& 2p_k\cdot p_l \delta_{\alpha}^{\omega} \, .
    \label{eq:p1p2+p2p1}
\end{eqnarray}

In the following, 
\begin{equation}
    c \equiv \cos\left(\frac{\theta}{2}\right),  \quad  
    s \equiv \sin\left(\frac{\theta}{2}\right)e^{i\phi},
\end{equation}
where $\theta$ is the polar angle and $\phi$ is the azimuthal angle. 

The basic right-angle spinor is given by
\begin{align}
    |\mathbf{i}\rangle^I_\alpha &=& 
    \left(\begin{array}{cc}
    \sqrt{E+p}\ c & -\sqrt{E-p}\ s^* \\
    \sqrt{E+p}\ s & \sqrt{E-p}\ c
    \end{array}\right). 
    \label{eq:|i> def} 
\end{align}
The left angle spinor is obtained with the epsilon tensor as
\begin{align}
    \langle\mathbf{i}|^{\alpha I} & = 
       \epsilon^{\alpha\beta}|\mathbf{i}\rangle_\beta^I
       \nonumber\\
       &= \left(\begin{array}{cc}0&1\\-1&0\end{array}\right)
       \left(\begin{array}{cc}\sqrt{E+p}\ c & -\sqrt{E-p}\ s^* \\
    \sqrt{E+p}\ s   & \sqrt{E-p}\ c \end{array}\right)
    \nonumber\\
    &= \left(\begin{array}{cc}
    \sqrt{E+p}\ s & \sqrt{E-p}\ c \\
    -\sqrt{E+p}\ c & \sqrt{E-p}\ s^* 
    \end{array}\right).
\end{align}
The angle spinors with lower spin indices can also be obtained with the epsilon tensor (but from the right) as
\begin{align}
    |\mathbf{i}\rangle_{\alpha I}   &= 
       |\mathbf{i}\rangle_\alpha^J\epsilon_{JI}
       \nonumber\\
       &= \left(\begin{array}{cc}\sqrt{E+p}\ c & -\sqrt{E-p}\ s^* \\
    \sqrt{E+p}\ s   & \sqrt{E-p}\ c \end{array}\right)
    \left(\begin{array}{cc}0&-1\\1&0\end{array}\right)
    \nonumber\\
    &= \left(\begin{array}{cc}
    -\sqrt{E-p}\ s^* & -\sqrt{E+p}\ c  \\
    \sqrt{E-p}\ c & -\sqrt{E+p}\ s   
    \end{array}\right)
\end{align}
and
\begin{align}
    \langle\mathbf{i}|^{\alpha}_{I} & = 
    \langle\mathbf{i}|^{\alpha J} \epsilon_{JI}
    \nonumber\\
    &= \left(\begin{array}{cc}
    \sqrt{E+p}\ s & \sqrt{E-p}\ c \\
    -\sqrt{E+p}\ c & \sqrt{E-p}\ s^* 
    \end{array}\right)
    \left(\begin{array}{cc}0&-1\\1&0\end{array}\right)
    \nonumber\\
    &= \left(\begin{array}{cc}
    \sqrt{E-p}\ c & -\sqrt{E+p}\ s  \\
    \sqrt{E-p}\ s^* & \sqrt{E+p}\ c  
    \end{array}\right)
    \nonumber\\
    &= \epsilon^{\alpha\beta}
    |\mathbf{i}\rangle_{\alpha I}.
\end{align}

The basic left square spinor is given by the conjugate of Eq.~(\ref{eq:|i> def})
\begin{align}
    \lbrack\mathbf{i}|_{\dot{\alpha}I} &=& 
    \left(\begin{array}{cc}
    \sqrt{E+p}\ c & -\sqrt{E-p}\ s \\
    \sqrt{E+p}\ s^* & \sqrt{E-p}\ c 
    \end{array}\right).
    \label{eq:|i] def}
\end{align}
The right square spinor is obtained with the epsilon tensor as
\begin{align}
    |\mathbf{i}\rbrack^{\dot{\alpha}}_I & = 
       \epsilon^{\dot{\alpha}\dot{\beta}} \lbrack \mathbf{i} |_{\dot{\beta}I}
       \nonumber\\
       &= \left(\begin{array}{cc}0&1\\-1&0\end{array}\right)
       \left(\begin{array}{cc}
    \sqrt{E+p}\ c & -\sqrt{E-p}\ s \\
    \sqrt{E+p}\ s^* & \sqrt{E-p}\ c \end{array}\right)
    \nonumber\\
    &= \left(\begin{array}{cc}
    \sqrt{E+p}\ s^* & \sqrt{E-p}\ c \\
    -\sqrt{E+p}\ c & \sqrt{E-p}\ s
    \end{array}\right).
\end{align}
The square spinors with upper spin indices can also be obtained with the epsilon tensor (but from the right) as
\begin{align}
    \lbrack\mathbf{i}|_{\dot{\alpha}}^I   &= 
       \lbrack \mathbf{i} |_{\dot{\alpha}J} \epsilon^{JI}
       \nonumber\\
    &= \left(\begin{array}{cc}
    \sqrt{E+p}\ c & -\sqrt{E-p}\ s \\
    \sqrt{E+p}\ s^* & \sqrt{E-p}\ c \end{array}\right)
    \left(\begin{array}{cc}0&1\\-1&0\end{array}\right)
    \nonumber\\
    &= \left(\begin{array}{cc}
    \sqrt{E-p}\ s & \sqrt{E+p}\ c  \\
    -\sqrt{E-p}\ c & \sqrt{E+p}\ s^*  
    \end{array}\right)
\end{align}
and
\begin{align}
    |\mathbf{i}\rbrack^{\dot{\alpha}I} & = 
    |\mathbf{i}\rbrack^{\dot{\alpha}}_J \epsilon^{JI} 
    \nonumber\\
    &= \left(\begin{array}{cc}
    \sqrt{E+p}\ s^* & \sqrt{E-p}\ c \\
    -\sqrt{E+p}\ c & \sqrt{E-p}\ s
    \end{array}\right)
    \left(\begin{array}{cc}0&1\\-1&0\end{array}\right)
    \nonumber\\
    &= \left(\begin{array}{cc}
    -\sqrt{E-p}\ c & \sqrt{E+p}\ s^* \\
    -\sqrt{E-p}\ s & -\sqrt{E+p}\ c 
    \end{array}\right)
    \\
    &= \epsilon^{\dot{\alpha}\dot{\beta}}
    \lbrack\mathbf{i}|_{\dot{\beta}}^I.
\end{align}

The helicity spinors are the non-zero column of these.  We have
\begin{align}
    \lvert \mathrm{i}\rangle &= \left(\begin{array}{c}
    \sqrt{2E}\ c \\
    \sqrt{2E}\ s
    \end{array}\right)
    \\
    \langle \mathrm{i}\rvert &= \left(\begin{array}{c}
    \sqrt{2E}\ s \\
    -\sqrt{2E}\ c
    \end{array}\right)
    \\
    \lvert \mathrm{i}\rbrack &= \left(\begin{array}{c}
    \sqrt{2E}\ s^* \\
    -\sqrt{2E}\ c
    \end{array}\right)
    \\
    \lbrack \mathrm{i}\rvert &= \left(\begin{array}{c}
    \sqrt{2E}\ c \\
    \sqrt{2E}\ s^*
    \end{array}\right) .
\end{align}

\subsection{Inner and Outer Products}

When we multiply $\langle\mathbf{j}\rvert^I$ and $\lvert\mathbf{j}\rangle^J$ using matrix notation, we have to first transpose the matrix on the left so that the spin index $I$ is the row and the Lorentz index $\alpha$ is the column (normally the spin index is the column and the Lorentz index is the row).  We obtain
\begin{align}
\langle\mathbf{jj}\rangle^{IJ} 
    &= \left(\begin{array}{cc}
    \sqrt{E+p}\ s & -\sqrt{E+p}\ c\\
    \sqrt{E-p}\ c & \sqrt{E-p}\ s^*
    \end{array}\right)
    \nonumber\\
    &\quad \times
    \left(\begin{array}{cc}
    \sqrt{E+p}\ c & -\sqrt{E-p}\ s^*\\
    \sqrt{E+p}\ s & \sqrt{E-p}\ c
    \end{array}\right)
    \nonumber\\
    &= -m\epsilon^{IJ},
    \\
\langle\mathbf{jj}\rangle^{I}_{\ J} 
    &= \left(\begin{array}{cc}
    \sqrt{E+p}\ s & -\sqrt{E+p}\ c\\
    \sqrt{E-p}\ c & \sqrt{E-p}\ s^*
    \end{array}\right)
    \nonumber\\
    &\quad \times
    \left(\begin{array}{cc}
    -\sqrt{E-p}\ s^* & -\sqrt{E+p}\ c\\
    \sqrt{E-p}\ c & -\sqrt{E+p}\ s
    \end{array}\right)
    \nonumber\\
    &= -m\delta^{I}_{J},
    \\
\langle\mathbf{jj}\rangle_{I}^{\ J} 
    &= \left(\begin{array}{cc}
    \sqrt{E-p}\ c &  \sqrt{E-p}\ s^* \\
    -\sqrt{E+p}\ s & \sqrt{E+p}\ c  
    \end{array}\right)
    \nonumber\\
    &\quad \times
    \left(\begin{array}{cc}
    \sqrt{E+p}\ c & -\sqrt{E-p}\ s^*\\
    \sqrt{E+p}\ s & \sqrt{E-p}\ c
    \end{array}\right)
    \nonumber\\
    &= m\delta_{I}^{J},
    \\
\langle\mathbf{jj}\rangle_{IJ} 
    &= \left(\begin{array}{cc}
    \sqrt{E-p}\ c &  \sqrt{E-p}\ s^* \\
    -\sqrt{E+p}\ s & \sqrt{E+p}\ c  
    \end{array}\right)
    \nonumber\\
    &\quad \times
    \left(\begin{array}{cc}
    -\sqrt{E-p}\ s^* & -\sqrt{E+p}\ c\\
    \sqrt{E-p}\ c & -\sqrt{E+p}\ s
    \end{array}\right)
    \nonumber\\
    &= m\epsilon_{IJ},
\end{align}
and
\begin{align}
    \lbrack\mathbf{jj}\rbrack_{IJ}
    &= \left(\begin{array}{cc}
    \sqrt{E+p}\ c & \sqrt{E+p}\ s^* \\
    -\sqrt{E-p}\ s & \sqrt{E-p}\ c 
    \end{array}\right)
    \nonumber\\
    &\quad \times
    \left(\begin{array}{cc}
    \sqrt{E+p}\ s^* & \sqrt{E-p}\ c \\
    -\sqrt{E+p}\ c & \sqrt{E-p}\ s
    \end{array}\right)
    \nonumber\\
    &= -m_j\epsilon_{IJ} 
    \label{eq:[jj]},\\
    \lbrack\mathbf{jj}\rbrack_{I}^{\ J}
    &= \left(\begin{array}{cc}
    \sqrt{E+p}\ c & \sqrt{E+p}\ s^* \\
    -\sqrt{E-p}\ s & \sqrt{E-p}\ c 
    \end{array}\right)
    \nonumber\\
    &\quad \times
    \left(\begin{array}{cc}
    -\sqrt{E-p}\ c & \sqrt{E+p}\ s^* \\
    -\sqrt{E-p}\ s & -\sqrt{E+p}\ c 
    \end{array}\right)
    \nonumber\\
    &= -m_j\delta_{I}^{J},
    \\
    \lbrack\mathbf{jj}\rbrack^{I}_{\ J}
    &= \left(\begin{array}{cc}
    \sqrt{E-p}\ s & -\sqrt{E-p}\ c \\
    \sqrt{E+p}\ c & \sqrt{E+p}\ s^*  
    \end{array}\right)
    \nonumber\\
    &\quad \times
    \left(\begin{array}{cc}
    \sqrt{E+p}\ s^* & \sqrt{E-p}\ c \\
    -\sqrt{E+p}\ c & \sqrt{E-p}\ s
    \end{array}\right)
    \nonumber\\
    &= m_j\delta^{I}_{J} ,
    \\
    \lbrack\mathbf{jj}\rbrack^{IJ}
    &= \left(\begin{array}{cc}
    \sqrt{E-p}\ s & -\sqrt{E-p}\ c \\
    \sqrt{E+p}\ c & \sqrt{E+p}\ s^*  
    \end{array}\right)
    \nonumber\\
    &\quad \times
    \left(\begin{array}{cc}
    -\sqrt{E-p}\ c & \sqrt{E+p}\ s^* \\
    -\sqrt{E-p}\ s & -\sqrt{E+p}\ c 
    \end{array}\right)
    \nonumber\\
    &= m_j\epsilon^{IJ} .
\end{align}

On the other hand, when we multiply $\lvert\mathbf{j}\rangle^I\lbrack\mathbf{j}\rvert_I$, we need to transpose the second spinor to make the spin index the row index and the Lorentz index the column index.  This gives us the momentum.
\begin{align}
    |\mathbf{j}\rangle^I_{\alpha}\lbrack\mathbf{j}|_{\dot{\beta}I} 
    &=
    \left(\begin{array}{cc}
    \sqrt{E+p}\ c & -\sqrt{E-p}\ s^* \\
    \sqrt{E+p}\ s & \sqrt{E-p}\ c
    \end{array}\right)
    \nonumber\\
    &\quad \times 
    \left(\begin{array}{cc}
    \sqrt{E+p}\ c & \sqrt{E+p}\ s^* \\
    -\sqrt{E-p}\ s & \sqrt{E-p}\ c 
    \end{array}\right)
    \nonumber\\
    &=
    \left(\begin{array}{cc}
    p^0 + p^3                   & p^1 - i p^2 \\
    p^1 + i p^2                 & p^0 - p^3 \end{array} \right)
    = p_{\alpha\dot{\beta}},
\end{align}
where we have used $\cos^2\left(\frac{\theta}{2}\right)-\sin^2\left(\frac{\theta}{2}\right) = \cos(\theta)$ in the top left and the bottom right and $2\cos\left(\frac{\theta}{2}\right)\sin\left(\frac{\theta}{2}\right)=\sin(\theta)$ and $e^{i\phi}=\cos(\phi)+i\sin(\phi)$ in the top right and bottom left.  Similarly,
\begin{align}
|\mathbf{j}\rangle_I^{\alpha}\lbrack\mathbf{j}|_{\dot{\beta}}^{I} 
    &=
    \left(\begin{array}{cc}
    -\sqrt{E-p}\ s^* & -\sqrt{E+p}\ c  \\
    \sqrt{E-p}\ c & -\sqrt{E+p}\ s   
    \end{array}\right)
    \nonumber\\
    &\quad \times 
    \left(\begin{array}{cc}
    \sqrt{E-p}\ s & -\sqrt{E-p}\ c  \\
    \sqrt{E+p}\ c & \sqrt{E+p}\ s^*  
    \end{array}\right)
    \nonumber\\
    &=
    \left(\begin{array}{cc}
    -p^0 - p^3   & -p^1 + i p^2 \\
    -p^1 - i p^2 & -p^0 + p^3 
    \end{array} \right)
    = -p_{\alpha\dot{\beta}},
\end{align}
\begin{align}
|\mathbf{i}\rbrack_I^{\dot{\alpha}}\langle\mathbf{i}|^{\alpha I} 
    &= 
    \left(\begin{array}{cc}
    \sqrt{E+p}\ s^* & \sqrt{E-p}\ c \\
    -\sqrt{E+p}\ c & \sqrt{E-p}\ s
    \end{array}\right)
    \nonumber\\
    &\quad \times
    \left(\begin{array}{cc}
    \sqrt{E+p}\ s & -\sqrt{E+p}\ c \\
    \sqrt{E-p}\ c & \sqrt{E-p}\ s^* 
    \end{array}\right)
    \nonumber\\
    &= 
    \left(\begin{array}{cc}
    p^0-p^3 & -p^1+i p^2\\
    -p^1-i p^2 & p^0+p^3
    \end{array}\right)
    = p^{\dot{\alpha}\alpha},
\end{align}

\begin{align}
|\mathbf{i}\rbrack^{\dot{\alpha}I}\langle\mathbf{i}|^{\alpha}_{I} 
    &= 
    \left(\begin{array}{cc}
    -\sqrt{E-p}\ c & \sqrt{E+p}\ s^* \\
    -\sqrt{E-p}\ s & -\sqrt{E+p}\ c 
    \end{array}\right)
    \nonumber\\
    &\quad \times
    \left(\begin{array}{cc}
    \sqrt{E-p}\ c &  \sqrt{E-p}\ s^* \\
    -\sqrt{E+p}\ s & \sqrt{E+p}\ c  
    \end{array}\right)
    \nonumber\\
    &= 
    \left(\begin{array}{cc}
    -p^0+p^3   &  p^1-i p^2\\
     p^1+i p^2 & -p^0-p^3
    \end{array}\right)
    = -p^{\dot{\alpha}\alpha}.
\end{align}

We now do momentum times spinors, which gives mass times spinors.  We only show the explicit cases for upper spin indices but have tests for the relations with lower spin indices as well.
\begin{align}
    p_{i}|\mathbf{i}\rbrack 
    &= 
    \left(\begin{array}{cc}p^0+p^3&p^1-i p^2\\p^1+i p^2&p^0-p^3\end{array}\right)
    \nonumber\\
    &\quad\times
    \left(\begin{array}{cc}
    -\sqrt{E-p}\ c & \sqrt{E+p}\ s^* \\
    -\sqrt{E-p}\ s & -\sqrt{E+p}\ c 
    \end{array}\right)
    %\nonumber\\
    %&= -m_i
    %\left(\begin{array}{cc}
    %\sqrt{E+p}\ c & -\sqrt{E-p}\ s^* \\
    %\sqrt{E+p}\ s & \sqrt{E-p}\ c
    %\end{array}\right)
    \nonumber\\
    &= -m_i|\mathbf{i}\rangle\label{eq:pi|i]},
\end{align}
\begin{align}
    p_j|\mathbf{j}\rangle 
    &= \left(\begin{array}{cc}p^0-p^3&-p^1+i p^2\\-p^1-i p^2&p^0+p^3\end{array}\right)
    \nonumber\\
    &\quad \times
    \left(\begin{array}{cc}
    \sqrt{E+p}\ c & -\sqrt{E-p}\ s^* \\
    \sqrt{E+p}\ s & \sqrt{E-p}\ c
    \end{array}\right)
    \nonumber\\
    &= -m_j
    \left(\begin{array}{cc}
    -\sqrt{E-p}\ c & \sqrt{E+p}\ s^* \\
    -\sqrt{E-p}\ s & -\sqrt{E+p}\ c 
    \end{array}\right)
    \nonumber\\
    &= -m_j|\mathbf{j}\rbrack
    \label{eq:pi|i>}.
\end{align}

Once again, transposing the spinor (on both sides) of the following two relations,
\begin{align}
    \lbrack\mathbf{i}|p_i 
    &= 
    \left(\begin{array}{cc}
    \sqrt{E-p}\ s & -\sqrt{E-p}\ c  \\
    \sqrt{E+p}\ c & \sqrt{E+p}\ s^*  
    \end{array}\right)
    \nonumber\\
    &\quad \times
    \left(\begin{array}{cc}p^0-p^3&-p^1+i p^2\\-p^1-i p^2&p^0+p^3\end{array}\right)
    %\nonumber\\
    %&= m_i \left(\begin{array}{cc}
    %\sqrt{E+p}\ s & -\sqrt{E+p}\ c \\
    %\sqrt{E-p}\ c & \sqrt{E-p}\ s^* 
    %\end{array}\right)
    \nonumber\\
    &= m_i\langle\mathbf{i}|,
    \label{eq:[i|pi}
\end{align}
\begin{align}
    \langle\mathbf{i}|p_{i}
    &= 
    \left(\begin{array}{cc}
    \sqrt{E+p}\ s & -\sqrt{E+p}\ c \\
    \sqrt{E-p}\ c & \sqrt{E-p}\ s^* 
    \end{array}\right)
    \nonumber\\
    &\quad \times
    \left(\begin{array}{cc}p^0+p^3&p^1-i p^2\\p^1+i p^2&p^0-p^3\end{array}\right)
    %\nonumber\\
    %&= m_i\left(\begin{array}{cc}
    %\sqrt{E-p}\ s & -\sqrt{E-p}\ c  \\
    %\sqrt{E+p}\ c & \sqrt{E+p}\ s^*  
    %\end{array}\right)
    \nonumber\\
    &= m_i\lbrack\mathbf{i}|\label{eq:<i|pi}, 
\end{align}
where we have used
\begin{align}
    & \sqrt{E+p}\left[p^0s+p^3s-p^1c-ip^2c \right]
    \nonumber\\
    &= \sqrt{E+p}\left[
    p^0s+p\left(\cos\theta\ s -\sin\theta\ c\ e^{i\phi}\right)
    \right]
    \nonumber\\
    &= \sqrt{E+p}\left[
    p^0s+ps\left(
        (2c^2-1) - 2 c^2
    \right)\right]
    \nonumber\\
    &= \sqrt{E+p}(E - p) s
    \nonumber\\
    &= m_i\sqrt{E-p}\ s,
\end{align}
\begin{align}
    & \sqrt{E+p}\left[
    p^0c -p^1s+ip^2s-p^3c
    \right]
    \nonumber\\
    &= \sqrt{E+p}\left[
    p^0c -p\left(\sin\theta e^{-i\phi} s+\cos\theta c\right)
    \right]
    \nonumber\\
    &= \sqrt{E+p}\left[
    p^0c -p c\left(2s^*s +(1-2s^*s)\right)
    \right]
    \nonumber\\
    &= \sqrt{E+p}\left(E -p \right)c
    \nonumber\\
    &= m_i\sqrt{E -p }\ c,
\end{align}
\begin{align}
    & \sqrt{E-p}\left[
    p^0c+p^3c+p^1s^*+ip^2s^*
    \right]
    \nonumber\\
    &= \sqrt{E-p}\left[
    p^0c+p\left(\cos\theta c+\sin\theta e^{i\phi}s^*\right)
    \right]
    \nonumber\\
    &= \sqrt{E-p}\left[
    p^0c+p c\left(1-2ss^*+2 ss^*\right)
    \right]
    \nonumber\\
    &= \sqrt{E-p}\left(E+p\right)c
    \nonumber\\
    &= m_i\sqrt{E+p}\ c,
\end{align}
\begin{align}
    & \sqrt{E-p}\left[
    p^0s^*+p^1c-ip^2c-p^3s^*
    \right]
    \nonumber\\
    &= \sqrt{E-p}\left[
    p^0s^*+p\left(\sin\theta e^{-i\phi}c-\cos\theta s^*\right)
    \right]
    \nonumber\\
    &= \sqrt{E-p}\left[
    p^0s^*+ps^*\left(2c^2-(2c^2-1) \right)
    \right]
    \nonumber\\
    &= \sqrt{E-p}\left(E+p\right)s^*
    \nonumber\\
    &= m_i\sqrt{E+p}\ s^*,
\end{align}
and the double-angle identity was used on the third line of each of these.

\subsection{Lorentz and Spin/Helicity Algebras}

A review of the algebra of the generators of the Lorentz group and for the spin and helicity little group can be found in \cite{Weinberg:1995mt} and the appendix of \cite{Christensen:2013aua}.  We give the relevant generators here and their transformations of the states.  

We follow the convention of defining the spin along the direction of motion for our spinors.  This component is also known as the helicity of the particle, even for a massive particle.  However, the algebras of massive and massless particles are not the same.  Massive particles transform under rotations under the full spin group and their spin-$\pm\frac{1}{2}$ components mix with each under.  Massless particles, on the other hand, have helicity states that do not mix with each other under rotations.  They remain distinct and the consequence of a rotation is, at most, a change of their phase.  In fact, the helicity states of a massless particle can be considered distinct particles, whereas, the helicity states of a massive particle cannot since they are mixed by a rotation.

As mentioned, our spinors are eigenstates of the generator that is along the direction of motion and we combine the other two generators to form raising and lowering operators, which raise and lower the spin (or annihilate the massless helicity spinors).   They all satisfy the spin algebra of $SU(2)$, namely
\begin{align}
    \left[ \mathcal{J}^{(+)} , \mathcal{J}^{(-)} \right] &= 2\mathcal{J}^{(3)}
    \\
    \left[ \mathcal{J}^{(3)} , \mathcal{J}^{(\pm)} \right] &= \pm\mathcal{J}^{(\pm)}
\end{align}
for the Lorentz generators  and
\begin{align}
    \left[ J^{(+)} , J^{(-)} \right] &= 2J^{(3)}
    \\
    \left[ J^{(3)} , J^{(\pm)} \right] &= \pm J^{(\pm)}
\end{align}
for the spin generators.

We begin with the spin generators.  They are
\begin{align}
J^{(3)\ \mathrm{J}}_{\quad \mathrm{K}} &=
    \frac{-1}{2}\left(\begin{array}{cc}
        1 & 0 \\
        0 & -1
    \end{array}\right)
\\
J^{(+)\ \mathrm{J}}_{\quad \mathrm{K}} &=
    -e^{i \phi}\left(\begin{array}{cc}
        0 & 0\\
        1 & 0
    \end{array}\right)
\\
J^{(-)\ \mathrm{J}}_{\quad \mathrm{K}} &=
    -e^{-i \phi}\left(\begin{array}{cc}
    0 & 1\\
    0 & 0
    \end{array}\right)
\end{align}
(acting on spin indices, from the right in matrix notation).  The form of these generators is due to the fact that the spin-$-\frac{1}{2}$ component is the left column while the spin-$+\frac{1}{2}$ component is the right column.  These are equivalent with the usual Pauli matrices and related to them by a similarity transformation.  These act on the spinors with upper spin indices.  The form that acts on lower spin indices can be obtained by applying $\epsilon^{\mathrm{IJ}}$ and $\epsilon_{\mathrm{IJ}}$.  They are
\begin{align}
J^{(3)\mathrm{K}}_{\quad \ \ \mathrm{J}} &=
    \frac{1}{2}\left(\begin{array}{cc}
        1 & 0 \\
        0 & -1
    \end{array}\right)
\\
J^{(+)\mathrm{K}}_{\quad \ \ \mathrm{J}} &=
    e^{i \phi}\left(\begin{array}{cc}
        0 & 1\\
        0 & 0
    \end{array}\right)
\\
J^{(-)\mathrm{K}}_{\quad \ \ \mathrm{J}} &=
    e^{-i \phi}\left(\begin{array}{cc}
    0 & 0\\
    1 & 0
    \end{array}\right) .
\end{align}
These generators only act on spin spinors.  The helicity spinors do not transform under this symmetry group.  

We next introduce the Lorentz generators acting on $\lvert \mathrm{i}\rangle_\alpha$ and $\lvert \mathbf{i}\rangle_\alpha$, where the spin index on the second case can be either upper or lower.  They are
\begin{align}
\mathcal{J}^{(3)\ \beta}_{\quad \alpha} &=
    \frac{-1}{2}\left(
\begin{array}{cc}
 c_\theta &
   s_\theta e^{-i \phi } \\
   s_\theta e^{i \phi }  & -c_\theta \\
\end{array}
\right)
\\
\mathcal{J}^{(+)\ \beta}_{\quad \alpha} &=
    \frac{-e^{-\eta}}{2}\left(
\begin{array}{cc}
 -s_\theta & e^{-i \phi
   } (c_\theta-1) \\
 e^{i \phi } (c_\theta+1) & s_\theta \\
\end{array}
\right)
\\
\mathcal{J}^{(-)\ \beta}_{\quad \alpha} &=
    \frac{-e^{\eta}}{2}\left(
\begin{array}{cc}
 -s_\theta & e^{-i \phi
   } (c_\theta+1) \\
 e^{i \phi } (c_\theta-1) & s_\theta \\
\end{array}
\right)
\end{align}
(acting on Lorentz indices, from the left in matrix notation)
where $c_\theta=\cos\theta$, $s_\theta=\sin\theta$ and 
\begin{equation}
    \eta = \frac{1}{2}\ln\left(\frac{E+p}{E-p}\right).
\end{equation}
With these, we find
\begin{align}
    \mathcal{J}^{(3)\ \beta}_{\quad \alpha}\lvert \mathbf{j}\rangle_\beta^{\mathrm{J}} &= 
    \lvert \mathbf{j}\rangle_\alpha^{\mathrm{K}}J^{(3)\ \mathrm{J}}_{\quad \mathrm{K}}
\\
    \mathcal{J}^{(\pm)\ \beta}_{\quad \alpha}\lvert \mathbf{j}\rangle_\beta^{\mathrm{J}} &= 
    \lvert \mathbf{j}\rangle_\alpha^{\mathrm{K}}J^{(\pm)\ \mathrm{J}}_{\quad \mathrm{K}}
\end{align}
and
\begin{align}
    \mathcal{J}^{(3)\ \beta}_{\quad \alpha}\lvert \mathbf{j}\rangle_{\beta \mathrm{J}} &= 
    \lvert \mathbf{j}\rangle_{\alpha \mathrm{K}}J^{(3)\mathrm{K}}_{\quad \ \ \mathrm{J}}
\\
    \mathcal{J}^{(\pm)\ \beta}_{\quad \alpha}\lvert \mathbf{j}\rangle_{\beta \mathrm{J}} &= 
    \lvert \mathbf{j}\rangle_{\alpha \mathrm{K}}J^{(\pm)\mathrm{K}}_{\quad \ \ \mathrm{J}}
\end{align}
for the right-angle spin spinors and 
\begin{align}
    \mathcal{J}^{(3)\ \beta}_{\quad \alpha}\lvert \mathrm{j}\rangle_\beta &= 
    -\frac{1}{2}\lvert \mathrm{j}\rangle_\alpha 
\\
    \mathcal{J}^{(\pm)\ \beta}_{\quad \alpha}\lvert \mathrm{j}\rangle_\beta &= 0
\end{align}
for the right-angle helicity-$-\frac{1}{2}$ spinors, where we used $c_\theta=2c^2-1=1-2\lvert s\rvert^2$, $s_\theta =2c s^*e^{i\phi}=2c s e^{-i\phi}$ and $\exp(-\eta) =$ $ \sqrt{(E-p)/(E+p)}=0$ for the helicity spinors.

In order for the Lorentz generators to act on $\langle\mathrm{i}\rvert^\alpha$, we need to raise and lower their Lorentz indices using $\epsilon^{\alpha\beta}$ and $\epsilon_{\alpha\beta}$, to obtain
\begin{align}
    \mathcal{J}^{(3)\alpha}_{\quad \ \ \beta} &= 
    \frac{1}{2}\left(
\begin{array}{cc}
 c_\theta & e^{i \phi }
   s_\theta \\
 e^{-i \phi } s_\theta &
   -c_\theta \\
\end{array}
\right)
\\
    \mathcal{J}^{(+)\alpha}_{\quad \ \ \beta} &= \frac{e^{-\eta}}{2}\left(
\begin{array}{cc}
 -s_\theta & e^{i \phi }
   (c_\theta+1) \\
 e^{-i \phi } (c_\theta-1) & s_\theta \\
\end{array}
\right)
\\
    \mathcal{J}^{(-)\alpha}_{\quad \ \ \beta} &= \frac{e^{\eta}}{2}\left(
\begin{array}{cc}
 -s_\theta & e^{i \phi }
   (c_\theta-1) \\
 e^{-i \phi } (c_\theta+1) & s_\theta \\
\end{array}
\right) .
\end{align}
With these, we find
\begin{align}
    \mathcal{J}^{(3)\alpha}_{\quad \ \ \beta} \langle \mathbf{j}\rvert^{\beta \mathrm{J}} &= 
     \langle \mathbf{j}\rvert^{\alpha \mathrm{K}} J^{(3)\ \mathrm{J}}_{\quad \mathrm{K}}
\\
    \mathcal{J}^{(\pm)\alpha}_{\quad \ \ \beta}  \langle \mathbf{j}\rvert^{\beta \mathrm{J}} &= 
     \langle \mathbf{j}\rvert^{\alpha \mathrm{K}} J^{(\pm)\ \mathrm{J}}_{\quad \mathrm{K}}
\end{align}
and
\begin{align}
    \mathcal{J}^{(3)\alpha}_{\quad \ \ \beta} \langle \mathbf{j}\rvert^{\beta}_{\mathrm{J}} &= 
     \langle \mathbf{j}\rvert^{\alpha}_{\mathrm{K}} J^{(3)\mathrm{K}}_{\quad \ \ \mathrm{J}}
\\
    \mathcal{J}^{(\pm)\alpha}_{\quad \ \ \beta} \langle \mathbf{j}\rvert^{\beta}_{\mathrm{J}} &= 
     \langle \mathbf{j}\rvert^{\alpha}_{\mathrm{K}} J^{(\pm)\mathrm{K}}_{\quad \ \ \mathrm{J}}
\end{align}
for left-angle spin spinors and 
\begin{align}
    \mathcal{J}^{(3)\alpha}_{\quad \ \ \beta} \langle\mathrm{j}\rvert^\beta &= -\frac{1}{2}\langle\mathrm{j}\rvert^\alpha\\
    \mathcal{J}^{(\pm)\alpha}_{\quad \ \ \beta} \langle\mathrm{j}\rvert^\beta &= 0,
\end{align}
for left-angle helicity-$-\frac{1}{2}$ spinors.  

The Lorentz generators of rotations for the square spinors are given by
\begin{align}
\mathcal{J}^{(3)\ \dot{\beta}}_{\quad \dot{\alpha}} &=
    \frac{1}{2}\left(
\begin{array}{cc}
 c_\theta & e^{i \phi }
   s_\theta \\
 e^{-i \phi } s_\theta &
   -c_\theta \\
\end{array}
\right)
\\
\mathcal{J}^{(+)\ \dot{\beta}}_{\quad \dot{\alpha}} &=
    \frac{e^{\eta}}{2}\left(
\begin{array}{cc}
 -s_\theta & e^{i \phi }
   (c_\theta+1) \\
 e^{-i \phi } (c_\theta-1) & s_\theta \\
\end{array}
\right)
\\
\mathcal{J}^{(-)\ \dot{\beta}}_{\quad \dot{\alpha}} &=
    \frac{e^{-\eta}}{2}\left(
\begin{array}{cc}
 -s_\theta & e^{i \phi }
   (c_\theta-1) \\
 e^{-i \phi } (c_\theta+1) & s_\theta \\
\end{array}
\right)
\end{align}
with 
\begin{align}
    \mathcal{J}^{(3)\ \dot{\beta}}_{\quad \dot{\alpha}} \lbrack \mathbf{j}\rvert_{\dot{\beta}}^{\mathrm{J}} &= 
    \lbrack \mathbf{j}\rvert_{\dot{\alpha}}^{\mathrm{K}} J^{(3)\ \mathrm{J}}_{\quad \mathrm{K}}
\\
    \mathcal{J}^{(\pm)\ \dot{\beta}}_{\quad \dot{\alpha}} \lbrack \mathbf{j}\rvert_{\dot{\beta}}^{\mathrm{J}} &= 
    \lbrack \mathbf{j}\rvert_{\dot{\alpha}}^{\mathrm{K}} J^{(\pm)\ \mathrm{J}}_{\quad \mathrm{K}}
\end{align}
and
\begin{align}
    \mathcal{J}^{(3)\ \dot{\beta}}_{\quad \dot{\alpha}} \lbrack \mathbf{j}\rvert_{\dot{\beta} \mathrm{J}} &= 
    \lbrack \mathbf{j}\rvert_{\dot{\alpha} \mathrm{K}} J^{(3)\mathrm{K}}_{\quad \ \ \mathrm{J}}
\\
    \mathcal{J}^{(\pm)\ \dot{\beta}}_{\quad \dot{\alpha}} \lbrack \mathbf{j}\rvert_{\dot{\beta} \mathrm{J}} &= 
    \lbrack \mathbf{j}\rvert_{\dot{\alpha} \mathrm{K}} J^{(\pm)\mathrm{K}}_{\quad \ \ \mathrm{J}}
\end{align}
for the left-square spin spinors and
\begin{align}
    \mathcal{J}^{(3)\ \dot{\beta}}_{\quad \dot{\alpha}} \lbrack\mathrm{j}\rvert_{\dot{\beta}} &= +\frac{1}{2}\lbrack\mathrm{j}\rvert_{\dot{\alpha}}\\
    \mathcal{J}^{(\pm)\ \dot{\beta}}_{\quad \dot{\alpha}} \lbrack\mathrm{j}\rvert_{\dot{\beta}} &= 0
\end{align}
for the left-square helicity-$+\frac{1}{2}$ spinors.  

Raising and lowering the Lorentz indices, we have
\begin{align}
    \mathcal{J}^{(3)\dot{\alpha}}_{\quad \ \ \dot{\beta}} &= 
    \frac{-1}{2}\left(
\begin{array}{cc}
 c_\theta & e^{-i \phi }
   s_\theta \\
 e^{i \phi } s_\theta &
   -c_\theta \\
\end{array}
\right)
\\
    \mathcal{J}^{(+)\dot{\alpha}}_{\quad \ \ \dot{\beta}} &= \frac{-e^{\eta}}{2}\left(
\begin{array}{cc}
 -s_\theta & e^{-i \phi
   } (c_\theta-1) \\
 e^{i \phi } (c_\theta+1) & s_\theta \\
\end{array}
\right)
\\
    \mathcal{J}^{(-)\dot{\alpha}}_{\quad \ \ \dot{\beta}} &= \frac{-e^{-\eta}}{2}\left(
\begin{array}{cc}
 -s_\theta & e^{-i \phi
   } (c_\theta+1) \\
 e^{i \phi } (c_\theta-1) & s_\theta \\
\end{array}
\right) ,
\end{align}
with 
\begin{align}
    \mathcal{J}^{(3)\dot{\alpha}}_{\quad \ \ \dot{\beta}} \lvert\mathbf{j}\rbrack^{\dot{\beta} \mathrm{J}} &= 
    \lvert\mathbf{j}\rbrack^{\dot{\alpha} \mathrm{K}} J^{(3)\ \mathrm{J}}_{\quad \mathrm{K}}
\\
    \mathcal{J}^{(\pm)\dot{\alpha}}_{\quad \ \ \dot{\beta}} \lvert\mathbf{j}\rbrack^{\dot{\beta} \mathrm{J}} &= 
    \lvert\mathbf{j}\rbrack^{\dot{\alpha} \mathrm{K}} J^{(\pm)\ \mathrm{J}}_{\quad \mathrm{K}}
\end{align}
and
\begin{align}
    \mathcal{J}^{(3)\dot{\alpha}}_{\quad \ \ \dot{\beta}} \lvert\mathbf{j}\rbrack^{\dot{\beta} \mathrm{J}} &= 
    \lvert\mathbf{j}\rbrack^{\dot{\alpha} \mathrm{K}} J^{(3)\mathrm{K}}_{\quad \ \ \mathrm{J}}
\\
    \mathcal{J}^{(\pm)\dot{\alpha}}_{\quad \ \ \dot{\beta}} \lvert\mathbf{j}\rbrack^{\dot{\beta} \mathrm{J}} &= 
    \lvert\mathbf{j}\rbrack^{\dot{\alpha} \mathrm{K}} J^{(\pm)\mathrm{K}}_{\quad \ \ \mathrm{J}}
\end{align}
and
\begin{align}
    \mathcal{J}^{(3)\dot{\alpha}}_{\quad \ \ \dot{\beta}} \lvert\mathrm{j}\rbrack^{\dot{\beta}} &= +\frac{1}{2}\lvert\mathrm{j}\rbrack^{\dot{\alpha}}\\
    \mathcal{J}^{(\pm)\dot{\alpha}}_{\quad \ \ \dot{\beta}} \lvert\mathrm{j}\rbrack^{\dot{\beta}} &= 0.
\end{align}

We use all these identities to validate our implementation, as we discuss in Sec.~\ref{sec:validation:boost}.

\section{\label{app:testing:SM}Some Details for the Implemented SM $\mathbf{2\to2}$ Processes}
In this appendix, we give some further details for the SM $2\to2$ processes.  We include the file names for each process in Table~\ref{tab:testing:SM:filenames}, where the position of the file name corresponds with the position in Table~\ref{tab:user-guide:SM Processes}. 
 
\begin{table*}
  \begin{center}
    \renewcommand{\arraystretch}{1.25}
    \setlength{\tabcolsep}{10pt}
    \begin{tabular}{|lll|}
       \hline \multicolumn{3}{|l|}{$q,\bar{q}\to q,\bar{q}$ \ Neutral } \\\hline
       \url{uucc.cpp} & \url{ucuc.cpp} &\\
       \url{uuuu.cpp} & \url{uuuu2.cpp} &\\
       \url{uuss.cpp} & \url{usus.cpp} &\\
       \url{uudd.cpp} & \url{udud.cpp} &\\
       \url{ddss.cpp} & \url{dsds.cpp} &\\
       \url{dddd.cpp} & \url{dddd2.cpp} &\\
       \hline \multicolumn{3}{|l|}{$q,\bar{q}\to l,\bar{l}$ \ Neutral }\\\hline
       \url{uuee.cpp} & \url{ueue.cpp} &\\
       \url{uunn.cpp} & \url{unnu.cpp} &\\
       \url{ddee.cpp} & \url{deed.cpp} &\\
       \url{ddnn.cpp} & \url{dnnd.cpp} &\\
       \hline \multicolumn{3}{|l|}{$l,\bar{l}\to l,\bar{l}$ \ Neutral }\\\hline
       \url{eemm.cpp} & \url{emem.cpp} &\\
       \url{eeee.cpp} & \url{eeee2.cpp} &\\
       \url{eenmnm.cpp} & \url{enmenm.cpp} &\\
       \url{eenene.cpp} & \url{enenee.cpp} &\\
       \url{nenenmnm.cpp} & \url{nenmnenm.cpp} &\\
       \url{nenenene.cpp} & \url{nenenene2.cpp} &\\
       \hline \multicolumn{3}{|l|}{$f,\bar{f}\to f,\bar{f}$ \ Charged }\\\hline
       \url{udtb.cpp} & \url{ubdt.cpp} &\\
       \url{udnl.cpp} & \url{ulnd.cpp} & \\
       \url{menn.cpp} & \url{mnen.cpp} &\\
       \hline \multicolumn{3}{|l|}{$l,\bar{l}\to b,\bar{b}$ \ Neutral }\\\hline
       \url{eehh.cpp} & \url{eheh.cpp} &\\
       \url{eeAh.cpp} & \url{eAeh.cpp} &\\
       \url{eeZh.cpp} & \url{eZeh.cpp} &\\
       \url{eeAA.cpp} & \url{eAAe.cpp} & \url{AAee.cpp} \\
       \url{AZee.cpp} & \url{AeZe.cpp} &\\
       \url{eeZZ.cpp} & \url{eZZe.cpp} &\\
       \url{eeWW.cpp} & \url{eWWe.cpp} &\\
       \url{nnZh.cpp} & \url{nZnh.cpp} &\\
       \url{nnZZ.cpp} & \url{nZZn.cpp} &\\
       \url{nnWW.cpp} & \url{nWWn.cpp} &\\
       \hline \multicolumn{3}{|l|}{$q,\bar{q}\to b,\bar{b}$ \ Neutral without $g$}\\\hline
       \url{uuhh.cpp} & \url{uhuh.cpp} &\\
       \url{uuAh.cpp} & \url{uAuh.cpp} &\\
       \url{uuZh.cpp} & \url{uZuh.cpp} &\\
       \url{uuAA.cpp} & \url{AAuu.cpp} & \url{uAAu.cpp} \\
       \url{AZuu.cpp} & \url{AuZu.cpp} & \\
       \hline
    \end{tabular}
    \begin{tabular}{|lll|}
       \hline \multicolumn{3}{|l|}{$q,\bar{q}\to b,\bar{b}$ \ Neutral without $g$ Continued }\\\hline
       \url{uuZZ.cpp} & \url{uZZu.cpp} &\\
       \url{uuWW.cpp} & \url{uWWu.cpp} &\\
       \url{ddhh.cpp} & \url{dhdh.cpp} &\\
       \url{ddAh.cpp} & \url{dAdh.cpp} &\\
       \url{ddZh.cpp} & \url{dZdh.cpp} &\\
       \url{ddAA.cpp} & \url{AAdd.cpp} & \url{dAAd.cpp} \\
       \url{AZdd.cpp} & \url{AdZd.cpp} &\\
       \url{ddZZ.cpp} & \url{dZZd.cpp} &\\
       \url{ddWW.cpp} & \url{WdWd.cpp} &\\
       \hline \multicolumn{3}{|l|}{$q,\bar{q}\to b,\bar{b}$ \ Neutral with $g$}\\\hline
       \url{uguh.cpp} & \url{hguu.cpp} &\\
       \url{gZuu.cpp} & \url{guZu.cpp} &\\
       \url{gAuu.cpp} & \url{guAu.cpp} &\\
       \url{gguu.cpp} & \url{gugu.cpp} &\\
       \url{dgdh.cpp} & \url{hgdd.cpp} &\\
       \url{gZdd.cpp} & \url{gdZd.cpp} &\\      \url{gAdd.cpp} & \url{dAgd.cpp} &\\
       \url{ggdd.cpp} & \url{dggd.cpp} &\\
       \hline \multicolumn{3}{|l|}{$f,\bar{f}\to b,\bar{b}$ \ Charged} \\\hline
       \url{enWh.cpp} & \url{hnWe.cpp} &\\
       \url{enAW.cpp} & \url{AnWe.cpp} &\\
       \url{enZW.cpp} & \url{ZnWe.cpp} &\\      \url{udWh.cpp} & \url{uhWd.cpp} &\\
       \url{udAW.cpp} & \url{AWud.cpp} &\\ 
       \url{dWgu.cpp} & \url{gWud.cpp} &\\
       \url{udZW.cpp} & \url{WdZu.cpp} &\\
       \hline \multicolumn{3}{|l|}{$b,\bar{b}\to b,\bar{b}$}  \\\hline
       \url{hhhh.cpp} &&\\
       \url{hhZZ.cpp} & \url{hZZh.cpp} &\\
       \url{hhWW.cpp} & \url{hWWh.cpp} &\\
       \url{AhWW.cpp} & \url{AWWh.cpp} &\\
       \url{ZhWW.cpp} & \url{ZWWh.cpp} &\\       \url{AAWW.cpp} & \url{AWAW.cpp} &\\
       \url{AZWW.cpp} & \url{AWZW.cpp} &\\
       \url{ZZZZ.cpp} &&\\
       \url{ZZWW.cpp} & \url{ZWZW.cpp} &\\
       \url{WWWW.cpp} & \url{WWWW2.cpp} &\\
       \url{gggg.cpp} &&\\
       &&\\
       \hline
    \end{tabular}
  \end{center}
  \caption{This is a list of the file names for the SM processes implemented.  They are found in the \url{SM} directory.  The processes should be clear from the naming, but correspond directly to the processes presented in Table~\ref{tab:user-guide:SM Processes}.}
  \label{tab:testing:SM:filenames}
\end{table*}

For each crossing-symmetry set of processes, we include a table of the masses and momenta used for the validations.  These masses and momenta were taken from the first process in the list.  The masses and momenta for the other processes related by crossing symmetry are typically similar, but are sometimes different, mainly to accommodate threshold momenta for each direction.  We attempted to include a variety of masses and momenta including some where each particle mass played a significant role in the calculation, as well as when the momentum was far above threshold and near threshold.  Agreement was found for all values described here.

When photons or gluons are in the initial state, we test their squared amplitude both without summing over the initial photon or gluon polarization and with averaging over them.  So, for example, when we test the process $\gamma,Z\to e,\bar{e}$, we test the polarized process $\gamma^+,Z\to e,\bar{e}$ as well as the unpolarized process $\gamma,Z\to e,\bar{e}$ where the photon helicities are summed over.  The masses and momenta for the tests are the same.

For other details, we refer the reader to Sec.~\ref{sec:validation:2to2}.  The order for the processes is the same as in Table~\ref{tab:user-guide:SM Processes}.  

\subsection{$\mathbf{q,\bar{q}\to q,\bar{q}}$ Neutral}
In this subsection, we consider processes with four quarks via a neutral channel.

\begin{center}
\begin{tabular}{|l|l|l|l|l|}
    \multicolumn{5}{c}{$\mathbf{u,\bar{u}\to c,\bar{c}}$ and $\mathbf{u,c\to u,c}$}\\
    \hline
    $m_u$ & $m_c$ & $m_h$ & $M_W$ & $p_{in}$  \\
    \hline\hline
    0.0042 & 1.23 & 125 & 80.385 & 250 \\
    0.0042 & 1.23 & 125 & 80.385 & 1.25 \\
    1.23 & 0.0042 & 125 & 80.385 & 0.005 \\
    1.2 & 1.23 & 125 & 80.385 & 0.3 \\
    1.2 & 1.23 & 125 & 2.11 & 0.3 \\
    1.2 & 1.23 & 125 & 0.006 & 0.3 \\
    1.2 & 1.23 & 3.125 & 2.11 & 0.3 \\
    1.2 & 1.23 & 3.125 & 0.006 & 0.3 \\
    \hline
\end{tabular}
\end{center}

\begin{center}
\begin{tabular}{|l|l|l|l|}
    \multicolumn{4}{c}{$\mathbf{u,\bar{u}\to u,\bar{u}}$ and $\mathbf{u,u\to u,u}$}\\
    \hline
    $m_u$ & $m_h$ & $M_W$ & $p_{in}$  \\
    \hline\hline
    0.0042 & 125 & 80.385 & 250 \\
    0.0042 & 125 & 80.385 & 0.005 \\
    1.2 & 125 & 80.385 & 0.3 \\
    1.2 & 125 & 2.11 & 0.3 \\
    1.2 & 125 & 0.006 & 0.3 \\
    1.2 & 3.125 & 2.11 & 0.3 \\
    1.2 & 3.125 & 0.006 & 0.3 \\
    \hline
\end{tabular}
\end{center}

\begin{center}
\begin{tabular}{|l|l|l|l|l|}
\multicolumn{5}{c}{$\mathbf{u,\bar{u}\to s,\bar{s}}$ and $\mathbf{u,s\to u,s}$}\\
    \hline
    $m_u$ & $m_s$ & $m_h$ & $M_W$ & $p_{in}$  \\
    \hline\hline
    0.0042 & 1.23 & 125 & 80.385 & 250 \\
    0.0042 & 1.23 & 125 & 80.385 & 1.25 \\
    1.23 & 0.0042 & 125 & 80.385 & 0.005 \\
    1.2 & 1.23 & 125 & 80.385 & 0.3 \\
    1.2 & 1.23 & 125 & 2.11 & 0.3 \\
    1.2 & 1.23 & 125 & 0.006 & 0.3 \\
    1.2 & 1.23 & 3.125 & 2.11 & 0.3 \\
    1.2 & 1.23 & 3.125 & 0.006 & 0.3 \\
    \hline
\end{tabular}
\end{center}

\begin{center}
\begin{tabular}{|l|l|l|l|l|}
    \multicolumn{5}{c}{$\mathbf{u,\bar{u}\to d,\bar{d}}$ and $\mathbf{u,d\to u,d}$}\\
    \hline
    $m_u$ & $m_d$ & $m_h$ & $M_W$ & $p_{in}$  \\
    \hline\hline
    0.0042 & 1.23 & 125 & 80.385 & 250 \\
    0.0042 & 1.23 & 125 & 80.385 & 1.25 \\
    1.23 & 0.0042 & 125 & 80.385 & 0.005 \\
    1.2 & 1.23 & 125 & 80.385 & 0.3 \\
    1.2 & 1.23 & 125 & 2.11 & 0.3 \\
    1.2 & 1.23 & 125 & 0.006 & 0.3 \\
    1.2 & 1.23 & 3.125 & 2.11 & 0.3 \\
    1.2 & 1.23 & 3.125 & 0.006 & 0.3 \\
    \hline
\end{tabular}
\end{center}

\begin{center}
\begin{tabular}{|l|l|l|l|l|}
    \multicolumn{5}{c}{$\mathbf{d,\bar{d}\to s,\bar{s}}$ and $\mathbf{d,s\to d,s}$}\\
    \hline
    $m_d$ & $m_s$ & $m_h$ & $M_W$ & $p_{in}$  \\
    \hline\hline
    0.0042 & 1.23 & 125 & 80.385 & 250 \\
    0.0042 & 1.23 & 125 & 80.385 & 1.25 \\
    1.23 & 0.0042 & 125 & 80.385 & 0.005 \\
    1.2 & 1.23 & 125 & 80.385 & 0.3 \\
    1.2 & 1.23 & 125 & 2.11 & 0.3 \\
    1.2 & 1.23 & 125 & 0.006 & 0.3 \\
    1.2 & 1.23 & 3.125 & 2.11 & 0.3 \\
    1.2 & 1.23 & 3.125 & 0.006 & 0.3 \\
    \hline
\end{tabular}
\end{center}

\begin{center}
\begin{tabular}{|l|l|l|l|l|}
    \multicolumn{4}{c}{$\mathbf{d,\bar{d}\to d,\bar{d}}$ and $\mathbf{d,d\to d,d}$}\\
    \hline
    $m_d$ & $m_h$ & $M_W$ & $p_{in}$  \\
    \hline\hline
    0.0042 & 125 & 80.385 & 250 \\
    0.0042 & 125 & 80.385 & 0.005 \\
    1.2 & 125 & 80.385 & 0.3 \\
    1.2 & 125 & 2.11 & 0.3 \\
    1.2 & 125 & 0.006 & 0.3 \\
    1.2 & 3.125 & 2.11 & 0.3 \\
    1.2 & 3.125 & 0.006 & 0.3 \\
    \hline
\end{tabular}
\end{center}

\subsection{$\mathbf{q,\bar{q}\to l,\bar{l}}$ Neutral}
Next, we consider amplitudes with two quarks and two leptons via a neutral channel.

\begin{center}
\begin{tabular}{|l|l|l|l|l|}
    \multicolumn{5}{c}{$\mathbf{u,\bar{u}\to e,\bar{e}}$ and $\mathbf{u,e\to u,e}$}\\
    \hline
    $m_u$ & $m_e$ & $m_h$ & $M_W$ & $p_{in}$  \\
    \hline\hline
    0.0042 & 0.0005 & 125 & 80.385 & 250 \\
    0.0042 & 1.23 & 125 & 80.385 & 1.25 \\
    1.23 & 0.0042 & 125 & 80.385 & 0.005 \\
    1.2 & 1.23 & 125 & 80.385 & 0.3 \\
    1.2 & 1.23 & 125 & 2.11 & 0.3 \\
    1.2 & 1.23 & 125 & 0.006 & 0.3 \\
    1.2 & 1.23 & 3.125 & 2.11 & 0.3 \\
    1.2 & 1.23 & 3.125 & 0.006 & 0.3 \\
    \hline
\end{tabular}
\end{center}

\begin{center}
\begin{tabular}{|l|l|l|}
    \multicolumn{3}{c}{$\mathbf{u,\bar{u}\to \nu_e,\bar{\nu}_e}$ and $\mathbf{u,\nu_\mu\to \nu_\mu,u}$}\\
    \hline
    $m_u$ & $M_W$ & $p_{in}$  \\
    \hline\hline
    0.0042 & 80.385 & 250 \\
    0.0042 & 80.385 & 0.001 \\
    0.0042 & 0.015 & 0.001 \\
    0.0042 & 0.0006 & 0.001 \\
    \hline
\end{tabular}
\end{center}

\begin{center}
\begin{tabular}{|l|l|l|l|l|}
    \multicolumn{5}{c}{$\mathbf{d,\bar{d}\to e,\bar{e}}$ and $\mathbf{d,e\to e,d}$}\\
    \hline
    $m_d$ & $m_e$ & $m_h$ & $M_W$ & $p_{in}$  \\
    \hline\hline
    0.0042 & 0.0005 & 125 & 80.385 & 250 \\
    0.0042 & 1.23 & 125 & 80.385 & 1.25 \\
    1.23 & 0.0042 & 125 & 80.385 & 0.005 \\
    1.2 & 1.23 & 125 & 80.385 & 0.3 \\
    1.2 & 1.23 & 125 & 2.11 & 0.3 \\
    1.2 & 1.23 & 125 & 0.006 & 0.3 \\
    1.2 & 1.23 & 3.125 & 2.11 & 0.3 \\
    1.2 & 1.23 & 3.125 & 0.006 & 0.3 \\
    \hline
\end{tabular}
\end{center}

\begin{center}
\begin{tabular}{|l|l|l|l|}
    \multicolumn{3}{c}{$\mathbf{d,\bar{d}\to \nu_e,\bar{\nu}_e}$ and $\mathbf{d,\nu_\mu\to \nu_\mu,d}$}\\
    \hline
    $m_d$ & $M_W$ & $p_{in}$  \\
    \hline\hline
    0.0042 & 80.385 & 250 \\
    0.0042 & 80.385 & 0.001 \\
    0.0042 & 0.015 & 0.001 \\
    0.0042 & 0.0006 & 0.001 \\
    \hline
\end{tabular}
\end{center} 

\subsection{$\mathbf{l,\bar{l}\to l,\bar{l}}$ Neutral}
Next, we consider processes with four leptons via a neutral channel.

\begin{center}
\begin{tabular}{|l|l|l|l|l|}
    \multicolumn{5}{c}{$\mathbf{e,\bar{e}\to \mu,\bar{\mu}}$ and $\mathbf{e,\mu\to e,\mu}$}\\
    \hline
    $m_e$ & $m_\mu$ & $m_h$ & $M_W$ & $p_{in}$  \\
    \hline\hline
    0.0005 & 0.105 & 125 & 80.385 & 250 \\
    0.0005 & 0.105 & 125 & 80.385 & 0.11 \\
    0.105 & 0.0005 & 125 & 80.385 & 0.005 \\
    0.1 & 0.105 & 125 & 80.385 & 0.05 \\
    0.1 & 0.105 & 125 & 0.11 & 0.05 \\
    0.1 & 0.105 & 125 & 0.006 & 0.05 \\
    0.1 & 0.105 & 0.125 & 0.11 & 0.05 \\
    0.1 & 0.105 & 0.125 & 0.006 & 0.05 \\
    \hline
\end{tabular}
\end{center}

\begin{center}
\begin{tabular}{|l|l|l|l|}
    \multicolumn{4}{c}{$\mathbf{e,\bar{e}\to e,\bar{e}}$ and $\mathbf{e,e\to e,e}$}\\
    \hline
    $m_e$ & $m_h$ & $M_W$ & $p_{in}$  \\
    \hline\hline
    0.0005 & 125 & 80.385 & 250 \\
    0.0005 & 125 & 80.385 & 0.005 \\
    0.1 & 125 & 80.385 & 0.05 \\
    0.1 & 125 & 0.11 & 0.05 \\
    0.1 & 125 & 0.006 & 0.05 \\
    0.1 & 0.125 & 0.11 & 0.05 \\
    0.1 & 0.125 & 0.006 & 0.05 \\
    \hline
\end{tabular}
\end{center}

\begin{center}
\begin{tabular}{|l|l|l|}
    \multicolumn{3}{c}{$\mathbf{e,\bar{e}\to \nu_\mu,\bar{\nu}_\mu}$ and $\mathbf{e,\nu_\mu\to e,\nu_\mu}$}\\
    \hline
    $m_e$ & $M_W$ & $p_{in}$  \\
    \hline\hline
    0.0005 & 80.385 & 250 \\
    0.0005 & 80.385 & 0.001 \\
    0.0005 & 0.015 & 0.001 \\
    0.0005 & 0.0006 & 0.001 \\
    \hline
\end{tabular}
\end{center}

\begin{center}
\begin{tabular}{|l|l|l|}
    \multicolumn{3}{c}{$\mathbf{e,\bar{e}\to \nu_e,\bar{\nu}_e}$ and $\mathbf{e,\nu_e\to \nu_e,e}$}\\
    \hline
    $m_e$ & $M_W$ & $p_{in}$  \\
    \hline\hline
    0.0005 & 80.385 & 250 \\
    0.0005 & 80.385 & 0.001 \\
    0.0005 & 0.015 & 0.001 \\
    0.0005 & 0.0006 & 0.001 \\
    \hline
\end{tabular}
\end{center}

\begin{center}
\begin{tabular}{|l|l|}
    \multicolumn{2}{c}{$\mathbf{\nu_e,\bar{\nu}_e\to \nu_\mu,\bar{\nu}_\mu}$ and $\mathbf{\nu_e,\nu_\mu\to \nu_e,\nu_\mu}$}\\
    \hline
    $M_W$ & $p_{in}$  \\
    \hline\hline
    80.385 & 250 \\
    80.385 & 0.001 \\
    0.015 & 0.001 \\
    0.0006 & 0.001 \\
    \hline
\end{tabular}
\end{center} 

\begin{center}
\begin{tabular}{|l|l|}
    \multicolumn{2}{c}{$\mathbf{\nu_e,\bar{\nu}_e\to \nu_e,\bar{\nu}_e}$ and $\mathbf{\nu_e,\nu_e\to \nu_e,\nu_e}$}\\
    \hline
    $M_W$ & $p_{in}$  \\
    \hline\hline
    80.385 & 250 \\
    80.385 & 0.001 \\
    0.015 & 0.001 \\
    0.0006 & 0.001 \\
    \hline
\end{tabular}
\end{center}

\subsection{$\mathbf{f,\bar{f}\to f,\bar{f}}$ Charged}
Next we consider four-fermion amplitudes that have a charged channel.

\begin{center}
\begin{tabular}{|l|l|l|l|l|l|}
    \multicolumn{6}{c}{$\mathbf{u,\bar{d}\to t,\bar{b}}$ and $\mathbf{u,b\to d,t}$}\\
    \hline
    $m_u$ & $m_d$ & $m_t$ & $m_b$ & $M_W$ & $p_{in}$  \\
    \hline\hline
    0.0042 & 0.0075 & 172.5 & 4.25 & 80.385 & 250 \\
    0.0042 & 0.0075 & 172.5 & 4.25 & 80.385 & 173 \\
    0.0042 & 0.0075 & 0.005 & 0.0047 & 80.385 & 250 \\
    0.0042 & 0.0075 & 0.005 & 0.0047 & 80.385 & 0.001 \\
    0.0042 & 0.0075 & 0.005 & 0.0047 & 0.006 & 250 \\
    0.0042 & 0.0075 & 0.005 & 0.0047 & 0.006 & 0.001 \\
    0.0042 & 0.0075 & 0.005 & 0.0047 & 0.0006 & 0.001 \\
    \hline
\end{tabular}
\end{center}

\begin{center}
\begin{tabular}{|l|l|l|l|l|}
    \multicolumn{5}{c}{$\mathbf{u,\bar{d}\to \nu_\tau,\bar{\tau}}$ and $\mathbf{u,\tau\to \nu_\tau,d}$}\\
    \hline
    $m_u$ & $m_d$ & $m_\tau$ & $M_W$ & $p_{in}$  \\
    \hline\hline
    0.0042 & 0.0075 & 1.777 & 80.385 & 250 \\
    0.0042 & 0.0075 & 1.777 & 80.385 & 1.8 \\
    0.0042 & 0.0075 & 0.005 & 80.385 & 250 \\
    0.0042 & 0.0075 & 0.005 & 80.385 & 0.001 \\
    0.0042 & 0.0075 & 0.005 & 0.006 & 250 \\
    0.0042 & 0.0075 & 0.005 & 0.006 & 0.001 \\
    0.0042 & 0.0075 & 0.005 & 0.0006 & 0.001 \\
    \hline
\end{tabular}
\end{center}

\begin{center}
\begin{tabular}{|l|l|l|l|}
    \multicolumn{4}{c}{$\mathbf{\mu,\bar{e}\to \bar{\nu}_e,\nu_\mu}$ and $\mathbf{\mu,\nu_e\to e,\nu_\mu}$}\\
    \hline
    $m_e$ & $m_\mu$ & $M_W$ & $p_{in}$  \\
    \hline\hline
    0.0005 & 0.106 & 80.385 & 250 \\
    0.0005 & 0.106 & 80.385 & 0.05 \\
    0.106 & 0.0005 & 80.385 & 250 \\
    0.106 & 0.0005 & 80.385 & 0.05 \\
    0.105 & 0.106 & 80.385 & 250 \\
    0.105 & 0.106 & 80.385 & 0.05 \\
    105 & 106 & 0.1 & 500 \\
    105 & 106 & 0.1 & 5 \\
    0.105 & 0.106 & 0.11 & 250 \\
    0.105 & 0.106 & 0.11 & 0.005 \\
    \hline
\end{tabular}
\end{center}

\subsection{$\mathbf{l,\bar{l}\to b,\bar{b}}$ Neutral}
We next move on to amplitudes with two leptons and two bosons, via a neutral channel.

\begin{center}
\begin{tabular}{|l|l|l|l|}
    \multicolumn{4}{c}{$\mathbf{e,\bar{e}\to h,h}$ and $\mathbf{e,h\to e,h}$}\\
    \hline
    $m_e$ & $m_h$ & $M_W$ & $p_{in}$  \\
    \hline\hline
    0.0005 & 125 & 80.385 & 250 \\
    0.0005 & 125 & 80.385 & 126 \\
    0.1 & 0.15 & 80.385 & 0.2 \\
    0.1 & 0.15 & 0.11 & 0.2 \\
    0.1 & 0.15 & 0.006 & 0.2 \\
    \hline
\end{tabular}
\end{center}

\begin{center}
\begin{tabular}{|l|l|l|l|}
    \multicolumn{4}{c}{$\mathbf{e,\bar{e}\to \gamma,h}$ and $\mathbf{e,\gamma\to e,h}$}\\
    \hline
    $m_e$ & $m_h$ & $M_W$ & $p_{in}$  \\
    \hline\hline
    0.0005 & 125 & 80.385 & 250 \\
    0.0005 & 125 & 80.385 & 125.1 \\
    125.1 & 125 & 80.385 & 95 \\
    125 & 0.0005 & 80.385 & 125.1 \\
    \hline
\end{tabular}
\end{center}

\begin{center}
\begin{tabular}{|l|l|l|l|}
    \multicolumn{4}{c}{$\mathbf{e,\bar{e}\to Z,h}$ and $\mathbf{e,Z\to e,h}$}\\
    \hline
    $m_e$ & $m_h$ & $M_W$ & $p_{in}$  \\
    \hline\hline
    0.0005 & 125 & 80.385 & 250 \\
    0.0005 & 125 & 80.385 & 125.1 \\
    125.1 & 125 & 80.385 & 95 \\
    125 & 0.0005 & 80.385 & 125.1 \\
    \hline
\end{tabular}
\end{center}

\begin{center}
\begin{tabular}{|l|l|}
    \multicolumn{2}{c}{$\mathbf{e,\bar{e}\to \gamma,\gamma}$ and $\mathbf{e,\gamma\to \gamma,e}$ and $\mathbf{\gamma,\gamma\to e,\bar{e}}$}\\
    \hline
    $m_e$ & $p_{in}$  \\
    \hline\hline
    0.0005 & 250 \\
    0.0005 & 0.0001 \\
    \hline
\end{tabular}
\end{center}

\begin{center}
\begin{tabular}{|l|l|l|}
    \multicolumn{3}{c}{$\mathbf{\gamma,Z\to \bar{e},e}$ and $\mathbf{\gamma,e\to Z,e}$}\\
    \hline
    $m_e$ & $M_W$ & $p_{in}$  \\
    \hline\hline
    0.0005 & 80.385 & 250 \\
    0.0005 & 80.385 & 125.1 \\
    125 & 80.385 & 250 \\
    125 & 80.385 & 125.1 \\
    \hline
\end{tabular}
\end{center}

\begin{center}
\begin{tabular}{|l|l|l|l|}
    \multicolumn{4}{c}{$\mathbf{e,\bar{e}\to Z,Z}$ and $\mathbf{e,Z\to Z,e}$}\\
    \hline
    $m_e$ & $m_h$ & $M_W$ & $p_{in}$  \\
    \hline\hline
    0.0005 & 125 & 80.385 & 250 \\
    0.0005 & 125 & 80.385 & 125.1 \\
    125.1 & 125 & 80.385 & 95 \\
    125 & 0.0005 & 80.385 & 125.1 \\
    \hline
\end{tabular}
\end{center}

\begin{center}
\begin{tabular}{|l|l|l|l|}
    \multicolumn{4}{c}{$\mathbf{e,\bar{e}\to W,\bar{W}}$ and $\mathbf{e,W\to W,e}$}\\
    \hline
    $m_e$ & $m_h$ & $M_W$ & $p_{in}$  \\
    \hline\hline
    0.0005 & 125 & 80.385 & 250 \\
    0.0005 & 125 & 80.385 & 2500 \\
    0.0005 & 125 & 80.385 & 125.1 \\
    125.1 & 125 & 80.385 & 95 \\
    125 & 0.0005 & 80.385 & 125.1 \\
    \hline
\end{tabular}
\end{center}

\begin{center}
\begin{tabular}{|l|l|l|}
    \multicolumn{3}{c}{$\mathbf{\nu_e,\bar{\nu}_e\to Z,h}$ and $\mathbf{\nu_e,Z\to \nu_e,h}$}\\
    \hline
    $m_h$ & $M_W$ & $p_{in}$  \\
    \hline\hline
    125 & 80.385 & 250 \\
    125 & 80.385 & 125.1 \\
    0.0005 & 80.385 & 125.1 \\
    \hline
\end{tabular}
\end{center}

\begin{center}
\begin{tabular}{|l|l|}
    \multicolumn{2}{c}{$\mathbf{\nu_e,\bar{\nu}_e\to Z,Z}$ and $\mathbf{\nu_e,Z\to Z,\nu_e}$}\\
    \hline
    $M_W$ & $p_{in}$  \\
    \hline\hline
    80.385 & 250 \\
    80.385 & 125.1 \\
    80.385 & 95 \\
    8 & 125.1 \\
    \hline
\end{tabular}
\end{center}

\begin{center}
\begin{tabular}{|l|l|l|}
    \multicolumn{3}{c}{$\mathbf{\nu_e,\bar{\nu}_e\to W,\bar{W}}$ and $\mathbf{\nu_e,W\to W,\nu_e}$}\\
    \hline
    $m_e$ & $M_W$ & $p_{in}$  \\
    \hline\hline
    0.0005 & 80.385 & 250 \\
    0.0005 & 80.385 & 125.1 \\
    125.1 & 80.385 & 95 \\
    125 & 8 & 125.1 \\
    \hline
\end{tabular}
\end{center}

\subsection{$\mathbf{q,\bar{q}\to b,\bar{b}}$ Neutral Without Gluons}
We next consider processes with two quarks and two bosons that are not gluons via a neutral channel.

\begin{center}
\begin{tabular}{|l|l|l|l|}
    \multicolumn{4}{c}{$\mathbf{u,\bar{u}\to h,h}$ and $\mathbf{u,h\to u,h}$}\\
    \hline
    $m_u$ & $m_h$ & $M_W$ & $p_{in}$  \\
    \hline\hline
    0.0042 & 125 & 80.385 & 250 \\
    0.0042 & 125 & 80.385 & 126 \\
    0.1 & 0.15 & 80.385 & 0.2 \\
    0.1 & 0.15 & 0.11 & 0.2 \\
    0.1 & 0.15 & 0.006 & 0.2 \\
    \hline
\end{tabular}
\end{center}

\begin{center}
\begin{tabular}{|l|l|l|l|}
    \multicolumn{4}{c}{$\mathbf{u,\bar{u}\to \gamma,h}$ and $\mathbf{u,\gamma\to u,h}$}\\
    \hline
    $m_u$ & $m_h$ & $M_W$ & $p_{in}$  \\
    \hline\hline
    0.0042 & 125 & 80.385 & 250 \\
    0.0042 & 125 & 80.385 & 125.1 \\
    125.1 & 125 & 80.385 & 95 \\
    125 & 0.0005 & 80.385 & 125.1 \\
    \hline
\end{tabular}
\end{center}

\begin{center}
\begin{tabular}{|l|l|l|l|}
    \multicolumn{4}{c}{$\mathbf{u,\bar{u}\to Z,h}$ and $\mathbf{u,Z\to u,h}$}\\
    \hline
    $m_u$ & $m_h$ & $M_W$ & $p_{in}$  \\
    \hline\hline
    0.0042 & 125 & 80.385 & 250 \\
    0.0042 & 125 & 80.385 & 125.1 \\
    125.1 & 125 & 80.385 & 95 \\
    125 & 0.0005 & 80.385 & 125.1 \\
    \hline
\end{tabular}
\end{center}

\begin{center}
\begin{tabular}{|l|l|}
    \multicolumn{2}{c}{$\mathbf{u,\bar{u}\to \gamma,\gamma}$ and $\mathbf{\gamma,\gamma\to u,\bar{u}}$ and $\mathbf{u,\gamma\to \gamma,u}$}\\
    \hline
    $m_u$ & $p_{in}$  \\
    \hline\hline
    0.0042 & 250 \\
    0.0042 & 0.001 \\
    \hline
\end{tabular}
\end{center}

\begin{center}
\begin{tabular}{|l|l|l|}
    \multicolumn{3}{c}{$\mathbf{\gamma,Z\to \bar{u},u}$ and $\mathbf{\gamma,u\to Z,u}$}\\
    \hline
    $m_u$ & $M_W$ & $p_{in}$  \\
    \hline\hline
    0.0042 & 80.385 & 250 \\
    0.0042 & 80.385 & 125.1 \\
    125 & 80.385 & 250 \\
    125 & 80.385 & 125.1 \\
    \hline
\end{tabular}
\end{center}

\begin{center}
\begin{tabular}{|l|l|l|l|}
    \multicolumn{4}{c}{$\mathbf{u,\bar{u}\to Z,Z}$ and $\mathbf{u,Z\to Z,u}$}\\
    \hline
    $m_u$ & $m_h$ & $M_W$ & $p_{in}$  \\
    \hline\hline
    0.0042 & 125 & 80.385 & 250 \\
    0.0042 & 125 & 80.385 & 125.1 \\
    125.1 & 125 & 80.385 & 95 \\
    125 & 0.0005 & 80.385 & 125.1 \\
    \hline
\end{tabular}
\end{center}

\begin{center}
\begin{tabular}{|l|l|l|l|l|}
    \multicolumn{5}{c}{$\mathbf{u,\bar{u}\to W,\bar{W}}$ and $\mathbf{u,W\to W,u}$}\\
    \hline
    $m_u$ & $m_d$ & $m_h$ & $M_W$ & $p_{in}$  \\
    \hline\hline
    0.0042 & 0.0075 & 125 & 80.385 & 250 \\
    0.0042 & 0.0075 & 125 & 80.385 & 2500 \\
    0.0042 & 0.0075 & 125 & 80.385 & 125.1 \\
    125.1 & 0.0075 & 125 & 80.385 & 95 \\
    125 & 0.0075 & 0.0005 & 80.385 & 125.1 \\
    \hline
\end{tabular}
\end{center}

\begin{center}
\begin{tabular}{|l|l|l|l|}
    \multicolumn{4}{c}{$\mathbf{d,\bar{d}\to h,h}$ and $\mathbf{d,h\to d,h}$}\\
    \hline
    $m_d$ & $m_h$ & $M_W$ & $p_{in}$  \\
    \hline\hline
    0.0075 & 125 & 80.385 & 250 \\
    0.0075 & 125 & 80.385 & 126 \\
    0.1 & 0.15 & 80.385 & 0.2 \\
    0.1 & 0.15 & 0.11 & 0.2 \\
    0.1 & 0.15 & 0.006 & 0.2 \\
    \hline
\end{tabular}
\end{center}

\begin{center}
\begin{tabular}{|l|l|l|l|}
    \multicolumn{4}{c}{$\mathbf{d,\bar{d}\to \gamma,h}$ and $\mathbf{d,\gamma\to d,h}$}\\
    \hline
    $m_d$ & $m_h$ & $M_W$ & $p_{in}$  \\
    \hline\hline
    0.0075 & 125 & 80.385 & 250 \\
    0.0075 & 125 & 80.385 & 125.1 \\
    125.1 & 125 & 80.385 & 95 \\
    125 & 0.0005 & 80.385 & 125.1 \\
    \hline
\end{tabular}
\end{center}

\begin{center}
\begin{tabular}{|l|l|l|l|}
    \multicolumn{4}{c}{$\mathbf{d,\bar{d}\to Z,h}$ and $\mathbf{d,Z\to d,h}$}\\
    \hline
    $m_d$ & $m_h$ & $M_W$ & $p_{in}$  \\
    \hline\hline
    0.0075 & 125 & 80.385 & 250 \\
    0.0075 & 125 & 80.385 & 125.1 \\
    125.1 & 125 & 80.385 & 95 \\
    125 & 0.0005 & 80.385 & 125.1 \\
    \hline
\end{tabular}
\end{center}

\begin{center}
\begin{tabular}{|l|l|}
    \multicolumn{2}{c}{$\mathbf{d,\bar{d}\to \gamma,\gamma}$ and $\mathbf{\gamma,\gamma\to d,\bar{d}}$ and $\mathbf{d,\gamma\to \gamma,d}$}\\
    \hline
    $m_d$ & $p_{in}$  \\
    \hline\hline
    0.0075 & 250 \\
    0.0075 & 0.001 \\
    \hline
\end{tabular}
\end{center}

\begin{center}
\begin{tabular}{|l|l|l|}
    \multicolumn{3}{c}{$\mathbf{\gamma,Z\to \bar{d},d}$ and $\mathbf{\gamma,d\to Z,d}$}\\
    \hline
    $m_d$ & $M_W$ & $p_{in}$  \\
    \hline\hline
    0.0075 & 80.385 & 250 \\
    0.0075 & 80.385 & 125.1 \\
    125 & 80.385 & 250 \\
    125 & 80.385 & 125.1 \\
    \hline
\end{tabular}
\end{center}

\begin{center}
\begin{tabular}{|l|l|l|l|}
    \multicolumn{4}{c}{$\mathbf{d,\bar{d}\to Z,Z}$ and $\mathbf{d,Z\to Z,d}$}\\
    \hline
    $m_d$ & $m_h$ & $M_W$ & $p_{in}$  \\
    \hline\hline
    0.0075 & 125 & 80.385 & 250 \\
    0.0075 & 125 & 80.385 & 125.1 \\
    125.1 & 125 & 80.385 & 95 \\
    125 & 0.0005 & 80.385 & 125.1 \\
    \hline
\end{tabular}
\end{center}

\begin{center}
\begin{tabular}{|l|l|l|l|l|}
    \multicolumn{5}{c}{$\mathbf{d,\bar{d}\to W,\bar{W}}$ and $\mathbf{d,W\to W,d}$}\\
    \hline
    $m_u$ & $m_d$ & $m_h$ & $M_W$ & $p_{in}$  \\
    \hline\hline
    0.0042 & 0.0075 & 125 & 80.385 & 250 \\
    0.0042 & 0.0075 & 125 & 80.385 & 125.1 \\
    0.0042 & 125.1 & 125 & 80.385 & 95 \\
    0.0042 & 125 & 0.0005 & 80.385 & 125.1 \\
    \hline
\end{tabular}
\end{center} 

\subsection{$\mathbf{q,\bar{q}\to b,\bar{b}}$ Neutral With Gluons}
We next consider amplitudes with two quarks and two bosons that includes at least one gluon, via a neutral channel.

\begin{center}
\begin{tabular}{|l|l|l|l|}
    \multicolumn{4}{c}{$\mathbf{u,g\to u,h}$ and $\mathbf{h,g\to u,\bar{u}}$}\\
    \hline
    $m_u$ & $m_h$ & $M_W$ & $p_{in}$  \\
    \hline\hline
    0.0042 & 125 & 80.385 & 250 \\
    0.0042 & 125 & 80.385 & 125.1 \\
    125.1 & 125 & 80.385 & 95 \\
    125 & 0.0005 & 80.385 & 125.1 \\
    \hline
\end{tabular}
\end{center}

\begin{center}
\begin{tabular}{|l|l|l|}
    \multicolumn{3}{c}{$\mathbf{g,Z\to u,\bar{u}}$ and $\mathbf{g,u\to Z,u}$}\\
    \hline
    $m_u$ & $M_W$ & $p_{in}$  \\
    \hline\hline
    0.0042 & 80.385 & 250 \\
    0.0042 & 80.385 & 125.1 \\
    125 & 80.385 & 250 \\
    125 & 80.385 & 125.1 \\
    \hline
\end{tabular}
\end{center}

\begin{center}
\begin{tabular}{|l|l|}
    \multicolumn{2}{c}{$\mathbf{g,\gamma\to u,\bar{u}}$ and $\mathbf{g,u\to \gamma,u}$}\\
    \hline
    $m_u$ & $p_{in}$  \\
    \hline\hline
    0.0042 & 250 \\
    0.0042 & 0.006 \\
    \hline
\end{tabular}
\end{center}

\begin{center}
\begin{tabular}{|l|l|}
    \multicolumn{2}{c}{$\mathbf{g,g\to u,\bar{u}}$ and $\mathbf{g,u\to g,u}$}\\
    \hline
    $m_u$ & $p_{in}$  \\
    \hline\hline
    0.0042 & 250 \\
    0.0042 & 0.005 \\
    \hline
\end{tabular}
\end{center}

We next consider two down quarks.

\begin{center}
\begin{tabular}{|l|l|l|l|}
    \multicolumn{4}{c}{$\mathbf{d,g\to d,h}$ and $\mathbf{h,g\to d,\bar{d}}$}\\
    \hline
    $m_d$ & $m_h$ & $M_W$ & $p_{in}$  \\
    \hline\hline
    0.0075 & 125 & 80.385 & 250 \\
    0.0075 & 125 & 80.385 & 125.1 \\
    125.1 & 125 & 80.385 & 95 \\
    125 & 0.0005 & 80.385 & 125.1 \\
    \hline
\end{tabular}
\end{center}

\begin{center}
\begin{tabular}{|l|l|l|}
    \multicolumn{3}{c}{$\mathbf{g,Z\to d,\bar{d}}$ and $\mathbf{g,d\to Z,d}$}\\
    \hline
    $m_d$ & $M_W$ & $p_{in}$  \\
    \hline\hline
    0.0075 & 80.385 & 250 \\
    0.0075 & 80.385 & 125.1 \\
    125 & 80.385 & 250 \\
    125 & 80.385 & 125.1 \\
    \hline
\end{tabular}
\end{center}

\begin{center}
\begin{tabular}{|l|l|}
    \multicolumn{2}{c}{$\mathbf{g,\gamma\to d,\bar{d}}$ and $\mathbf{g,d\to \gamma,d}$}\\
    \hline
    $m_d$ & $p_{in}$  \\
    \hline\hline
    0.0075 & 250 \\
    0.0075 & 0.008 \\
    \hline
\end{tabular}
\end{center}

\begin{center}
\begin{tabular}{|l|l|}
    \multicolumn{2}{c}{$\mathbf{g,g\to d,\bar{d}}$ and $\mathbf{g,d\to g,d}$}\\
    \hline
    $m_d$ & $p_{in}$  \\
    \hline\hline
    0.0075 & 250 \\
    0.0075 & 0.008 \\
    \hline
\end{tabular}
\end{center}

\subsection{$\mathbf{f,\bar{f}\to b,\bar{b}}$ Charged}
Next, we consider processes with two fermions and two bosons via a charged channel.

\begin{center}
\begin{tabular}{|l|l|l|l|}
    \multicolumn{4}{c}{$\mathbf{\bar{e},\nu_e\to W,h}$ and $\mathbf{h,\nu_e\to W,e}$}\\
    \hline
    $m_e$ & $m_h$ & $M_W$ & $p_{in}$  \\
    \hline\hline
    0.0005 & 125 & 80.385 & 2500 \\
    0.0005 & 125 & 80.385 & 250 \\
    0.0075 & 125 & 80.385 & 125.1 \\
    125.1 & 125 & 80.385 & 95 \\
    125 & 0.0005 & 80.385 & 125.1 \\
    125 & 0.0005 & 0.0004 & 125.1 \\
    \hline
\end{tabular}
\end{center}

\begin{center}
\begin{tabular}{|l|l|l|}
    \multicolumn{3}{c}{$\mathbf{\bar{e},\nu_e\to \gamma,W}$ and $\mathbf{\gamma,\nu_e\to W,e}$}\\
    \hline
    $m_e$ & $M_W$ & $p_{in}$  \\
    \hline\hline
    0.0005 & 80.385 & 250 \\
    0.0005 & 80.385 & 81 \\
    80 & 80.385 & 250 \\
    80 & 80.385 & 1 \\
    80 & 1 & 250 \\
    80 & 1 & 1 \\
    \hline
\end{tabular}
\end{center}

\begin{center}
\begin{tabular}{|l|l|l|}
    \multicolumn{3}{c}{$\mathbf{\bar{e},\nu_e\to Z,W}$ and $\mathbf{Z,\nu_e\to W,e}$}\\
    \hline
    $m_e$ & $M_W$ & $p_{in}$  \\
    \hline\hline
    0.0005 & 80.385 & 250 \\
    0.0005 & 80.385 & 90 \\
    80 & 80.385 & 250 \\
    80 & 80.385 & 70 \\
    80 & 1 & 250 \\
    80 & 1 & 1 \\
    \hline
\end{tabular}
\end{center}

\begin{center}
\begin{tabular}{|l|l|l|l|l|}
    \multicolumn{5}{c}{$\mathbf{u,\bar{d}\to W,h}$ and $\mathbf{u,h\to W,d}$}\\
    \hline
    $m_u$ & $m_d$ & $m_h$ & $M_W$ & $p_{in}$  \\
    \hline\hline
    0.0042 & 0.0075 & 125 & 80.385 & 2500 \\
    0.0042 & 0.0075 & 125 & 80.385 & 250 \\
    0.0042 & 0.0075 & 125 & 80.385 & 125.1 \\
    125.1 & 130 & 125 & 80.385 & 1 \\
    125 & 130 & 0.0005 & 80.385 & 250 \\
    125 & 130 & 0.0005 & 80.385 & 1 \\
    0.004 & 130 & 0.0005 & 0.0004 & 250 \\
    0.004 & 130 & 0.0005 & 0.0004 & 1 \\
    0.004 & 0.007 & 0.0005 & 0.0004 & 2.5 \\
    0.004 & 0.007 & 0.0005 & 0.0004 & 0.001 \\
    \hline
\end{tabular}
\end{center}

\begin{center}
\begin{tabular}{|l|l|l|l|}
    \multicolumn{4}{c}{$\mathbf{\gamma,W\to u,\bar{d}}$ and $\mathbf{\bar{u},d\to \gamma,\bar{W}}$}\\
    \hline
    $m_u$ & $m_d$ & $M_W$ & $p_{in}$  \\
    \hline\hline
    0.0042 & 0.0075 & 80.385 & 250 \\
    0.0042 & 0.0075 & 80.385 & 81 \\
    80 & 0.0075 & 80.385 & 250 \\
    80 & 0.0075 & 80.385 & 1 \\
    80 & 0.0075 & 1 & 250 \\
    40 & 10 & 1 & 50 \\
    \hline
\end{tabular}
\end{center}

\begin{center}
\begin{tabular}{|l|l|l|l|}
    \multicolumn{4}{c}{$\mathbf{g,W\to u,\bar{d}}$ and $\mathbf{d,W\to g,u}$}\\
    \hline
    $m_u$ & $m_d$ & $M_W$ & $p_{in}$  \\
    \hline\hline
    0.0042 & 0.0075 & 80.385 & 250 \\
    0.0042 & 0.0075 & 80.385 & 81 \\
    80 & 0.0075 & 80.385 & 250 \\
    80 & 0.0075 & 80.385 & 1 \\
    80 & 0.0075 & 1 & 250 \\
    80 & 0.0075 & 1 & 50 \\
    0.0042 & 0.0042 & 80.385 & 250 \\
    \hline
\end{tabular}
\end{center}

\begin{center}
\begin{tabular}{|l|l|l|l|}
    \multicolumn{4}{c}{$\mathbf{\bar{u},d\to Z,\bar{W}}$ and $\mathbf{W,d\to Z,u}$}\\
    \hline
    $m_u$ & $m_d$ & $M_W$ & $p_{in}$  \\
    \hline\hline
    0.0042 & 0.0075 & 80.385 & 250 \\
    0.0042 & 0.0075 & 80.385 & 90 \\
    80 & 0.0075 & 80.385 & 250 \\
    80 & 0.0075 & 80.385 & 70 \\
    80 & 40 & 30 & 250 \\
    80 & 40 & 1 & 1 \\
    \hline
\end{tabular}
\end{center}

\subsection{$\mathbf{b,\bar{b}\to b,\bar{b}}$}
We end with processes with four bosons.

\begin{center}
\begin{tabular}{|l|l|l|}
    \multicolumn{3}{c}{$\mathbf{h,h\to h,h}$}\\
    \hline
    $m_h$ & $M_W$ & $p_{in}$  \\
    \hline\hline
    125 & 80.385 & 250 \\
    125 & 80.385 & 0.008 \\
    \hline
\end{tabular}
\end{center}

\begin{center}
\begin{tabular}{|l|l|l|}
    \multicolumn{3}{c}{$\mathbf{h,h\to Z,Z}$ and $\mathbf{h,Z\to Z,h}$}\\
    \hline
    $m_h$ & $M_W$ & $p_{in}$  \\
    \hline\hline
    125 & 80.385 & 250 \\
    125 & 80.385 & 125.1 \\
    125 & 80.385 & 1 \\
    5 & 80.385 & 95 \\
    5 & 80.385 & 250 \\
    0.0005 & 80.385 & 95 \\
    0.0005 & 80.385 & 250 \\
    0 & 80.385 & 95 \\
    0 & 80.385 & 250 \\
    \hline
\end{tabular}
\end{center}

\begin{center}
\begin{tabular}{|l|l|l|}
    \multicolumn{3}{c}{$\mathbf{h,h\to W,\bar{W}}$ and $\mathbf{h,W\to W,h}$}\\
    \hline
    $m_h$ & $M_W$ & $p_{in}$  \\
    \hline\hline
    125 & 80.385 & 250 \\
    125 & 80.385 & 125 \\
    125 & 80.385 & 1 \\
    5 & 80.385 & 95 \\
    5 & 80.385 & 250 \\
    0.0005 & 80.385 & 95 \\
    0.0005 & 80.385 & 250 \\
    0 & 80.385 & 95 \\
    0 & 80.385 & 250 \\
    \hline
\end{tabular}
\end{center}

\begin{center}
\begin{tabular}{|l|l|l|}
    \multicolumn{3}{c}{$\mathbf{\gamma,h\to W,\bar{W}}$ and $\mathbf{\gamma,W\to W,h}$}\\
    \hline
    $m_h$ & $M_W$ & $p_{in}$  \\
    \hline\hline
    125 & 80.385 & 250 \\
    125 & 80.385 & 125 \\
    125 & 80.385 & 40 \\
    0.0005 & 80.385 & 95 \\
    \hline
\end{tabular}
\end{center}

\begin{center}
\begin{tabular}{|l|l|l|}
    \multicolumn{3}{c}{$\mathbf{Z,h\to W,\bar{W}}$ and $\mathbf{Z,W\to W,h}$}\\
    \hline
    $m_h$ & $M_W$ & $p_{in}$  \\
    \hline\hline
    125 & 80.385 & 250 \\
    125 & 80.385 & 125 \\
    125 & 80.385 & 1 \\
    0.0005 & 80.385 & 95 \\
    \hline
\end{tabular}
\end{center}

\begin{center}
\begin{tabular}{|l|l|}
    \multicolumn{2}{c}{$\mathbf{\gamma,\gamma\to W,\bar{W}}$ and $\mathbf{\gamma,W\to \gamma,W}$}\\
    \hline
    $M_W$ & $p_{in}$  \\
    \hline\hline
    80.385 & 250 \\
    80.385 & 81 \\
    \hline
\end{tabular}
\end{center}

\begin{center}
\begin{tabular}{|l|l|}
    \multicolumn{2}{c}{$\mathbf{\gamma,Z\to W,\bar{W}}$ and $\mathbf{\gamma,W\to Z,W}$}\\
    \hline
    $M_W$ & $p_{in}$  \\
    \hline\hline
    80.385 & 250 \\
    80.385 & 125 \\
    80.385 & 60 \\
    \hline
\end{tabular}
\end{center}

\begin{center}
\begin{tabular}{|l|l|l|}
    \multicolumn{3}{c}{$\mathbf{Z,Z\to Z,Z}$}\\
    \hline
    $m_h$ & $M_W$ & $p_{in}$  \\
    \hline\hline
    125 & 80.385 & 250 \\
    125 & 80.385 & 2500 \\
    125 & 80.385 & 125 \\
    125 & 80.385 & 1 \\
    \hline
\end{tabular}
\end{center}

\begin{center}
\begin{tabular}{|l|l|l|}
    \multicolumn{3}{c}{$\mathbf{Z,Z\to W,\bar{W}}$ and $\mathbf{Z,W\to Z,W}$}\\
    \hline
    $m_h$ & $M_W$ & $p_{in}$  \\
    \hline\hline
    125 & 80.385 & 250 \\
    125 & 80.385 & 125 \\
    125 & 80.385 & 1 \\
    1 & 80.385 & 250 \\
    1 & 80.385 & 125 \\
    1 & 80.385 & 1 \\
    125 & 50 & 125 \\
    125 & 50 & 1 \\
    1 & 50 & 125 \\
    1 & 50 & 1 \\
    \hline
\end{tabular}
\end{center}

\begin{center}
\begin{tabular}{|l|l|l|}
    \multicolumn{3}{c}{$\mathbf{W,W\to W,W}$ and $\mathbf{W,\bar{W}\to W,\bar{W}}$}\\
    \hline
    $m_h$ & $M_W$ & $p_{in}$  \\
    \hline\hline
    125 & 80.385 & 250 \\
    125 & 80.385 & 125 \\
    125 & 80.385 & 1 \\
    1 & 80.385 & 250 \\
    1 & 80.385 & 125 \\
    1 & 80.385 & 1 \\
    125 & 50 & 125 \\
    125 & 50 & 1 \\
    1 & 50 & 125 \\
    1 & 50 & 1 \\
    \hline
\end{tabular}
\end{center}

\begin{center}
\begin{tabular}{|l|}
    \multicolumn{1}{c}{$\mathbf{g,g\to g,g}$}\\
    \hline
    $p_{in}$  \\
    \hline\hline
    250 \\
    \hline
\end{tabular}
\end{center}

\end{document}